\documentclass[%
superscriptaddress,
reprint,
preprintnumbers,
amsmath,amssymb,prx,longbibliography,nofootinbib]{revtex4-2}

\usepackage{bm}% bold math
\usepackage[colorlinks=true,linktocpage=true,linkcolor=blue,citecolor=blue,urlcolor=blue,breaklinks]{hyperref} 

\usepackage{cleveref} 

\usepackage[utf8]{inputenc} % To support utf8 characters
\usepackage[T1]{fontenc} % T1 font encoding
\usepackage{microtype}

\usepackage[mathscr]{euscript}
\usepackage{relsize}

\newcommand{\half}{\frac{1}{2}}
\newcommand{\nn}{\nonumber}

\newcommand{\lie}{\pounds}
\newcommand{\df}{\mathrm{d}}
\newcommand{\dow}{\partial} 

\usepackage{mathtools}

\usepackage{soul}

\usepackage{xparse}

\ExplSyntaxOn

\clist_map_inline:nn{A,B,C,D,E,F,G,H,I,J,K,L,M,N,O,P,Q,R,S,T,U,V,W,X,Y,Z}
{
  \expandafter\def\csname b#1\endcsname{\bm{#1}} % e.g. \bA -> \bm{A}
  
  \expandafter\def\csname c#1\endcsname{\mathcal{#1}} % e.g. \cA -> \mathcal{A}
  
  \expandafter\def\csname bc#1\endcsname{\bm{\mathcal{#1}}} % e.g. \bcA -> \bm{\mathcal{A}}
  
  \expandafter\def\csname s#1\endcsname{{\mathsmaller{#1}}} % e.g. \sA -> {\mathsmaller{A}}
  
  \expandafter\def\csname bb#1\endcsname{\mathbb{#1}} % e.g. \bbA -> \mathbb{A}
  
  \expandafter\def\csname rm#1\endcsname{\mathrm{#1}} % e.g. \rmA -> \mathrm{A}
  
  \expandafter\def\csname sc#1\endcsname{\mathscr{#1}} % e.g. \scA -> \mathscr{A}
  
  \expandafter\def\csname sf#1\endcsname{\mathsf{#1}} % e.g. \sfA -> \mathsf{A}
  
  \expandafter\def\csname f#1\endcsname{\mathfrak{#1}} % e.g. \fA -> \mathfrak{A}
}

\ExplSyntaxOff

\newcommand{\lB}{\left [}
\newcommand{\rB}{\right ]}
\newcommand{\lb}{\left (}
\newcommand{\rb}{\right )}

\newcommand\ext{\text{ext}}

\newcommand\eqb{\text{eqb}}
\newcommand\loc{\text{loc}}

\usepackage{blindtext}
\usepackage[bbgreekl]{mathbbol}

\renewcommand*{\arraystretch}{1.2}
\usepackage{tabularx}

\newcommand\tightmath[1]{{%
    \everymath{\medmuskip=2.5mu minus 2.5mu\thickmuskip=3mu minus 3mu}%
    {#1}%
}}

\renewcommand{\thesection}{\arabic{section}}

\makeatletter
\renewcommand{\p@subsection}{}
\renewcommand{\p@subsubsection}{}
\makeatother

\makeatletter
\def\l@subsubsection#1#2{}
\makeatother

\numberwithin{equation}{section}

\begin{document} 

\title{Approximate higher-form symmetries, topological defects, \\ and dynamical phase transitions}

\author{Jay Armas}\email{j.armas@uva.nl}
\author{Akash Jain}\email{ajain@uva.nl}

\affiliation{Institute for Theoretical Physics, University of Amsterdam, 1090
  GL Amsterdam, The Netherlands}
\affiliation{Dutch Institute for Emergent Phenomena, 1090 GL Amsterdam, The Netherlands}

%\date{\today}

\begin{abstract}
Higher-form symmetries are a valuable tool for classifying topological phases of matter. However, emergent higher-form symmetries in interacting many-body quantum systems are not typically exact due to the presence of topological defects. In this paper, we develop a systematic framework for building effective theories with approximate higher-form symmetries, i.e. higher-form symmetries that are weakly explicitly broken. We focus on a continuous U(1) $q$-form symmetry and study various patterns of symmetry breaking. This includes spontaneous or explicit breaking of higher-form symmetries, as well as pseudo-spontaneous symmetry breaking patterns where the higher-form symmetry is both spontaneously and explicitly broken. We uncover a web of dualities between such phases and highlight their role in describing the presence of dynamical higher-form vortices. In order to study the out-of-equilibrium dynamics of these phases of matter, we formulate respective hydrodynamic theories and study the spectra of excitations exhibiting higher-form charge relaxation and Goldstone relaxation effects. We show that our framework is able to describe various phase transitions due to proliferation of vortices or defects. This includes the melting transition in smectic crystals, the plasma phase transition from polarised gases to magnetohydrodynamics, the spin-ice transition, the superfluid to neutral fluid transition and the Meissner effect in superconductors, among many others. 

\end{abstract} 

\pacs{Valid PACS appear here}

\maketitle

{
\parskip=0.1\baselineskip \advance\parskip by 0pt plus 0.1pt
\setcounter{tocdepth}{1}
\tableofcontents
}

\section{The higher-form of life}

Identifying the symmetries underlying fundamental interactions and emergent collective phenomena continues to be one of the most important and interesting problems in theoretical physics. Symmetries provide a powerful characterisation of phases of matter at all length scales, paving the way for constraining effective field theories governing the dynamics of a large variety of physical systems. Examples include the quark-gluon plasma, astrophysical plasma, quantum many-body systems, active matter, and liquid crystals. 

In recent years, exotic generalised notions of symmetry have received considerable attention, including higher-form symmetries, higher-group symmetries, subsystem symmetries, and non-invertible symmetries; see~\cite{Cordova:2022ruw} for a review. They have been particularly useful in the context of topological phases of matter for which conventional notions of symmetry cannot provide an understanding in terms of the Landau paradigm~\cite{McGreevy:2022oyu}.  An interesting case to highlight is that of U(1) $q$-form symmetries~\cite{Gaiotto:2014kfa}, denoted as U(1)$_q$, characterised by a conserved $(q+1)$-form current $J$ and associated non-local order parameters constructed from $q$-dimensional charged operators. 
The best explored example of such symmetries comes from electromagnetism in 3 spatial dimensions, which has two 1-form symmetries: a ``magnetic'' U(1)$_1$ symmetry whose charged objects are 't Hooft lines (magnetic field lines) and another ``electric'' U(1)$_1$ symmetry in the absence of free charges whose charged objects are Wilson lines (electric field lines). 
Higher-form symmetries not only allow for classifying novel phases of matter but also provide new organising principles for phases with conventional symmetries. This includes phases of hot electromagnetism such as magnetohydrodynamics, characterised by magnetic U(1)$_1$ symmetry, and polarised plasma, characterised by magnetic and electric $\rmU(1)_1\times\rmU(1)_1$ symmetry~\cite{Armas:2018atq, Armas:2018zbe}. 
Other examples include the theory of elasticity in 2 spatial dimensions which can be recast in terms of a U(1)$_1$ symmetry~\cite{Armas:2019sbe} and the theory of superfluidity which can be written in terms of a $\rmU(1)_0 \times \rmU(1)_1$ symmetry~\cite{Delacretaz:2019brr}. A systematic exploration of the applications of generalised symmetries is an exciting frontier of modern physics.

The broad scope of applications of higher-form symmetries makes them an ideal tool for classifying phases of matter and studying phases transitions. However, most systems in nature do not have exact higher-form symmetries. In fact most systems do not have exact conventional 0-form symmetries either. For instance, the crystalline phase of matter is characterised by a pseudo-spontaneous pattern of symmetry breaking, in which translation symmetry is both spontaneously as well as weakly explicitly broken due to the presence of impurities. In a recent series of papers~\cite{Armas:2021vku, Armas:2022vpf}, we showed that even when 0-form symmetries are only approximate due to weak explicit breaking, it is still possible to significantly constrain the low energy effective theory and extend the Landau paradigm to realistic situations. The primary goal of this paper is to generalise this framework to approximate higher-form symmetries covering a plethora of physical systems.

Approximate higher-form symmetries are ubiquitous in nature. For instance, consider a smectic crystal in two spatial dimensions, which is characterised by a Goldstone field $\phi$ arising due to spontaneous breaking of translational symmetry along one of the spatial directions. If the translation order is exact, $\phi$ is a smooth field and satisfies $\partial_{[\mu}\partial_{\nu]}\phi=0$. This ``Bianchi-identity'' for the Goldstone field can be redecorated as a 1-form conservation equation $\dow_\mu \tilde J^{\mu\nu}=0$ for a 2-form current $\tilde J^{\mu\nu} = \epsilon^{\lambda\mu\nu}\partial_\lambda \phi$. 
Therefore a smectic crystal has a global 1-form symmetry with an associated conserved charge that counts the number of lattice lines piercing a given codimension-1 surface. The generalisation to isotropic crystals is straight-forward and results in a 1-form symmetry in every spatial direction~\cite{Armas:2019sbe}. 
According to the theory of melting of two-dimensional crystals~\cite{1979PhRvB..19.2457N, 1980PhRvB..22.2514Z}, increasing temperature leads to translational disorder due to spontaneous formation of localised topological defects called dislocations.
This causes the Goldstone field $\phi$ become singular, modifying the Bianchi identity to  $2\partial_{[\mu}\partial_{\nu]}\phi = - \tilde\ell \epsilon_{\lambda\mu\nu}\tilde L^\lambda$, where $\tilde L^\mu$ is the ``dislocation current'' and $\tilde\ell$ is a small parameter that controls the strength of dislocations. This leads to a violation of the 1-form conservation law $\dow_\mu \tilde J^{\mu\nu}= - \tilde\ell \tilde L^\nu$. 
Thus, as a \emph{phase of matter}, a 2-dimensional smectic crystal with dislocations is characterised by an approximate 1-form symmetry and an emergent topological 0-form symmetry $\dow_\mu \tilde L^\mu=0$ arising from the modified Bianchi identity. The conserved charge associated with this latter symmetry counts the number of dislocations. If we melt the crystal by proliferating dislocations, the 1-form symmetry is strongly violated, giving rise to a fluid with spontaneously-restored translational symmetry.
In the absence of external gauge fields, the same construction holds for superfluids with vortices in two-spatial dimensions, in which case the Goldstone field $\phi$ arises due to spontaneous symmetry breaking of a U(1)$_0$ symmetry instead.

\begin{figure*}[t]
    \centering
    \includegraphics[width=\textwidth]{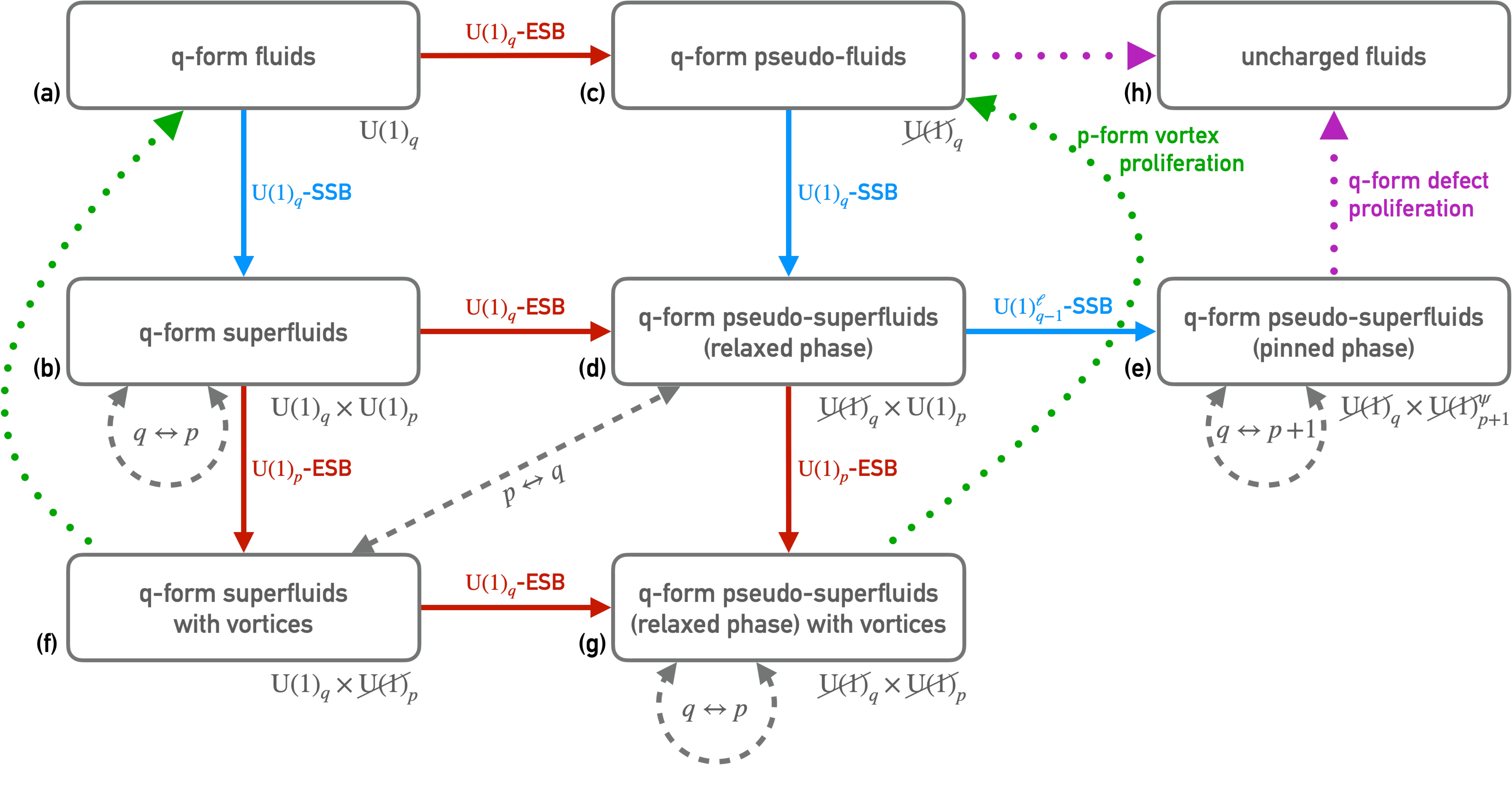}
    \caption{Phases with approximate higher-form symmetry.}
    \label{fig:chart}
\end{figure*}

Returning to the example of electromagnetism, the polarised plasma phase (also, free electromagnetism in vacuum) has electric and magnetic $\rmU(1)_1\times\rmU(1)_1$ symmetry, with the associated currents $J^{\mu\nu} = - 1/g^2\,\xi^{\mu\nu} + {\cal M}^{\mu\nu}$ and $\tilde J^{\mu\nu} = \half\epsilon^{\mu\nu\rho\sigma}\xi_{\rho\sigma}$. Here $\xi_{\mu\nu} = 2\dow_{[\mu}\phi_{\nu]}$ denotes the electromagnetic field strength, $\phi_\mu$ the dynamical gauge field or photon, ${\cal M}^{\mu\nu}$ the polarisation tensor, and $g$ the electromagnetic coupling constant.\footnote{We use the notation $\xi_{\mu\nu}$ and $\phi_\mu$ for the electromagnetic field strength and gauge field instead of the conventional $F_{\mu\nu}$ and $A_\mu$ because, as we shall discuss further, the polarised plasma phase of electromagnetism can be understood as a 1-form superfluid with the 1-form $\phi_\mu$ playing the role of the associated Goldstone and $\xi_{\mu\nu}$ its ``superfluid velocity''.} 
Explicitly breaking the magnetic U(1)$_1$ symmetry leads to magnetic monopoles in the theory, causing $\dow_\mu \tilde J^{\mu\nu} = - \tilde\ell \tilde L^\mu$, where $\tilde L^\nu$ can be seen as the magnetic monopole current. While no fundamental monopoles have been observed in nature, this model is still useful for the phenomenology of emergent magnetic monopoles observed in spin ice~\cite{Castelnovo:2007qi} and anomalous Hall effect~\cite{Fang:2003ir}. 

On the other hand, explicit breaking of the electric U(1)$_1$ symmetry can be understood as introducing free charges in the theory that screen the Wilson lines. This results in the violation of the associated conservation law (Maxwell's equations) $\dow_\mu J^{\mu\nu} = -\ell L^\nu$, where $L^\mu$ can be seen as the electric charge current. 
In fact, this phase can be further fine-grained depending on the fate of the emergent topological U(1)$^\ell_0$ symmetry $\dow_\mu L^\mu = 0$; the superscript $\ell$ is to distinguish this from the original higher-form global symmetry. For electromagnetism, this is precisely the dynamical U(1) symmetry associated with conservation of electric charges.
If this emergent symmetry is spontaneously-unbroken, we reside in the Coulomb phase of electromagnetism with massless photon $\phi_\mu$. Whereas, if it is spontaneously broken leading to a Goldstone phase $\phi_\ell$, we reside in the Higgs phase where the photon $\phi_\mu$ eats the Goldstone to become $\psi_\mu = \ell(\phi_\mu - \dow_\mu\phi_\ell)$ and acquires a mass, resulting in the theory of superconductivity. In terms of symmetries, the Higgs phase is characterised by an explicitly broken $\rmU(1)_1\times\rmU(1)_{2}^\psi$ symmetry, where the non-conserved current associated with the latter part of the symmetry group is merely $\tilde J^{\mu\nu\rho}_\psi = \epsilon^{\lambda\mu\nu\rho}\psi_\lambda$, which satisfies the approximate conservation equation $\dow_\mu \tilde J^{\mu\nu\rho}_\psi = -\ell\tilde J^{\nu\rho}$. The superscript $\psi$ is to distinguish this symmetry from the original higher-form symmetries in the system. The original magnetic U(1)$_1$ symmetry now arises as an emergent topological symmetry due to the breaking of U(1)$^\psi_2$ symmetry. 

\vspace{1em}
\noindent
\emph{Phases of approximate higher-form symmetry.}---Various phases of matter described above can be organised using spontaneous and explicit breaking of higher-form symmetries; see \cref{fig:chart}. We start with a $q$-form fluid with a U(1)$_q$ symmetry in \cref{fig:chart}(a),\footnote{At finite temperature, for $q>0$, this symmetry needs be partially-spontaneously broken in the time-direction to allow for a finite $q$-form density~\cite{Armas:2018atq, Armas:2018zbe}.} endowed with some $(q+1)$-form conserved current $J$. This can describe ordinary charged fluids for $q=0$, smectic crystals for $q=d-1$, and magnetohydrodynamics for $q=d-2$, where $d$ is the number of spatial dimensions.
We can spontaneously break this U(1)$_q$ symmetry by introducing a $q$-form Goldstone field $\phi$, giving rise to a U(1)$_q$ superfluid in \cref{fig:chart}(b). This phase also has a new emergent U(1)$_p$ symmetry, with \tightmath{$p=d-1-q$} and the $(p+1)$-form current $\tilde J = \star\df\phi$, which is conserved due to the Bianchi identity associated with $\phi$. Here $\star$ denotes the spacetime Hodge duality operation. This phase describes a U(1)$_0$ superfluid for $q=0$ and polarised plasma for $q=1$.

Starting from a $q$-form fluid, we can also break the U(1)$_q$ symmetry explicitly by introducing a $q$-form defect current $L$, resulting in a $q$-form pseudo-fluid in \cref{fig:chart}(c) with relaxed $q$-form charges. This can describe particle number violating interactions in a relaxed charged fluid for $q=0$, dislocations in a smectic crystal for $q=d-1$, and magnetic monopoles in magnetohydrodynamics for $q=d-2$.
One can also break the U(1)$_q$ symmetry both explicitly and spontaneously, colloquially called pseudo-spontaneous symmetry breaking, which results in a $q$-form pseudo-superfluid in \cref{fig:chart}(d) where $q$-form charges are relaxed but $q$-form Goldstone $\phi$ is not relaxed. For $q=0$, this describes an ordinary relaxed superfluid, whereas for $q=1$, this describes electromagnetism in the Coulomb phase. 
In fact, explicitly breaking the U(1)$_q$ symmetry gives rise to an emergent explicitly-unbroken topological U(1)$^\ell_{q-1}$ symmetry associated with the defect current $L$. If we further spontaneously break this emergent symmetry by introducing a $(q-1)$-form Goldstone $\phi_\ell$, we arrive at a pinned U(1)$_q$ pseudo-superfluid in \cref{fig:chart}(e) where $q$-form charges and $q$-form Goldstone $\phi$ are both relaxed. This phase is characterised by a massive pseudo-Goldstone field $\psi = \ell(\phi-\df\phi_\ell)$ and describes an ordinary pinned superfluid for $q=0$~\cite{Armas:2021vku} and the Higgs phase of electromagnetism or a superconductor for $q=1$. 

Starting from the U(1)$_q$ superfluid in \cref{fig:chart}(b), we can instead explicitly break the U(1)$_p$ symmetry. The associated $p$-form defects are understood as vortices in a $q$-form superfluid in \cref{fig:chart}(f). For $q=0$, this phase describes ordinary superfluid vortices, while for $q=1$, this is the theory of magnetic monopoles in electromagnetism. Finally, we can envision breaking both U(1)$_q$ and U(1)$_p$ symmetries together and describe a theory of $q$-form psuedo-superfluids with vortices. For $q=1$, this would be electromagnetism with both electric and magnetic free charges (monopoles). As is well-known, this is not something we can do consistently at zero temperature in Maxwell's electromagnetism, but as it turns out, we can indeed realise this possibility at finite temperature as we shall see in the course of our discussion.

The U(1)$_q$ and U(1)$_p$ higher-form symmetries of a $q$-form (pseudo-)superfluid actually have a mixed anomaly between them, which plays a crucial role while organising the phase space of approximate higher-form symmetries. Similarly, there is also a mixed anomaly between the U(1)$_q$ and U(1)$_{p+1}^\psi$ higher-form symmetries of the U(1)$_q$ pseudo-superfluid in the pinned phase. The respective anomaly coefficient in both these cases is nothing but the charge $c_\phi$ of the U(1)$_q$ Goldstone field.

% The generalization of this discussion to spontaneously broken U(1)$_q$ higher-form symmetries with associated q-form Goldstone field allows to capture many more physical systems with approximate symmetries including spin ices with emergent magnetic monopoles and polarized gasses with free electric charges viewed as topological defects. 
% In fact, analogously to the case discussed above, if the spontaneously broken U(1)$_q$ symmetry is also weakly broken in the case q>0, there is an emergent U(1)$_{q-1}$ symmetry associated with the conservation of topological defects.
% In such generic situations two phases can be distinguished. If the emergent U(1)$_{q-1}$ symmetry is not spontaneously broken such phases can be viewed as Coulomb phases of U(1)$_{q-1}$  gauge theory.

\vspace{1em}
\noindent
\emph{Dualities.}---Looking at the symmetry structure of the phase space in \cref{fig:chart}, we can identify a few dualities. First of all, a $q$-form superfluid in \cref{fig:chart}(b) is dual to a $p$-form superfluid, and consequently is self-dual when $p=q$ in $d=2q+1$ spatial dimensions. This is just the generalisation of the electromagnetic duality in $3$ spatial dimensions in the absence of free charges that exchanges electric and magnetic fields. In 2 spatial dimensions, this also realises the electromagnetism/superfluid duality that exchanges electric fields for superfluid velocity and magnetic fields for superfluid charge. Under the same duality, the relaxed phase of a $q$-form pseudo-superfluid in \cref{fig:chart}(d) maps to a $p$-form superfluid with vortices \cref{fig:chart}(f), and vice-versa, and a relaxed $q$-form superfluid with vortices in \cref{fig:chart}(g) maps to its $p$-form version with the roles of $p$-form vortices and $q$-form defects exchanged. This maps electric and magnetic free charges to each-other in the context of electromagnetic duality. In the context of electromagnetism/superfluid duality, this maps free electric charges to superfluid vortices (also called the particle/vortex duality) and free magnetic charges, if present, to the sources of superfluid charge relaxation. 

Interestingly, we also have a different duality in the pinned phase of a $q$-form pseudo-superfluid in \cref{fig:chart}(e) with its $(p+1)$-form version, making it self dual in $d=2q$ spatial dimensions. This means that a superconductor is self-dual in 2 spatial dimensions, with electric fields exchanged with massive gauge fields and free electric charges exchanged with magnetic fields. In 1 spatial dimension, this also yields a duality between superconductors and pinned superfluids, mapping electric fields to massive pseudo-Goldstones, free electric charges to superfluid velocity, massive gauge fields to superfluid charges, and magnetic fields to charge relaxation sources. We will explore these dualities in more detail in the bulk of the paper.

% Due to electromagnetic duality and more generally particle/vortex duality in two-spatial dimensions, or brane/defect duality in general dimensions, phases of matter with this symmetry breaking pattern describe dynamical vortices/topological defects.  
% On the other hand, if the emergent U(1)$_{q-1}$ symmetry is spontaneously broken the phase of matter is akin to the Higgs phase of gauge theory. In this context, the q-form Goldstone field acquires a mass. In the case of U(1)$_0$ symmetries massive Goldstones describe pinned Wigner crystals or pinned charge density waves, among other systems. 
% In the 1-form case, for instance, it describes a superconductor for which, due to the Anderson-Higgs mechanism, the photon field becomes massive. 

% In this paper we also approach such phases systematically for arbitrary q-form symmetries. In fact, given the presence of external gauge fields to which these field theories couple to, it is possible to view these systems as being endowed with an anomalous $\text{U(1)}_q \times \text{U(1)}_p$ global symmetry with $q+p=d-1$ and $d$ the number of spatial dimensions. This web of dualities is depicted in Figure XXX.

% Anomaly

\vspace{1em}
\noindent
\emph{Phase transitions.}---In this work, our interest lies not only in classifying the phases of matter according to their higher-form symmetry breaking pattern, but also to understand the out-of-equilibrium dynamics and transitions between these phases. 
To this aim, we draw motivation from our previous work on approximate 0-form symmetries and vortices~\cite{Armas:2021vku, Armas:2022vpf}, and construct hydrodynamic theories at finite temperature for phases with spontaneously and/or explicitly broken higher-form symmetry. This allows us to identify potential phase transitions as guided by the proliferation of $p$-form defects (vortices) or $q$-form defects (charge relaxation sources), which have been illustrated in \cref{fig:chart}. The first class of such transitions is the transition from $q$-form superfluids (or $q$-form pseudo-superfluids in relaxed phase) to $q$-form fluids (or $q$-form pseudo-fluids). This is mediated by the proliferation of $p$-form defects (vortices) in the $q$-form superfluid phase and restores the previously spontaneously-broken (approximate) U(1)$_q$ symmetry. The obvious application of this for $q=0$ is the transition from ordinary 0-form superfluids (with relaxation) to ordinary 0-form fluids (with relaxation), mediated by vortices in the scalar superfluid phase. Taking $q=d-2$, the phase transition from polarised plasma to magnetohydrodynamics also falls in this class, mediated by free electric charges playing the role of vortices of the $(d-2)$-form dual magnetic photon. Theoretically, taking $q=1$, we can also get a transition from the polarised plasma phase to electrohydrodynamics, mediated by magnetic monopoles playing the role of vortices of the 1-form photon.

There is a similar class of transitions from a $q$-form fluid to a fluid with no higher-form symmetry, mediated by $q$-form defects. For $q=0$, these are transitions from charged to neutral fluids due to the proliferation of charge-violating interactions. For $q=d-1$, these also include the melting phase transition from crystals to fluids, mediated by $(d-1)$-form dislocations.\footnote{Note that this does not necessarily mean that melting in arbitrary spatial dimensions is guided by the proliferation of vortices. While this is known to be true in 2 spatial dimensions, in arbitrary dimensions this just means that proliferation of vortices would contribute to melting but might not be the primary mechanism as we increase temperature.} Finally, for $q=d-2$, these include transitions from magnetohydrodynamics to neutral fluids due to the proliferation of $(d-2)$-dimensional magnetic monopoles. A phenomenological application of the final case is the spin-ice phase transition. 

The final class of phase transitions is from the pinned phase of a U(1)$_q$ superfluid to a neutral fluid. This is also mediated by the proliferation of $q$-form defects, but due to spontaneously broken $U(1)^\ell_{q-1}$ symmetry, also gaps out the massive $q$-form pseudo-Goldstone field $\psi = \ell(\phi - \df\phi_\ell)$ from the theory. This can be used to model the Meissner effect in superconductors, where the gapped massive photon removes all electromagnetic field excitations from the material, effectively leaving only a theory of neutral excitations inside the superconductor.

% These include well studied transitions such as the melting of smectic crystals, superfluid to fluid transitions, and superconductor to fluid transitions, but also less studied transitions such as the dynamics of ionization and the transition from polarized gases to plasmas governed by magnetohydrodynamics or the transition from spin ice to fluid phases via the proliferation of magnetic monopoles. 

\vspace{1em}
\noindent
\emph{Damping-attenuation relations.}---
For these broad class of phases of matter, we use our hydrodynamic theory to investigate the linearised spectrum of excitations and find novel physical effects related to charge relaxation, Goldstone relaxation, and pinning effects. In particular, we find that all phases with explicitly broken higher-form symmetries feature modes satisfying damping-attenuation relations of the kind
\begin{equation}
    \Gamma = D k_0^2.
\end{equation}
Here $\Gamma$ is the damping or relaxation rate of the mode, $D$ its attenuation or diffusion constant, and $1/k_0$ characterises some finite hydrostatic correlation length associated with the degrees of freedom that carry this mode. Such relations are a hallmark feature of systems with symmetries that are both spontaneously as well as explicitly broken, i.e. pseudo-spontaneously broken, and have already been derived in a variety of systems with 0-form symmetries such as pinned superfluids and pinned crystals~\cite{Armas:2021vku, Delacretaz:2021qqu}.\footnote{For $q=0$, no such damping-attenuation relation exists for relaxed 0-form pseudo-fluids; see \cref{fig:chart}(c). This is because the U(1)$_0$ symmetry in this phase is only explicitly broken and not spontaneously broken. However for $q>0$, this phase features partial-spontaneous breaking of the U(1)$_q$ symmetry in the time-direction, yielding the damping-attenuation relations.} 

We find such relations in the propagation of $p$-form defects (vortices) in $q$-form superfluids, including vortices in ordinary 0-form superfluids, and $q$-form defects in the relaxed phase of $q$-form pseudo-superfluids. In this context, $k^2_0 = \ell^2\chi_\ell/\chi$ is the inverse-squared correlation length of $q$-form defects, with $\ell^2\chi_\ell$ being their susceptibility and $\chi$ the susceptibility of $q$-form charge, and similarly for the $p$-form defects (vortices). The susceptibility for $p$-form charge is inverse of the superfluid density.
Using $q=1$, the same relation also arises for the propagation of free electric charges in electromagnetism with $k_0$ being the inverse Debye-screening length, $\ell^2\chi_\ell$ the susceptibility of free charges, and $\chi$ the permittivity of electric fields, and similarly for magnetic charges in the presence of magnetic monopoles. In the pinned phase of $q$-form pseudo-superfluids, we still get the damping-attenuation relations for $q$-form defects, but we get another one for the propagation of superfluid velocity (or $p$-form charge). In this context, $(k^\phi_0)^2 = \ell^2m^2\tilde\chi$ represents the pinning momenta squared, where $\tilde\chi$ is the susceptibility of $p$-form charge (or inverse the superfluid density) and $m$ is the mass of the pseudo-Goldstone. Correspondingly in superconductivity, $k_0^\phi$ is the inverse London penetration depth, $\tilde\chi$ the permeability of magnetic fields, and $m$ the mass of the massive gauge fields.

% Perhaps the $q=1$ case is better to relate

% \begin{equation}
%     \dow_i E^i = \rho/\epsilon_0.
%     \qquad
%     \dow_t \rho 
%     = \sigma \dow_i (\dow^i\mu - E^i)
%     = D \dow^2 n - \sigma/\epsilon_0 n.
% \end{equation}
% \begin{equation}
%     \Gamma = \sigma/\epsilon_0 = \chi/\epsilon_0 D.
% \end{equation}

\vspace{1em}
\noindent
\emph{Organisation of the paper.}---This paper is organised as follows. We start in \cref{sec:higher-form-general} with a review of the basics of higher-form symmetries and discuss how to explicitly break them by introducing appropriate background sources.
In \cref{sec:zero-temperature-higher-form} we discuss different patterns of spontaneous and explicit breaking of higher-form symmetries at zero temperature using an action principle.
% We begin by spontaneously breaking a global U(1)$_q$ symmetry and writing an effective action for the resulting higher-form superfluid in terms of the associated q-form Goldstone field. We note that spontaneously breaking the U(1)$_q$ symmetry leads to an emergent "magnetic" symmetry U(1)$_p$ in the absence of external background gauge fields. In the presence of external fields, this allows us to view a U(1)$_q$ superfluid as a phase of matter with an anomalous global $\text{U(1)}_q\times \text{U(1)}_p$ symmetry. In this same section we discuss a \emph{pseudo-spontaneous} pattern of symmetry breaking in which the original U(1)$_q$ global symmetry is both spontaneously and explicitly broken. 
% In this context, we distinguish between the Coulomb phase and the Higgs phase, with the former describing vortices via particle/vortex or brane/defect duality and the later describing pseudo-Goldtones (i.e. massive Goldstone fields). 
In \cref{sec:higher-form-finite-temperature} we consider the same phase-space of spontaneously and explicitly broken higher-form symmetries in thermal equilibrium using thermal partition functions.
An important result that rises from this section is that, at finite temperature, higher-form symmetries must at least be partially-spontaneously broken in the time-direction to allow for nonzero thermodynamic density of the respective higher-form charge.
In \cref{sec:higher-form-hydro}, we outline the hydrodynamic theory for $q$-form (pseudo-)fluids with temporal-(pseudo-)spontaneous symmetry breaking and in \cref{sec:higher-form-Coulomb,sec:higher-form-Higgs} we discuss $q$-form (pseudo-)superfluids with complete-(pseudo-)spontaneous symmetry breaking in the relaxed and pinned phases respectively. We employ the hydrodynamic theories in these sections to compute the linearised mode spectra in the respective phases and explore transitions between different phases. Finally, in \cref{sec:outlook} we end with a discussion of our results. We also provide several appendices. As the paper relies heavily on the differential forms we have summarised our conventions in \cref{app:form-conventions}. In \cref{app:anomaly-inflow}, we give details of the anomaly-inflow mechanism for higher-form symmetries used in the core of the paper. In \cref{app:correlators} we give the details of various retarded correlation functions derived from our hydrodynamic construction.

\section{Introduction to higher-form symmetries}
\label{sec:higher-form-general}

We start our discussion with a pedagogical overview of systems with higher-form symmetries~\cite{Gaiotto:2014kfa}. There is a vast amount of work in the literature exploring the intricacies of higher-form symmetries in great detail; see~\cite{Cordova:2022ruw} for a recent review. We will only touch upon certain practical aspects of higher-form symmetries that are relevant for studying out-of-equilibrium dynamics. In particular, we will introduce a controlled perturbative procedure to break higher-form symmetries and discuss its physical implications. 

\subsection{Higher-form symmetries}

\begin{figure}[t]
    \centering
    \includegraphics[width=0.5\linewidth]{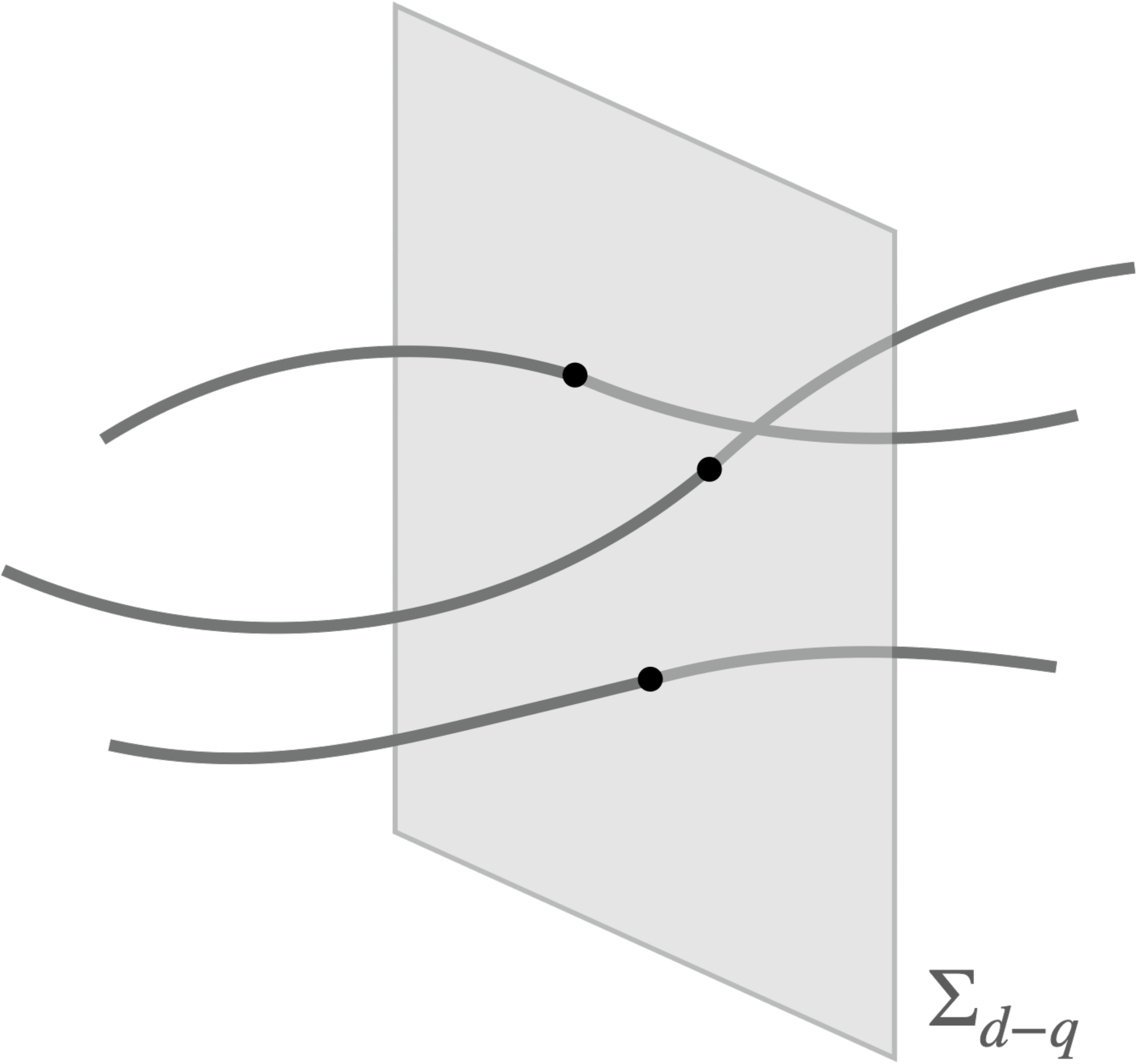}
    \caption{A snapshot of a system with a $q$-form U(1)$_q$ symmetry. The charged operators are $q$-dimensional extended objects (strings in this illustration). The total U(1)$_q$ charge can be obtained by counting the number of operators crossing a spacelike surface $\Sigma_{d-q}$. This charge remains conserved under smooth deformations of $\Sigma_{d-q}$.}
    \label{fig:strings}
\end{figure}

A physical system is said to admit a continuous $q$-form U(1) generalised global symmetry~\cite{Gaiotto:2014kfa}, which we refer to as U(1)$_q$, if it admits a \tightmath{$(q+1)$}-form conserved current $J$ satisfying a set of conservation
equations
\begin{align}
  % \dow_\mu J^{\mu\nu_1\ldots\nu_q} &= 0 \nn\\
  % \text{or}\quad 
  \df{\star J} &= 0.
  \label{eq:higherform-cons}
\end{align}
For $q=0$, we recover the ordinary ``0-form'' U(1)$_0$ symmetries, with the familiar conservation equation $\dow_\mu J^\mu = 0$. 
Just like point-particles are charged under a U(1)$_0$ symmetry, the operators charged under a U(1)$_q$ symmetry are $q$-dimensional extended objects. The total conserved charge associated with a U(1)$_0$ symmetry can be defined by integrating the current $J^\mu$ over a spacelike hypersurface $\Sigma_d$, which counts the number of particle worldlines crossing the hypersurface. This charge remains ``conserved'' under smooth deformations of $\Sigma_d$, in particular time-translations. Similarly, the total conserved charge associated with a U(1)$_q$ symmetry can be defined by integrating the associated current over a \tightmath{$(d-q)$}-dimensional spacelike surface $\Sigma_{d-q}$, i.e.
\begin{equation}
    Q[\Sigma_{d-q}] 
    = \int_{\Sigma_{d-q}} {\star J},
    % = \int \df \Sigma_{\mu_1\ldots\mu_{q+1}}
    % J^{\mu_1\ldots\mu_{q+1}},
\end{equation}
which counts the number of charged operator worldsheets that intersect $\Sigma_{d-q}$; see \cref{fig:strings}. 
The fact that this charge is conserved amounts to the statement that $Q[\Sigma_{d-q}]$ is invariant under smooth deformations of the surface $\Sigma_{d-q}$ over which it is defined, keeping the boundary fixed.

Note that the notion of higher-form conservation is qualitatively more general than its 0-form counterpart. 
In particular, the U(1)$_q$ charge $Q[\Sigma_{d-q}]$ is conserved not only under time-translations of the surface $\Sigma_{d-q}$ over which it is defined, but also under translations in any of the $q$ spatial directions transverse to $\Sigma_{d-q}$. This can also be seen by decomposing the U(1)$_q$ conservation equations \eqref{eq:higherform-cons} into space and time components as
\begin{subequations}
\begin{align}
    \dow_t J^{tk_1\ldots k_{q}}
    + \dow_i J^{ik_1\ldots k_{q}} &= 0, 
    \label{eq:cons-dyn} \\
    \dow_i J^{tik_2\ldots k_{q}} &= 0.
    \label{eq:cons-gauss}
\end{align}
\label{eq:non-cov-cons}%
\end{subequations}
The first equation here is the true ``conservation'' equation, telling us that the rate of change of the $q$-form density is given by the divergence of the \tightmath{$(q+1)$}-form flux. 
Whereas the second equation, which is absent for $q=0$, is like a ``Gauss constraint'' for the divergence of the $q$-form density and is responsible for the invariance of the charge $Q[\Sigma_{d-q}]$ under transverse spatial deformations of the defining surface. Note that the time-derivative of \cref{eq:cons-gauss} is trivially zero due to \cref{eq:cons-dyn}. This means that it is sufficient to impose the Gauss constraint as a boundary condition on some initial time-slice, after which it is automatically satisfied at all later times. Since this constraint needs to be satisfied even for equilibrium configurations, it will play an important role for us when considering thermal systems with higher-form symmetries.

\begin{subequations}
To probe a U(1)$_q$ symmetry in a field theory, following common lore from 0-form symmetries, it is convenient to introduce a \tightmath{$(q+1)$}-form background gauge field $A$ coupled to the associated current $J$, with the standard coupling action
\begin{equation}
    \int \delta A \wedge \star J.
\end{equation}
The conservation equation
\eqref{eq:higherform-cons} can then be understood as the Noether conservation laws associated with the $q$-form background transformations
\begin{align}
  A &\to A + \df\Lambda.
  \label{eq:Atrans}
\end{align} 
We can also define the associated background field strength tensor $F=\df A$,
which is invariant under these background gauge transformations. 
\end{subequations}

\subsection{Approximate higher-form symmetries}
\label{sec:appx_higher_form}

\begin{figure}[t]
    \centering
    \includegraphics[width=0.75\linewidth]{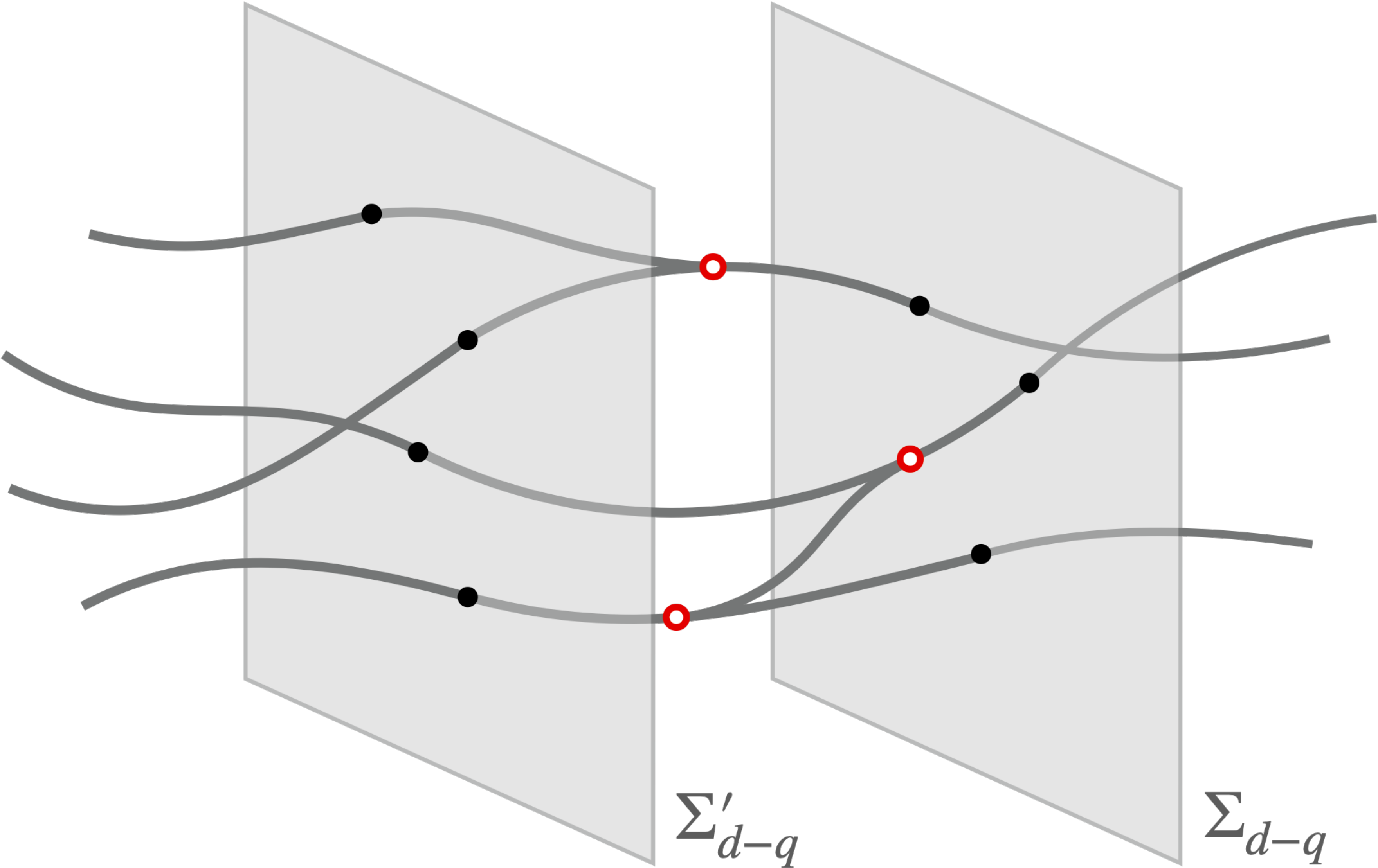}
    \caption{A snapshot of a system with an \emph{approximate} $q$-form U(1)$_q$ symmetry. The charged $q$-dimensional hypersurfaces (strings in this illustration) are no longer conserved and can be created or annihilated at \tightmath{$(q-1)$}-dimensional defects (red points in this illustration). The number of charged operators crossing a spacelike hypersurface $\Sigma_{d-q}$ is no longer conserved under a smooth deformation of $\Sigma_{d-q}$ to $\Sigma'_{d-q}$. However, the defects themselves furnish a U(1)$_{q-1}$ symmetry and are conserved.}
    \label{fig:strings-defected}
\end{figure}

The description of a system in terms of higher-form symmetries can still be useful when the symmetry is only approximate. One might hope to get some control over the problem by studying the system as a perturbative expansion around the symmetry-invariant point.
For a system that respects a U(1)$_q$ global symmetry only approximately, the
conservation equations \eqref{eq:higherform-cons} modify to include an arbitrary source term
\begin{subequations}
\begin{align}
  % \dow_\mu J^{\mu\nu_1\ldots\nu_q}
  % &= -\ell L^{\nu_1\ldots\nu_q} \nn\\
  % \text{or}\quad
  \df{\star J} &= (-)^{q+1} \ell\,{\star L},
  % \nn\\
  % \quad\text{or}\quad
  % \df{\star J}_{(q+1)}
  % &= (-1)^{q+1} \ell\, {\star L}_{(q)},
    \label{eq:higherform-noncons}
\end{align}
where $L$ is the $q$-form ``defect'' current and $\ell$ is a small parameter controlling the strength of explicit symmetry breaking. 
The factor of $(-)^{q+1}$ is introduced for later convenience.
What this means is that the $q$-dimensional operators charged under the U(1)$_q$ symmetry are no longer conserved and can be locally sourced by defects; see \cref{fig:strings-defected}. A similar discussion for approximate U(1)$_0$ symmetries appeared in our recent paper~\cite{Armas:2021vku}. However, approximate higher-form symmetries are qualitatively distinct because the defect current itself furnishes an emergent unbroken U(1)$^\ell_{q-1}$ topological symmetry and follows the conservation equation
\begin{align}
  % \dow_\mu L^{\mu\nu_1\ldots\nu_{q-1}} &= 0 \nn\\
  %   \text{or}\quad
  \df{\star L} &= 0,
    \label{eq:source-cons}
\end{align}
\end{subequations}
which is a direct consequence of \cref{eq:higherform-noncons}. We use the superscript ``$\ell$'' to distinguish this global symmetry from the original U(1)$_q$ symmetry.
The associated conserved charge is topological, i.e. the total number of defects integrated over a \tightmath{$(d-q+1)$}-dimensional spacelike surface $\Sigma_{d-q+1}$  only depends on its boundary
\begin{align}
    \ell Q_\ell[\Sigma_{d-q+1}]
    &= \int_{\Sigma_{d-q+1}} \ell\,{\star L} \nn\\
    % &= (-)^{q+1}\int_{\Sigma_{d-q+1}} \df{\star J} \nn\\
    &= (-)^{q+1}\int_{\dow\Sigma_{d-q+1}} {\star J} \nn\\
    &= (-)^{q+1} Q[\dow\Sigma_{d-q+1}].
    \label{eq:topological_charge}
\end{align}
Note that no such conservation appears for $q=0$.

It is instructive to see \cref{eq:higherform-noncons} in components, giving rise to the explicitly broken version of \cref{eq:non-cov-cons}, i.e.
\begin{subequations}
\begin{align}
    \dow_t J^{tk_1\ldots k_{q}} + \dow_i J^{ik_1\ldots k_{q}} 
    &= -\ell L^{k_1\ldots k_{q}}, 
    \label{eq:cons-dyn-broken} \\
    \dow_iJ^{tik_2\ldots k_q}
    &= \ell L^{tk_2\ldots k_q}.
    \label{eq:cons-gauss-broken}
\end{align}
\label{eq:non-cov-cons-broken}%
\end{subequations}
We see that the defect density arises as a source term in the Gauss constraint, acting as points where the $q$-form operators can begin or end; see \cref{fig:strings-defected}. On the other hand, the flux of defects causes the number of $q$-form operators to not be conserved in time.\footnote{Some aspects of explicit symmetry breaking of higher-form symmetries were discussed in specific models in the context of holography \cite{Grozdanov:2018fic, Davison:2022vqh}.}

\begin{subequations}
Following the approach of~\cite{Armas:2021vku}, it is convenient to artificially restore the U(1)$_q$ symmetry by introducing a $q$-form background source $\Phi$ in the theory for the defect current $\ell L$. We choose the coupling between $\Phi$ and the defect current $L$ to be
\begin{equation}
    \int \ell\, \delta\Phi \wedge \star L~.
\end{equation}
The factor of $\ell$ here ensures that the dependence on $\Phi$ drops out from the effective theory in the defect-free limit $\ell\to 0$.
To obtain the conservation equations \eqref{eq:higherform-noncons}, the new background field must transform under the U(1)$_q$ background transformations as
\begin{align}
    \Phi &\to \Phi - \Lambda.
    \label{eq:Phi-trans}
\end{align}
From this point of view, the background explicitly breaks the original U(1)$_q$ symmetry by
picking up a preferred set of $q$-form phases $\Phi$. 
\end{subequations}
Replacing $\Lambda$ with $-\df\Lambda_\ell$, or by a constant $-a_\ell$ for $q=0$, we can obtain the U(1)$^\ell_{q-1}$ background transformations responsible for the defect conservation equation \eqref{eq:source-cons}, i.e.
\begin{align}
    \Phi &\to \Phi
    + \begin{cases}
        \df\Lambda_\ell, &\quad \text{for } q > 0, \\
        a_\ell, &\quad \text{for } q=0,
    \end{cases}
    \label{eq:U1defect-trans}
\end{align}
while leaving the gauge fields $A$ invariant.
We can see that, for $q>0$, $\Phi$ acts as the background gauge field associated with the U(1)$^\ell_{q-1}$ global symmetry. 
% In foresight, we define the $q=0$ version of this transformation to be constant shifts of the background phase $\Phi \to \Phi + a_\ell$. 
We can construct a field strength associated with $\Phi$ as $\Xi = \df \Phi + A$, 
which remains invariant under both $q$-form and \tightmath{$(q-1)$}-form background gauge transformations.

\section{Approximate higher-form symmetries at zero temperature}
\label{sec:zero-temperature-higher-form}

As a warm-up exercise, we consider systems with approximate higher-form symmetries at zero temperature. The simplest field theories that realise higher-form symmetries are theories in which these symmetries are spontaneously broken, giving rise to a $q$-form Goldstone field $\phi$. For ordinary U(1)$_0$ global symmetries, this effective theory is understood as describing a superfluid. On the other hand, the spontaneously broken phase of a higher U(1)$_q$ global symmetry describes a higher-form superfluid or a U(1)$^\loc_{q-1}$ gauge theory in the absence of free charges, with the $q$-form $\phi$ playing the role of the associated dynamical gauge field. We use the superscript ``loc'' to remind ourselves that this symmetry is local and not global.
When the U(1)$_q$ symmetry is further explicitly broken, the U(1)$_q$ superfluid can exist in one of two phases, relaxed or pinned, depending on the symmetry breaking pattern of the emergent topological U(1)$^\ell_{q-1}$ symmetry in \cref{eq:source-cons}. These correspond to the Coulomb and Higgs phases of U(1)$^\loc_{q-1}$ gauge theory respectively. In the relaxed phase, the $q$-form charges are relaxed but the Goldstone field $\phi$ is massless, so the superfluid velocity is not relaxed. On the other hand, in the pinned phase, the Goldstone field $\phi$ is massive and hence the superfluid velocity is also relaxed; see e.g.~\cite{Armas:2021vku} for the relevant 0-form discussion. We also discuss vortices which, given the definition we will introduce later on, can only exist in the relaxed phase of the U(1)$_q$ pseudo-superfluid (see figure \ref{fig:chart}) and also have the physical effect of relaxing the superfluid velocity without pinning the Goldstone field $\phi$, i.e. without giving it a mass.

% As we shall discuss below, when we explicitly break the U(1)$_q$ global symmetry, the system can exist in one of two phases, Coulomb or Higgs, familiar from gauge theories. The Coulomb phase is characterised by the introduction of free charges in the U(1)$^\loc_{q-1}$ gauge theory screening the electric fields, whereas the Higgs phase is characterised by a massive gauge field screening both electric and magnetic fields.

\subsection{Spontaneous symmetry breaking and superfluids}
\label{sec:SSB-zeroT}

When a continuous U(1)$_q$ global symmetry of a physical system is spontaneously broken in the ground state, the low-energy effective description admits a $q$-form Goldstone phase field $\phi$, transforming as a shift under the U(1)$_q$ transformations
\begin{equation}
    \phi \to \phi - c_\phi\Lambda.
    \label{eq:trans-phi}
\end{equation}
Here $c_\phi$ denotes the constant charge of the Goldstone.\footnote{We discussed 1-form superfluids in a previous paper~\cite{Armas:2018zbe}, where we chose the charge of the Goldstone to be $c_\phi = 1$. In the work of~\cite{Delacretaz:2019brr}, the authors used $c_\phi = -a$.} It will also be convenient to define the associated $(q+1)$-form covariant derivative
\begin{equation}
    \xi = \df\phi + c_\phi A,
    \label{eq:xi-defn}
\end{equation}
which is invariant under the U(1)$_q$ background gauge transformations. Keeping with the terminology from spontaneously broken 0-form symmetries, we refer to $\xi$ as the superfluid velocity, even though it is only a vector when $q=0$. Due to the U(1)$_q$ symmetry, all the dependence on $\phi$ comes via $\xi$ in the effective theory. This results in the invariance of the theory under shifts of $\phi$ by an exact $q$-form or by a constant for $q=0$, i.e.
\begin{equation}
    \phi \to \phi
    - c_\phi 
    \begin{cases}
        \df \lambda, &\quad \text{for } q> 0, \\
        a,  &\quad \text{for } q=0.
    \end{cases}.
    \label{eq:gauge-symmetry}
\end{equation}
For $q>0$, this invariance can be interpreted as a U(1)$^\loc_{q-1}$ \emph{gauge} symmetry of the effective theory, with $\phi$ acting as the dynamical gauge field and $\xi$ its field strength. The special case of U(1)$_1$ superfluids describes Maxwell's electromagnetism with U(1)$^\loc_0$ gauge symmetry.

\subsubsection{Goldstone action}

We can write a Landau-Ginzburg-type effective action for the Goldstone as
\begin{equation} \label{eq:LG_goldstone}
    S = - \int \frac{f_s}{2}\, \xi \wedge {\star\xi}
    - \star{\cal L}_{\text{pol}}
    + \xi\wedge \tilde A,
\end{equation}
where $f_s$ denotes the constant superfluid density or the gauge coupling constant in the context of U(1)$^\loc_{q-1}$ gauge theories.
The ``matter Lagrangian'' ${\cal L}_{\text{pol}}$ contains contributions from additional matter fields or higher-derivative/higher-powers of $\xi$. The U(1)$_q$ global symmetry of the theory requires that all dependence on $\phi$ in ${\cal L}_{\text{pol}}$ must come via $\xi$. Therefore, we can parameterise its variation as 
\begin{equation}
    {\star{\delta\cal L}_{\text{pol}}} = 
    \delta\xi \wedge {\star{\cal M}},
    \label{eq:Lm-var-pol}
\end{equation}
for some \tightmath{$(q+1)$}-form ${\cal M}$.
In the context of gauge theories, this just means that we do not have any free U(1)$^\loc_{q-1}$ charges in the description; the matter Lagrangian ${\cal L}_{\text{pol}}$ is purely polarised with the polarisation tensor ${\cal M}$. We have also introduced a new \tightmath{$(d-q)$}-form source $\tilde A$ in the action that can be used to compute the correlation functions of the superfluid velocity $\xi$. The reason for the particular notation will become clear momentarily. 

Varying the action \eqref{eq:LG_goldstone} with respect to $\phi$, we can read off the respective equations of motion
\begin{subequations}
\begin{equation}
    - c_\phi\, \df{\star\!\lb f_s\xi - {\cal M} \rb}
    = c_\phi \tilde F,
    \label{eq:max-1}
\end{equation}
where $\tilde F = \df\tilde A$. There is also a Bianchi identity associated with $\xi$, which takes a similar form as above
\begin{equation}
    \df\xi = c_\phi F.
    \label{eq:max-2}
\end{equation}
\label{eq:max-eqns}%
\end{subequations}
In the context of U(1)$^\loc_{q-1}$ gauge theories, these are nothing but the Maxwell's equations and Bianchi identity associated with the dynamical field strength $\xi$.

\begin{subequations}
Looking at \cref{eq:max-eqns}, we can give a physical interpretation to the two higher-form background gauge fields. In the superfluid interpretation of this theory, $A$ is the background gauge field associated with the symmetry being spontaneously broken, with $F=\df A$ the associated field strength. On the other hand, $\tilde F = \df \tilde A$ can be Hodge-dualised to define
\begin{equation}
    \tilde F = (-)^q\,{\star K_\ext},
\end{equation}
where the $q$-form $K_\ext$ serves as a background source coupled to the Goldstone phase field $\phi$. Since $\phi$ is not U(1)$_q$-invariant, this source appears on the right-hand side of the U(1)$_q$ conservation equation \eqref{eq:max-1}. 
In the context of gauge theories, $K_\ext$ can be understood as the external electric current coupled to the dynamical gauge field $\phi$, showing up as a source in the Maxwell's equations \eqref{eq:max-1}. Similarly, we can define the dual version $\tilde K_\ext$ via
\begin{equation}
    F = (-)^p\,{\star \tilde K_\ext},
\end{equation}
which serves as the external magnetic current, sourcing the Bianchi identity \eqref{eq:max-2}.
\label{eq:external-currents}
\end{subequations}
Note that these background currents are conserved, i.e.
\begin{equation}
    \df{\star K_\ext} = \df{\star \tilde K_\ext} = 0.
\end{equation}

\subsubsection{Higher-form symmetries}

\begin{subequations}
A U(1)$_q$ superfluid or a U(1)$^\loc_{q-1}$ gauge theory has two higher-form symmetries.
The theory manifestly respects the original U(1)$_q$ ``electric'' global symmetry that was spontaneously broken, with the associated current
\begin{equation}
    J_{\text{cons}} = -c_\phi \lb f_s\xi - {\cal M} \rb
    - \tilde c_\phi {\star\tilde A},
\end{equation}
that can be obtained by varying the action \eqref{eq:LG_goldstone} with respect to the background field $A$.
Here we have defined $\tilde c_\phi = (-)^{pq+p+q} c_\phi$ which will be useful later. The associated conservation equations are precisely \cref{eq:max-1}. Interestingly, this theory also has a U(1)$_{p}$ ``magnetic'' global symmetry,
where \tightmath{$p=d-1-q$}, given by the associated current
\begin{equation}
    \tilde J_{\text{cons}} = \star\xi - c_\phi {\star A} = \star\df\phi,
    \label{eq:tildeJ-cons}
\end{equation}
\label{eq:cons-currents}%
\end{subequations}
with the associated conservation equation coming from the Bianchi identity \eqref{eq:max-2}. This corresponds to the conservation of $p$-dimensional ``equipotential'' surfaces of the phase field $\phi$.
We use the ``tilde'' to distinguish the quantities related to this second higher-form symmetry. The associated conserved charges are given as
\begin{subequations}
\begin{align}
    Q[\Sigma_{d-q}] 
    &= \int_{\Sigma_{d-q}}
    -c_\phi \lb f_s\,{\star\xi} - {\star\cal M} \rb -c_\phi \tilde A, 
    \label{eq:elec-charges}
    \\
    \tilde Q[\Sigma_{d-p}] 
    &= \int_{\Sigma_{d-p}} (-)^{pq+p+q}\xi - \tilde c_\phi A.
    \label{eq:mag-charges}
\end{align}
These charges remain unchanged under smooth deformations of the surfaces $\Sigma_{d-q}$ and $\Sigma_{d-p}$ over which they are defined, provided that we do not cross any free charges.
\label{eq:cons-operators}
\end{subequations}

In the context of gauge theories, switching off the background fields $A$ and $\tilde A$, the higher-form charged objects can be understood as the $q$-dimensional field lines of the electric displacement tensor $J^{ti\ldots} = -c_\phi f_s \xi^{ti\ldots} + c_\phi {\cal M}^{ti\ldots}$ and the $p$-dimensional field lines of the magnetic field $\tilde J^{ti\ldots} = \frac{1}{(q+1)!} \epsilon^{k\ldots ti\ldots}\xi_{k\ldots}$.
The associated conserved charges in \cref{eq:cons-operators} count the number of field lines passing the cross-sections $\Sigma_{d-q}$ and $\Sigma_{d-p}$ respectively.

The U(1)$_q$ electric symmetry is realised in the action \eqref{eq:LG_goldstone} via invariance under U(1)$_q$ symmetry transformations of the background gauge field $A$ given in \cref{eq:Atrans}, together with the shift \eqref{eq:trans-phi} of the phase field $\phi$. 
On the other hand, the action is invariant under U(1)$_p$ symmetry transformations of the associated gauge field $\tilde A$ as $\tilde A \to \tilde A + \df\tilde\Lambda$, but only in the absence of the U(1)$_q$ gauge field $A$.\footnote{The action \eqref{eq:LG_goldstone} only strictly respects the U(1)$_p$ if the spacetime is unbounded. This is because the U(1)$_p$ and U(1)$_q$ symmetries have a mixed anomaly, as we discuss below.} We could exchange the coupling term $\xi\wedge\tilde A$ in the action \eqref{eq:LG_goldstone} with $\df\phi\wedge \tilde A$ to make it invariant under the U(1)$_p$ symmetry, but doing this will violate the U(1)$_q$ symmetry that was originally manifest. 
Unfortunately, we cannot manifest both the higher-form global symmetries together in the action because the theory has a mixed 't Hooft anomaly. This is the reason why the U(1)$_q$ and U(1)$_p$ conserved currents in \cref{eq:cons-currents} and the respective charges in \cref{eq:cons-operators} are not invariant under the U(1)$_p$ and U(1)$_q$ global transformations respectively.

This anomaly can actually be treated using the anomaly inflow mechanism, by coupling the system to a $(d+2)$-dimensional bulk theory with a Chern-Simons-like topological Lagrangian (see appendix \ref{app:anomaly-inflow} for details)
\begin{equation}
    S_{\text{bulk}} = c_\phi\int_{\text{bulk}}
    \df A \wedge \tilde A. 
    \label{eq:bulk-lagrangian}
\end{equation}
One can check that the combined theory is invariant under both global symmetries. By varying the full action with respect to the two higher-form sources, we can read out the covariant versions of the two higher-form currents 
\begin{subequations}
\begin{align}
    J &= -c_\phi \lb f_s\xi - {\cal M} \rb,
    \label{eq:J-electric} \\
    \tilde J &= \star\xi.
    \label{eq:tildeJ}
\end{align}
\label{eq:gauge-invariant-currents}
\end{subequations}
They obey a set of anomalous conservation laws
\begin{subequations}
\begin{align}
    \df{\star J}
    &= c_\phi\, \tilde F, \\
    \df{\star \tilde J} 
    &= \tilde c_\phi\, F,
\end{align}
\label{eq:anomalous_cons}%
\end{subequations}
Therefore, an equivalent way to think about a U(1)$_q$ superfluid, or a U(1)$_{q-1}^\loc$ gauge theory without free charges, is as a system with an anomalous $\rmU(1)_q\times\rmU(1)_p$ global symmetry. 
This representation is appealing because it does not require one to make assumptions about the field content or the gauge symmetries of the underlying description, and relies solely on the physical global symmetry structure of the theory  and the associated conservation equations. 
It also allows us to systematically introduce dissipative phenomena into the model without having to resort to any of the microscopic details of the theory, as we shall explore later in our discussion.

% \hl{Can we make the precise statement that at zero temperature the U(1)q symmetry is necessary spontaneously broken and you can't get rid of the Goldstone?}

\subsubsection{Duality transformations}
\label{sec:duality-zeroT}

Note that the description of a U(1)$_q$ superfluid in terms of an anomalous $\rmU(1)_q\times\rmU(1)_p$ global symmetry is invariant under the exchange of $q\leftrightarrow p$, which suggests a duality with a U(1)$_p$ superfluid. 
This can be made precise by coupling the action \eqref{eq:LG_goldstone} to a 
\tightmath{$p$}-form Lagrange multiplier $\tilde\phi$ through a term like 
\begin{equation}
    S \sim \int \frac{1}{c_\phi}\df\tilde\phi\wedge
    \lb \xi - c_\phi A \rb,
    \label{eq:lagrange-mult}
\end{equation}
implementing the Bianchi identity \eqref{eq:max-2}. The full action is still invariant under the U(1)$_q$ symmetry, however for U(1)$_p$ symmetry, the Lagrange multiplier should shift as a phase $\tilde\phi\to \tilde\phi - \tilde c_\phi \tilde\Lambda$. Therefore, we can think of $\tilde\phi$ as a Goldstone for the U(1)$_p$ global symmetry.
Having introduced this term in the action, the superfluid velocity $\xi$ becomes an independent unconstrained degree of freedom in the Lagrangian with the classical equation of motion
\begin{equation}
    \xi = 
    -\frac{1/f_s}{c_\phi} \lb {\star\tilde\xi} - c_\phi {\cal M} \rb,
    \label{eq:xi-onshell}
\end{equation}
where $\tilde\xi = \df\tilde\phi + \tilde c_\phi \tilde A$.
Substituting this back into the action \eqref{eq:LG_goldstone} together with the Lagrange multiplier term \eqref{eq:lagrange-mult}, we are led to the same theory as before, but with the substitutions
% \begin{equation}
%     S = - \int \frac{1/f_s}{2c_\phi^2}\, {\tilde\xi}
%     \wedge {\star\tilde\xi}
%     - \star{\cal\tilde L}_m
%     + \tilde\xi\wedge A.
%     \label{eq:sf-action-q}
% \end{equation}
\begin{gather}
    \phi \leftrightarrow \tilde\phi, \qquad 
    A \leftrightarrow \tilde A, \qquad 
    \xi  \leftrightarrow \tilde\xi, \nn\\
    % \frac{1}{f_s c_\phi}{\cal M} \to {\star\cal M}, \nn\\
    c_\phi  \leftrightarrow \tilde c_\phi, \qquad 
     f_s  \leftrightarrow \frac{1}{f_s c_\phi^2}, \qquad 
     J \leftrightarrow \tilde J,
    \label{eq:EM-mappings}
\end{gather}
together with a transformation of the polarised matter Lagrangian
\begin{equation}
    \star{\cal L}_{\text{pol}}
     \leftrightarrow
    \star{\cal L}_{\text{pol}}
    - \frac{1}{2f_s}{\cal M}\wedge {\star\cal M}.
\end{equation}
Note that we also need to perform these transformations to the bulk anomaly-inflow action in \cref{eq:bulk-lagrangian}, giving rise to
\begin{equation}
    S_{\text{bulk}}
     \leftrightarrow S_{\text{bulk}}
    - c_\phi \int A\wedge \tilde A.
    \label{eq:bulk-lagrangian-dual}
\end{equation} 

Therefore, a U(1)$_q$ superfluid is dual to a U(1)$_p$ superfluid. This is just a realisation of the electromagnetic duality for higher-form gauge theories, which states that a U(1)$^\loc_{q-1}$ gauge theory is dual to a U(1)$^\loc_{p-1}$ gauge theory in the absence of free charges, with the electric and magnetic sectors exchanged. The dual Goldstone field $\tilde\phi$ in the context of electromagnetism is known as the dual/magnetic gauge field. Setting \tightmath{$p=q=1$} in \tightmath{$d=3$}, one recovers the well-known self-duality of electromagnetism in \tightmath{$(3+1)$}-dimensions. Such self-dualities exist for all U(1)$_{q-1}^\loc$ gauge theories (or U(1)$_q$ superfluids) in \tightmath{$d=2q+1$} spatial dimensions.

\subsection{Pseudo-spontaneous symmetry breaking and relaxed pseudo-superfluids}
\label{sec:SSB-zeroT-broken}

The setup becomes more interesting when we explicitly break one of the higher-form global symmetries of the superfluid, let's say the U(1)$_q$ electric symmetry. In this context, the global U(1)$_q$ symmetry is said to be 
\emph{pseudo-spontaneously} broken and the field $\phi$ is referred to as a \emph{pseudo-Goldstone} field. The effective theory for a pseudo-superfluid can additionally depend on a $q$-form misalignment tensor $\psi$ capturing the difference between the superfluid phase field $\phi$ and the background phase field $\Phi$ introduced in \cref{sec:appx_higher_form}, i.e.
\begin{equation}
    \psi = \ell\lb \phi - c_\phi\Phi \rb.
    \label{eq:misalignment}
\end{equation}
We have included a factor of $\ell$ in the definition of $\psi$ to make sure that it vanishes in the limit $\ell\to 0$, restoring the U(1)$_q$ symmetry.
A similar construction for U(1)$_0$ pseudo-superfluids appeared in our recent paper~\cite{Armas:2021vku}. 

\subsubsection{Relaxed phase}
\label{sec:Coulomb-zeroT}

A pseudo-superfluid can exist in two distinct phases depending on the fate of the U(1)$^\ell_{q-1}$ global symmetry associated with the conservation of defects given in \cref{eq:U1defect-trans}. The first of these is the \emph{relaxed phase}, where the U(1)$^\ell_{q-1}$ global symmetry is spontaneously intact. Another way to think about this phase is as follows: consider the U(1)$^\ell_{q-1}$ transformation with parameter $\Lambda_\ell = \lambda$ (or $a_\ell = a$ for $q=0$), together with a U(1)$_q$ transformation with parameter $\Lambda = \df\lambda$ (or $\Lambda = a$ for $q=0$). This combination leaves both the background fields $A$ and $\Phi$ invariant, but the phase field $\phi$ undergoes the U(1)$^\loc_{q-1}$ gauge transformation given in \cref{eq:gauge-symmetry}. This means that the effective theory must be invariant under gauge transformations of the misalignment tensor
\begin{equation}
    \psi \to \psi - \ell c_\phi
    \begin{cases}
        \df \lambda, &\qquad \text{for } q> 0, \\
        a,  &\qquad \text{for } q=0.
    \end{cases}
    \label{eq:psi-trans}
\end{equation}
Therefore, the relaxed phase of a pseudo-superfluid is where the U(1)$^\loc_{q-1}$ gauge symmetry in \cref{eq:gauge-symmetry} is respected. The physical picture one can keep in mind is that the U(1)$_q$ global symmetry is spontaneously as well explicitly broken, but the operator that condensed to spontaneously break the symmetry was itself invariant under U(1)$_q$ transformations. This means that the effective theory is still invariant under gauge shifts U(1)$^\loc_{q-1}$ of the Goldstone phase $\phi$ that take us from one ground-state of the condensate to another. 

The action describing this phase still takes the same schematic form as \eqref{eq:LG_goldstone}; to wit
\begin{equation}
    S = - \int \frac{f_s}{2}\, \xi \wedge {\star\xi}
    - \star{\cal L}_{e}
    + \xi\wedge \tilde A.
    \label{eq:sf-action-Coulomb}
\end{equation}
In this expression, the matter Lagrangian ${\cal L}_{\text{pol}}$ is replaced by ${\cal L}_e$, which can also depend on the misalignment tensor $\psi$ in addition to $\xi$. However, the dependence on $\psi$ cannot be arbitrary and must confirm to the gauge transformations in \cref{eq:psi-trans}. In other words, if we parametrise the variations of ${\cal L}_e$ as 
\begin{equation}
    {\star{\delta\cal L}_e} = 
    \delta\xi \wedge {\star{\cal M}}
    + \delta\psi \wedge {\star{\cal J}_e},
    \label{eq:Lm-var-free}
\end{equation}
we must have that ${\cal J}_e$ is conserved, i.e.  $\df{\star{\cal J}_e} = 0$, when all other matter fields have been taken onshell. Such a term does not exist for $q=0$, so naively it looks like there are no signatures of explicit symmetry breaking in this phase. However, upon including dissipative effects at finite temperature, we will see that explicit symmetry breaking still leaves physically measurable signatures in the theory such as charge relaxation. For $q>0$, this can be identified precisely as the Coulomb phase of the U(1)$_{q-1}^\loc$ gauge theory with ${\cal J}_e$ playing the role of free electric charge current. To wit, the equation of motion \eqref{eq:max-1} modifies to
\begin{equation}
    - c_\phi\, \df{\star\!\lb f_s\xi - {\cal M} \rb}
    = c_\phi \tilde F
    + (-)^q c_\phi \ell\,{\star{\cal J}_e}.
    \label{eq:max-1-Coulomb}
\end{equation}
The background field $\Phi$ serves as a source for ${\cal J}_e$ and, in the context of gauge theories, can be interpreted as a background gauge field coupled to the free electric charges in the theory, with $\Xi = \df\Phi + A$ the associated U(1)$_q$-invariant background field strength. The identity $\df\Xi = F$ yields the Bianchi identity for $\Xi$,
\begin{equation}
    \df\Xi = (-)^p\,{\star \tilde K_\ext},
    \label{eq:XiBianchi}
\end{equation}
which is sourced by the external magnetic current  $\tilde K_\ext$ defined in \cref{eq:external-currents}.

The U(1)$_q$ electric global symmetry is now explicitly broken by the defect current
\begin{equation}
    L = -c_\phi{\cal J}_e~,
    \label{eq:defect-current-Coulomb}
\end{equation}
which can be obtained by varying the action with respect to the background phase field $\Phi$. The conservation equations modify from \cref{eq:anomalous_cons} to
\begin{subequations}
\begin{align} 
    \df{\star J}
    &= c_\phi\, \tilde F + (-)^{q+1}\ell\,{\star L}, \\
    \df{\star \tilde J} 
    &= \tilde c_\phi\, F, \\
    \df{\star L} &= 0.
\end{align}
\label{eq:conservation_explicitbroken}%
\end{subequations}
The U(1)$_q$ charges (electric field lines) given in \cref{eq:elec-charges} are no longer conserved in the presence of $L$. However, the defect current $L$ itself furnishes a topologically conserved U(1)$^\ell_{q-1}$ charge, given in the spirit of \cref{eq:topological_charge} as 
\begin{align}
    \ell Q_\ell[\Sigma_{d-q+1}]
    &= -\ell c_\phi\int_{\Sigma_{d-q+1}} 
    {\star{\cal J}_e} \nn\\
    &= (-)^{q+1} Q[\dow\Sigma_{d-q+1}].
    \label{eq:Qell-Coulomb}
\end{align}
The value of this topological charge is given by the number of U(1)$_q$ charged objects (electric field lines) crossing the boundary of the region $\Sigma_{d-q+1}$ over which it is defined. The U(1)$_p$ charges (magnetic field lines) given in \cref{eq:elec-charges} are still conserved.

\subsubsection{Vortices}

Let us take a quick detour and consider what happens if we break the U(1)$_p$ magnetic global symmetry instead. These are related to the topological defects of the phase field (dynamical gauge field) $\phi$. Let us consider a superfluid with an explicitly unbroken U(1)$_q$ symmetry, which got spontaneously broken by the condensation of some charged operator $\Psi$ to the ground state value $\Psi_0$. 
The usual Higgs mechanism lore is to decompose the fluctuations of $\Psi$ around $\Psi_0$ into a massive magnitude $\eta$ and a massless Goldstone phase $\phi$. However, there is no fundamental principle guaranteeing that such a decomposition can be smoothly implemented all over spacetime. When this decomposition fails, the phase field $\phi$ in the effective theory can generically be singular, in a way that the physically observable superfluid velocity $\xi$ is still smooth. Borrowing superfluid terminology, we will refer to such configurations as \emph{vortices}; in the context of gauge theories, these would be the infamous \emph{magnetic monopoles}. In the presence of vortices $\df\df\phi\neq 0$. Formally, we can split the gradient $\df\phi$ into a defect-free and a defected part according to 
\begin{align}
    \df\phi = \df\bar\phi + \tilde\ell\, V,
    \label{eq:V-intro}
\end{align}
where $\bar\phi$ should roughly be thought of as the smooth part of the Goldstone field and $V$ as the new degrees of freedom required to describe the configurations of vortices. We have introduced a small parameter $\tilde\ell$ in the decomposition to control the strength of vortices. Plugging this into the definition of the superfluid velocity in \cref{eq:xi-defn}, we can see that the Bianchi identity \eqref{eq:max-2} modifies to
\begin{equation}
    \df\xi 
    = c_\phi F 
    + \tilde\ell\, {\df V}.
    \label{eq:max-2-broken}
\end{equation}
% where ${\cal J}_m = (-)^{q} {\star\df V}$.

% $\df\df\phi \equiv (-)^{pq+q+1} \tilde\ell\, {\star{\cal J}_m} \neq 0$, where ${\cal J}_m$ is identified as the vortex (magnetic monopole) current and $\tilde\ell$ is a small parameter controlling the strength of vortices. This results in a violation of the Bianchi identity

To add vortices to the action of U(1)$_q$ superfluids, we need to introduce a U(1)$_p$ background phase field $\tilde\Phi$, transforming as $\tilde\Phi\to\tilde\Phi - \tilde\Lambda$. This can also be understood as a background gauge field associated with the vortex (or magnetic monopole) current. The object $\tilde\Xi = \df\tilde\Phi + \tilde A$ serves as the associated field strength, whose Bianchi identity is sourced by the external electric current similar to \cref{eq:XiBianchi}, i.e.
\begin{equation}
    \df\tilde\Xi = (-)^q {\star K_\ext}.
\end{equation}
Together with the dual Goldstone field $\tilde\phi$ introduced around \cref{eq:lagrange-mult}, we can use this to define the dual misalignment tensor $\tilde\psi$ similar to \cref{eq:misalignment}, i.e.
\begin{equation}
    \tilde\psi = \tilde\ell \big( \tilde\phi 
    - \tilde c_\phi \tilde\Phi \big).
\end{equation}
Armed with this, we can append the action \eqref{eq:LG_goldstone} to allow for vortices as
\begin{align}
    S &= - \int \frac{f_s}{2}\, \xi \wedge {\star\xi}
    - \star{\cal L}_{\text{pol}}
    + \xi\wedge \tilde A \nn\\
    &\qquad\qquad
    - \frac{1}{c_\phi}\df\tilde\phi\wedge
    \lb \xi - c_\phi A \rb
    - \frac{(-)^{p}}{c_\phi} \tilde\psi 
    \wedge {\df V}.
    \label{eq:monopole-action}
\end{align}
The field $\tilde\phi$ now acts as a Lagrange multiplier for the broken Bianchi identity \eqref{eq:max-2-broken}. 

One can check that this action is manifestly invariant under both U(1)$_q$ and U(1)$_p$ background gauge transformations, up to the mixed anomaly discussed previously. However, the U(1)$_p$ conservation is now explicitly broken by the vortex current
\begin{equation}
    \tilde L = (-)^q {\star\df V},
    \label{eq:tildeL}
\end{equation}
that can be obtained by varying the action with respect to the source $\tilde\Phi$. Note that the object ${\star\df V}$ is conserved by construction, so the theory is invariant under transformations of the dual misalignment tensor $\tilde\psi$ similar to the one given in \cref{eq:psi-trans}. This implies that we are in the relaxed phase with respect to the explicitly broken U(1)$_p$ symmetry. The conservation equations in the presence of vortices take the form
\begin{subequations}
\begin{align} 
    \df{\star J}
    &= c_\phi\, \tilde F, \\
    \df{\star \tilde J} 
    &= \tilde c_\phi\, F
    + (-)^{p+1}\tilde\ell\,{\star\tilde L}, \\
    \df{\star\tilde L} 
    &= 0.
\end{align}
\label{eq:conservation_defect}%
\end{subequations}
The number of vortices crossing a surface $\Sigma_{d-p+1}$ is conserved and, in the absence of the U(1)$_q$ gauge field $A$, is given by the number of the U(1)$_p$ charged objects (magnetic field lines) crossing its boundary
\begin{align}
    \tilde\ell \tilde Q_\ell[\Sigma_{d-p+1}]
    &= (-)^{pq+q+1}\tilde\ell \int_{\Sigma_{d-p+1}} 
    \df V \nn\\
    &= (-)^{p+1} \tilde Q[\dow\Sigma_{d-p+1}].
    % \nn\\
    % &\hspace{-0.45em}\overset{A=0}{=} 
    % (-)^{p+1} \tilde Q[\dow\Sigma_{d-p+1}].
\end{align}

One can see that conservation equations \eqref{eq:conservation_defect} are almost identical to the ones we obtained for U(1)$_q$ pseudo-superfluids in \eqref{eq:conservation_explicitbroken}, on account of the electromagnetic dualities discussed towards the end of \cref{sec:SSB-zeroT}. Since the U(1)$_q$ symmetry is explicitly unbroken and all dependence on $\phi$ comes purely via $\xi$, we can proceed as before and integrate out $\xi$ from the action. The classical equations of motion for $\xi$ are still given by \cref{eq:xi-onshell}. Substituting this into the action \eqref{eq:monopole-action}, we recover the theory of a U(1)$_p$ pseudo-superfluid in relaxed phase
(i.e. a U(1)$_{p-1}^\loc$ gauge theory coupled to free ``electric'' charges), described by the action \eqref{eq:sf-action-Coulomb} with $p\leftrightarrow q$ and the substitutions
\begin{gather}
    \ell \leftrightarrow \tilde\ell, \qquad
    \Phi \leftrightarrow \tilde\Phi, \qquad 
    \psi \leftrightarrow \tilde\psi, \qquad 
    L \leftrightarrow \tilde L, \nn\\
    \star{\cal L}_e 
     \leftrightarrow \star{\cal L}_{\text{pol}}
    - \frac{1}{2f_s}{\cal M}\wedge {\star\cal M}
    + \frac{(-)^p}{c_\phi} \tilde\psi \wedge {\df V},
    \label{eq:EM-mappings-vor}
\end{gather}
together with the mappings in \cref{eq:EM-mappings}. This means that a U(1)$_q$ superfluid with vortices is dual to a U(1)$_p$ pseudo-superfluid in the relaxed phase, and vice-versa. Analogously, the statement of electromagnetic duality is that a U(1)$_{q-1}^\loc$ gauge theory with magnetically charged matter is dual to a U(1)$_{p-1}^\loc$ gauge theory with electrically charged matter, and vice versa. A corollary of this duality is that, just like magnetic charges can be understood as topological defects of the electric gauge field, free electric charges can also be understood as topological defects of the dual magnetic gauge field.

A special case of the above duality is that vortices in a U(1)$_0$ superfluid are dual to charged particles in a U(1)$_{d-2}^\loc$ gauge theory. In $d=2$, this is well-known \emph{particle/vortex duality} that relates vortices in a U(1)$_0$ superfluid to charged particles in U(1)$_{0}^\loc$ electromagnetism.

The natural question to consider now is whether it is possible to break both the U(1)$_q$ and U(1)$_p$ global symmetries simultaneously, i.e. to write down a theory of vortices for a U(1)$_q$ pseudo-superfluid. This is the same question as whether it is possible to write down a U(1)$^l_{q-1}$ gauge theory with both electrically and magnetically charged matter.
There seems to be no simple way of realising this possibility by means of a local action principle, because both the $q$-form Goldstone $\phi$ and the dual $p$-form Goldstone $\tilde\phi$ will need to be singular. However, as it turns out, it is possible to write down such effective theories at finite temperature in the presence of a preferred thermal rest frame, by making the spatial components of $\phi$ and $\tilde\phi$ singular, while keeping their time-components smooth.
Formally, this would lead the symmetric set of conservation equations 
\begin{subequations}
\begin{align} 
    \df{\star J}
    &= c_\phi\, \tilde F
    + (-)^{q+1}\ell\,{\star L}, \\
    \df{\star \tilde J} 
    &= \tilde c_\phi\, F
    + (-)^{p+1}\tilde\ell\,{\star\tilde L}, \\
    \df{\star L} 
    &= 0, \\
    \df{\star\tilde L} 
    &= 0.
\end{align}
\label{eq:conservation-Coulomb-symmetric}%
\end{subequations}
We will look at this scenario in more detail in \cref{sec:appx-SF-Coulomb-eqb}.

%\begin{align}
   % \df{\star J}
   % &= c_\phi\, \df\tilde A + (-)^{q+1}\ell\,{\star L}, \nn\\
   % \df{\star \tilde J} 
  %  &= \tilde c_\phi\, \df A.
%\end{align}

\subsection{Pseudo-spontaneous symmetry breaking and pinned pseudo-superfluids}
\label{sec:Higgs-zeroT}

Let us return to explicitly broken U(1)$_q$ global symmetry. 
The \emph{pinned phase} of a U(1)$_q$ pseudo-superfluid is defined to be the one where the U(1)$^\ell_{q-1}$ global symmetry in \cref{eq:U1defect-trans} is spontaneously broken, giving rise to a \tightmath{$(q-1)$}-form Goldstone $\phi_\ell$ transforming as $\phi_\ell \to \phi_\ell - c_\phi\Lambda_\ell$. For $q=0$, we would instead introduce a constant parameter $\alpha_\ell$ in the theory that shifts as $\alpha_\ell \to \alpha_\ell - c_\phi a_\ell$. One can actually redefine 
the misalignment tensor in \cref{eq:misalignment} to ``eat'' this Goldstone, i.e.\footnote{For $q=0$ case, we can actually redefine the background phase $\Phi \to \Phi - \alpha_\ell/c_\phi$ to make it invariant under ``U(1)$^\ell_{-1}$'' transformations in \cref{eq:U1defect-trans}. The misalignment tensor is then just given by $\psi = \ell(\phi-\Phi)$ as in our previous work~\cite{Armas:2021vku}.}
\begin{equation}
    \psi 
    = \ell
    \begin{cases}
        \phi - c_\phi\Phi - \df\phi_\ell, &\quad \text{for } q>0, \\
        \phi - c_\phi\Phi - \alpha_\ell, &\quad \text{for } q=0,
    \end{cases}
    \label{eq:psi-Higgs}
\end{equation}
and become entirely gauge-invariant. Consequently, the effective theory can now depend on $\psi$ arbitrarily. 

\subsubsection{Pinned phase}

The action  describing this phase remains the same in form as in \cref{eq:sf-action-Coulomb}, except that we can introduce a new background  coupling term for the gauge-invariant $\psi$, i.e.
\begin{equation}
    S = - \int \frac{f_s}{2}\, \xi \wedge {\star\xi}
    - \star{\cal L}_{e}
    + \xi \wedge \tilde A
    - (-)^q\psi \wedge \tilde A_\psi.
    \label{eq:sf-action-Higgs}
\end{equation}
We will return to the physical interpretation of the source $\tilde A_\psi$ in a bit. Furthermore, ${\cal J}_e$ defined in \cref{eq:Lm-var-free} is not required to be automatically conserved anymore. In particular, the matter Lagrangian $\star{\cal L}_e$ can contain a new mass term like $-m^2/2\, \psi\wedge{\star\psi}$, leading to a contribution like $-m^2\psi$ in ${\cal J}_e$. The parameter $\ell m$ can be understood as the mass parameter for $\phi$. Note that factors of $\ell$ appear in the action implicitly through the definition of $\psi$ in \cref{eq:misalignment}. For $q=0$, the new mass term gives rise to the phenomenology of \emph{pinned superfluids}; see~\cite{Armas:2021vku}. For $q>0$, on the other hand, we can understand this as the \emph{Higgs phase} of the U(1)$_{q-1}^\loc$ gauge theory with massive gauge fields $\phi$.
For $q=1$, this describes the theory of \emph{superconductivity}.
The quantity $k_0^\phi = \ell m/\sqrt{f_s}$ denotes the inverse correlation length of a pinned superfluid or the inverse London penetration depth of a superconductor.

The physical consequence of the pseudo-Goldstones (massive gauge fields) $\phi$ is that below the mass scale $\ell m$, they are too heavy to excite and the spectrum becomes trivial that is characteristic of superconductivity. In fact, despite there still existing an explicitly-unbroken U(1)$_p$ magnetic symmetry, there are no low-energy modes in the theory to carry the associated charge (magnetic fields for gauge theories) at macroscopically long distance and time scales. 
Controlling the $\ell$ parameter appropriately, allows us to systematically probe the regime near or above the gauge-field/pseudo-Goldstone mass scale.

Given the definition of the misalignment tensor $\psi$ in \cref{eq:psi-Higgs}, it satisfies a Bianchi identity
\begin{equation}
    \df\psi = \ell\xi - \ell c_\phi \Xi,
    \label{eq:Bianchi-psi}
\end{equation}
relating $\psi$ and $\xi$. The consequence of this identity is that the associated background sources $\tilde A_\psi$ and $\tilde A$ in the action \eqref{eq:sf-action-Higgs} feature a new U(1)$^\psi_{p+1}$ global symmetry 
\begin{align}
    \tilde A_\psi &\to \tilde A_\psi + \df \tilde\Lambda_\psi, \nn\\
    \tilde A &\to \tilde A - \ell \tilde\Lambda_\psi,
\end{align}
that is respected in the absence of the background defect field strength $\Xi = \df\Phi + A$. Note that the U(1)$_p$ gauge field $\tilde A$ shifts as a background phase, meaning that the U(1)$^\psi_{p+1}$ symmetry is explicitly broken; see \cref{eq:Phi-trans} for reference.
Correspondingly, $\tilde F = \df\tilde A$ appearing in the equations of motion for $\phi$ in \cref{eq:max-1-Coulomb} gets replaced with its new U(1)$^\psi_{p+1}$-invariant definition 
\begin{equation}
    \tilde F = \df \tilde A + \ell \tilde A_\psi.
    \label{eq:new-tildeF}
\end{equation}
The Bianchi identity \cref{eq:max-2} for $\xi$ remains the same as before. In fact, it follows as a consequence of the $\psi$ Bianchi identity in \cref{eq:Bianchi-psi}. This brings us to the physical interpretation of $\tilde A_\psi$. Specialising to the context of gauge theories, note that the external electric current $\star K_\ext$ defined in \cref{eq:external-currents} is no longer conserved in the presence of $\tilde A_\psi$. Instead, we get a new charge source term
\begin{equation}
    \df{\star K_\ext} = (-)^{q+1}\ell {\star Q_\ext},
\end{equation}
where we have defined the electric charge source $Q_\ext$ via
\begin{equation}
    \tilde F_\psi = -{\star Q_\ext}.
\end{equation}
Therefore $\tilde A_\psi$ contains information about the additional electric charges being pumped into the system causing the dynamical gauge field $\phi$ to acquire a mass. 

To make the action manifestly invariant under the new U(1)$^\psi_{p+1}$ global symmetry even in the presence of $\Xi$, we need to modify the anomaly-inflow Lagrangian in \cref{eq:bulk-lagrangian} to
\begin{equation}
    S_{\text{bulk}} = c_\phi\int_{\text{bulk}}
    \df A \wedge \tilde A
    + (-)^q \ell \Xi \wedge \tilde A_\psi. 
    \label{eq:bulk-S-higgs}
\end{equation}
Varying the new full action with respect to the associated background fields, we can read off the same currents $J$, $\tilde J$, and $L$ as defined in \cref{eq:gauge-invariant-currents,eq:defect-current-Coulomb}. There is also a new $(p+2)$-form U(1)$^\psi_{p+1}$ current coupled to $\tilde A_\psi$ given as
\begin{equation}
    \tilde J_\psi = (-)^{q+1}{\star\psi}.
    \label{eq:tildeJpsi}
\end{equation}
The full set of conservation equations is given as
\begin{subequations}
\begin{align}
    \df{\star J}
    &= c_\phi \tilde F + (-)^{q+1}\ell\,{\star L}, 
    \label{eq:cons-Higgs-1} \\
    \df{\star\tilde J_\psi} 
    &= 
    (-)^{p+1} \tilde c_\phi \ell\Xi
    + (-)^{p}\ell\,{\star\tilde J}, 
    \label{eq:cons-Higgs-2} \\
    \df{\star L} 
    &= (-)^q c_\phi \tilde F_\psi, 
    \label{eq:cons-Higgs-3} \\
    \df{\star \tilde J} 
    &= \tilde c_\phi F,
    \label{eq:cons-Higgs-4}
\end{align}
\label{eq:cons-Higgs}%
\end{subequations}
where $\tilde F_\psi = \df\tilde A_\psi$. Note that all four conservation equations are anomalous. We can define non-gauge-invariant conserved currents 
\begin{align}
    J_{\text{cons}} 
    &= J - \tilde c_\phi {\star\tilde A}, \nn\\
    \tilde J^\psi_{\text{cons}} 
    &= \tilde J_\psi + (-)^{q+1} c_\phi \ell{\star\Phi},  \nn\\
    L_{\text{cons}}
    &= L + (-)^{p} \tilde c_\phi {\star\tilde A_\psi}, \nn\\
    \tilde J_{\text{cons}} 
    &= \tilde J - c_\phi {\star A}~,
\end{align}
that satisfy the non-anomalous version of the conservation equations \eqref{eq:cons-Higgs}.

There are a few interesting features of the conservation equations that we should highlight. Firstly, we see that the U(1)$^\ell_{q-1}$ defect conservation equation \eqref{eq:cons-Higgs-3}, obtained by taking a differential of the U(1)$_q$ conservation equation \eqref{eq:cons-Higgs-1}, is now anomalous unlike the relaxed phase. This means that, instead of \cref{eq:Qell-Coulomb}, the topologically conserved U(1)$\ell_{q-1}$ defect charge now needs to be defined using the gauge-non-invariant conserved current and is given as
\begin{align}
    \ell Q_\ell[\Sigma_{d-q+1}]
    &= -\ell c_\phi\int_{\Sigma_{d-q+1}} 
    {\star{\cal J}_e} + (-)^q \tilde A_\psi \nn\\
    &= (-)^{q+1} Q[\dow\Sigma_{d-q+1}].
\end{align}
Furthermore, we can see that the U(1)$^\psi_{p+1}$ symmetry is explicitly broken by the U(1)$_p$ current (magnetic field lines) acting as defects. Correspondingly, the U(1)$_p$ conservation equation \eqref{eq:cons-Higgs-4} follows as a consequence of the approximate U(1)$^\psi_{p+1}$ conservation \eqref{eq:cons-Higgs-2}. Indeed, the U(1)$_p$ charge (magnetic field lines) defined in \cref{eq:mag-charges} becomes topological in the pinned phase; to wit
\begin{align}
    \ell\tilde Q[\Sigma_{d-p}] 
    &= \ell\int_{\Sigma_{d-p}} (-)^{pq+p+q}\xi - \tilde c_\phi A \nn\\
    &= (-)^{p} \tilde Q_\psi[\dow\Sigma_{d-p}]~,
\end{align}
where 
\begin{equation}
    \tilde Q_\psi[\Sigma_{d-p-1}]
    = (-)^{pq+q}\int_{\Sigma_{d-p-1}} \psi
    + c_\phi \ell \Phi,
\end{equation}
is the approximately conserved U(1)$^\psi_{p+1}$ charge operator.

% To wit, we can modify the U(1)$_p$ current in \cref{eq:tildeJ-cons} by a total-derivative background term as 
% \begin{align}
%     \tilde J'
%     = \tilde J_{\text{cons}} + c_\phi {\star\df\Phi}
%     = \tilde J + c_\phi{\star\Xi}.
% \end{align}
% The associated conserved charge is the same as \cref{eq:elec-charges}, provided that we take $\Phi$ to vanish at the boundary of the defining surface $\Sigma_{d-p}$. We find 
% \begin{align}
%     \ell\tilde Q'[\Sigma_{d-p}] 
%     &= \ell\int_{\Sigma_{d-p}} (-)^{pq+p+q}\xi - \tilde c_\phi \Xi \nn\\
%     &= (-)^{pq+p+q}\int_{\dow\Sigma_{d-q}} \psi.
%      \label{eq:Higgs-top-mag}
% \end{align}
% We see that the modified U(1)$_p$ charge is given entirely by the integral of $\psi$ at the boundary of $\Sigma_{d-p}$ and hence is topological.

\subsubsection{Duality transformations}

Notice that the conservation equations \eqref{eq:cons-Higgs} can be understood as corresponding to a system with an anomalous $\rmU(1)_q \times \rmU(1)^\psi_{p+1}$ global symmetry, with both the symmetry groups weakly explicitly broken. This suggests a duality under $q\leftrightarrow p+1$. This can be made precise coupling the action to Lagrange multipliers $\tilde\phi$ and $\tilde\phi_\psi$ imposing the Bianchi identities of $\xi$ and $\psi$ respectively. These take the form
\begin{align}
    S 
    &= \int 
    \frac{1}{c_\phi}\df\tilde\phi\wedge
    \lb \xi - c_\phi A \rb \nn\\
    &\qquad\qquad 
    + \frac{1}{c_\phi}
    \tilde\phi_\psi \wedge \lb \df \psi - \ell\xi + c_\phi \ell\Xi \rb.
    \label{eq:Lagrange-higgs}
\end{align}
Furthermore, we isolate the mass term from the matter Lagrangian and electric current to define
\begin{align}
    {\star{\cal L}_e} &= - \frac{m^2}{2}\psi\wedge{\star\psi} +
    {\star{\cal L}'_e}, \nn\\
    {\cal J}_e &= -m^2 \psi + {\cal J}'_e.
\end{align}
With these in place, $\xi$ and $\psi$ can be integrated out from the action to give
\begin{align}
    \xi 
    &= \frac{-1}{c_\phi f_s}{\star\tilde\xi}
    + \frac{1}{f_s} {\cal M}
    , \nn\\
    \psi 
    &= \frac{(-)^{p+1}}{c_\phi m^2}{\star\tilde\xi_\psi}
    + \frac{1}{m^2} {\cal J}'_e,
\end{align}
where we have defined 
\begin{align}
    \tilde\xi 
    &= \df\tilde\phi + \tilde c_\phi \tilde A 
    - \ell\tilde\phi_\psi, \nn\\
    \tilde\xi_\psi 
    &= \df\tilde\phi_\psi + \tilde c_\phi \tilde A_\psi.
    \label{eq:newdefs-xi}
\end{align}
Note that the definition of $\tilde\xi$ has been modified compared to \cref{sec:duality-zeroT} to make it U(1)$^\psi_{p+1}$-invariant. Plugging these back into the action \eqref{eq:sf-action-Higgs}, together with the Lagrange multiplier terms in \cref{eq:Lagrange-higgs}, we arrive at exactly the same theory as before but with substitutions
\begin{gather}
    \phi \leftrightarrow (-)^{p+1}\tilde\phi_\psi, \qquad 
    A \leftrightarrow \tilde A_\psi, \qquad
    \xi \leftrightarrow (-)^{p+1} \tilde\xi_\psi, \nn\\
    \ell\phi_\ell \leftrightarrow (-)^{p+1}\tilde\phi, \qquad
    \ell\Phi \leftrightarrow \tilde A, \qquad
    \psi \leftrightarrow (-)^p \tilde \xi, \nn\\
    c_\phi \leftrightarrow (-)^{p+1} \tilde c_\phi, \quad 
    f_s \to \frac{1}{m^2 c_\phi^2}, \quad 
    m^2 \to \frac{1}{f_s c_\phi^2}, \nn\\
    J \leftrightarrow \tilde J_\psi, \qquad 
    L \leftrightarrow \tilde J,
    \label{eq:Higgs-mappings}
\end{gather}
together with $q\leftrightarrow p+1$ and a transformation of the matter Lagrangian
\begin{equation}
    {\star{\cal L}'_e} \to 
    {\star{\cal L}'_e}
    - \frac{1}{2f_s} {\cal M} \wedge{\star\cal M}
    - \frac{1}{2m^2} {\cal J}'_e \wedge {\star{\cal J}'_e}.
\end{equation}
We also need to perform these transformations on the bulk anomaly-inflow Lagrangian defined in \cref{eq:bulk-S-higgs}, which results in
\begin{align}
    S_{\text{bulk}} 
    &\leftrightarrow 
    S_{\text{bulk}} 
    - c_\phi\int 
    A \wedge \tilde A
    + (-)^{q}
    \ell\Phi \wedge \tilde A_\psi.
\end{align}
Therefore, we are led to the conclusion that a U(1)$_q$ pseudo-superfluid in the pinned phase admits a dual description as a U(1)$_{p+1}$ pseudo-superfluid in the pinned phase. A corollary to this statement is that a U(1)$^\loc_{0}$ superconductor is self-dual in $d=2$ spatial dimensions and is dual to a pinned U(1)$_0$ superfluid in $d=1$ spatial dimensions.

\section{Approximate higher-form symmetries in thermal equilibrium}
\label{sec:higher-form-finite-temperature}
Our discussion thus far has focused on higher-form symmetries at zero temperature. When we couple the system to a thermal bath, it is no longer possible to base our description on a simple action principle due to the presence of stochastic thermal noise.\footnote{This is only partially true. In recent years, a new action principle approach for non-equilibrium thermal systems has been formulated, known as the Schwinger-Keldysh effective field theory~\cite{Crossley:2015evo, Jensen:2017kzi, Haehl:2018lcu}; see~\cite{Glorioso:2018wxw} for a review. Among other things, this framework requires a doubling of all degrees of freedom and a new KMS symmetry. At the level of classical hydrodynamic equations, this framework is equivalent to the conventional framework of dissipative hydrodynamics which we explore in this paper.} Things are under better control in thermal equilibrium, where one can talk about an equilibrium free energy or partition function instead of an effective action and use it to obtain the equilibrium values of the conserved currents. In this section, we will review higher-form symmetries in thermal equilibrium, drawing from our previous work~\cite{Armas:2018atq, Armas:2018zbe, Armas:2019sbe}. We will then proceed to weakly break these symmetries and discuss the repercussions thereof. 

\subsection{Temporal-pseudo-spontaneous symmetry breaking and pseudo-fluids}
\label{sec:fluids-eqb}

Consider a physical system with a spontaneously unbroken U(1)$_q$ global symmetry, meaning that there are no low-energy fields in the theory charged under the U(1)$_q$ symmetry. The thermal equilibrium state of such a system can be described by a partition function ${\cal Z}_\eqb$, expressed as a local functional of time-independent configurations of the U(1)$_q$ gauge field $A$. The equilibrium expectation values of the conserved U(1)$_q$ current $J_\eqb$ can be obtained by varying ${\cal Z}_\eqb$ with respect to $A$. The partition function is required to be invariant under time-independent U(1)$_q$ background gauge transformations
\begin{equation}
    A \to A + \df\Lambda 
    \qquad\text{s.t.}\qquad 
    \lie_\beta\Lambda = 0.
    \label{eq:eqb-gauge-trans}
\end{equation}
Here $\lie_\beta$ denotes the Lie derivative with respect to the thermal time vector $\beta^\mu = \delta^\mu_t/T_0$, with $T_0$ being the constant temperature of the thermal state. In non-covariant notation, the condition merely means $\dow_t\Lambda = 0$.
In general, the partition function can also depend on other thermodynamic variables relevant for the system such as temperature and chemical potentials, which we shall drop for now for clarity.

Let us start with the standard $q=0$ case for motivation. We can prepare a thermodynamic ensemble with a constant nonzero density $J^t_\eqb = \chi\mu_0$ using the partition function $T_0\ln {\cal Z}_\eqb = \half \int\df^dx\,\chi\,\mu^2$, with $\mu = \mu_0 + T_0\iota_\beta A$. 
Here $\mu_0$ is the equilibrium value of the chemical potential, $\chi$ the susceptibility, and $\iota_\beta$ denotes the exterior product with respect to $\beta^\mu$. The reason why this works is because the time-component of the gauge field $\iota_\beta A$ is invariant under time-independent gauge transformations. Unfortunately, as we can see from \cref{eq:eqb-gauge-trans}, this is no longer the case for $q\neq0$. In fact
\begin{equation}
    \iota_\beta A \to \iota_\beta A - \df \iota_\beta\Lambda
    \qquad\text{when}\qquad 
    \lie_\beta\Lambda = 0.
    \label{eq:A-inhomogeneous}
\end{equation}
This means that it is not possible to construct a thermodynamic ensemble with a spontaneously unbroken higher-form global symmetry that admits a nonzero $q$-form density in equilibrium.\footnote{It is still possible to construct higher-derivative theories with spontaneously unbroken higher-form global symmetry, with terms such as $\star\!\lb \df\iota_\beta A \wedge {\star\df\iota_\beta A} \rb$. However, the resultant $q$-form density is zero in the absence of background fields. See~\cite{Armas:2018zbe}.}
It is interesting to note that the gauge-parameters $\iota_\beta\Lambda$ are responsible for the Gauss-constraint components \eqref{eq:cons-gauss} of the U(1)$_q$ conservation equations. Since this constraint is non-trivial in equilibrium, we need a way to impose it in the free energy.

\subsubsection{Temporal-spontaneous symmetry breaking}
\label{sec:TSSB}

We can remedy this situation if we allow the global U(1)$_q$ symmetry to be spontaneously broken. We could bring in the full $q$-form Goldstone field $\phi$ from \cref{sec:SSB-zeroT}, but, for our purposes, it is sufficient to only spontaneously break the U(1)$_q$ symmetry in the time-direction. Note that this option is only available to us at finite temperature due to the existence of a preferred thermal rest frame, and turns out to apply to a wide array of physical systems including magnetohydrodynamics~\cite{Armas:2018atq, Armas:2018zbe} and viscoelastic crystals~\cite{Armas:2019sbe}.
To this end, we introduce a purely-spatial \tightmath{$(q-1)$}-form field $\varphi$ satisfying 
\begin{subequations}
\begin{equation}
    \iota_\beta\varphi = 0,
    \label{eq:varphi-condition}
\end{equation}
that transforms as
\begin{equation}
  \varphi \to \varphi - \iota_\beta \Lambda,
  \label{eq:varphi-trans}
\end{equation}
\end{subequations}
and hence can compensate for the inhomogeneous transformation in \cref{eq:A-inhomogeneous}. We can interpret $\varphi$ as a Goldstone field arising from the spontaneous breaking of the temporal-part of the U(1)$_q$ symmetry. In scenarios where the U(1)$_q$ symmetry is completely spontaneously broken, we could obtain $\varphi$ as the time-component of the full Goldstone field as $\varphi = \iota_\beta\phi/c_\phi$. We will return to this case in the next subsection. The utility of the temporal-spontaneous breaking of the U(1)$_q$ symmetry is that we can now define a chemical potential via
\begin{equation}
    \frac{\mu}{T_0}
    = \iota_\beta A 
    + \begin{cases}
        - \df\varphi, &\quad \text{for } q>0, \\
        \mu_0/T_0, &\quad \text{for } q=0,
    \end{cases}
    \label{eq:qform-mu}
\end{equation}
which is gauge-invariant in equilibrium.
It is easy to see that $\mu$ is also purely spatial in equilibrium, i.e. $\iota_\beta\mu = 0$, since $\lie_\beta\varphi = 0$.

In the presence of temporal-spontaneous breaking of the higher-form symmetry, the thermal partition function is given by a Euclidean path integral over all time-independent configurations of the temporal-Goldstone field, i.e. ${\cal Z}_\eqb = \int{\cal D}\varphi \exp S_\eqb$, where $S_\eqb$ denotes the Euclidean equilibrium effective action of the theory.\footnote{The quantity $-T_0 S_\eqb$ can be identified with the offshell free energy of the system.} Using the $q$-form chemical potential $\mu$, we can write down an expression for $S_\eqb$ as
\begin{equation}
    S_\eqb = \int_{\Sigma_\beta} \frac{\chi}{2}
    \mu \wedge {*\mu},
    \label{eq:freeE-string}
\end{equation}
which neatly generalises the thermal partition function from the $q=0$ case. The integral here is performed over a spatial slice $\Sigma_\beta$ normal to the thermal vector $\beta^\mu$.\footnote{In coordinates, this integral is just $\int_{\Sigma_\beta}{*f} = T_0^{-1} \int\df^dx\, f$.} Furthermore, ``$*$'' denotes the spatial Hodge duality operation, defined as $*\mu = \star(u\wedge\mu)$, with $u^\mu = T_0\beta^\mu$ denoting the equilibrium frame velocity. We can vary $S_\eqb$ with respect to the background U(1)$_q$ gauge field $A$ to obtain the respective current
\begin{equation}
    J_\eqb = u\wedge n,
    \label{eq:Jeqb-TSSB}
\end{equation}
where $n = \chi\mu$.
This current corresponds to a nonzero $q$-form density in equilibrium $J^{ti\ldots}_\eqb = n^{i\ldots}$, with zero flux. 

\begin{subequations}
The classical configuration equation for $\varphi$ obtained via extremising the equilibrium action precisely implements the Gauss-constraint \eqref{eq:cons-gauss} in equilibrium; to wit
\begin{equation}
    \df{*n} = 0.
    % \quad\overset{A=0}{\Longrightarrow}\quad 
    % \df{*\df\varphi} = 0,
    \label{eq:gauss-eqb}
\end{equation}
We can solve this constraint for a constant configuration of the chemical potential \tightmath{$\mu=\mu_0$}, or \tightmath{$\varphi = -\frac1q \iota_x \mu_0/T_0$} in the absence of background gauge fields, giving rise to the required nonzero $q$-form density. Depending on the boundary conditions, it is also possible to find more interesting solutions of the Gauss constraint; see~\cite{Armas:2018zbe} for more discussion on temporally-spontaneously broken higher-form symmetries. There is also a Bianchi identity associated with $\mu$ given as
\begin{equation}
    \df\mu = -\iota_u F.
    % \quad\Longrightarrow\quad 
    % \df\df\varphi = 0.
    \label{eq:bianchi-eqb}
\end{equation}
\end{subequations}
This is identically satisfied as long as $\varphi$ is smooth.

% The Gauss constraint also ensures that the equilibrium conserved charge
% \begin{equation}
%     Q_\eqb[\Sigma_{d-q}] = \chi\int_{\Sigma_{d-q}} {*\mu},
%     \label{eq:Q-eqb}
% \end{equation}
% is invariant under smooth deformations of $\Sigma_{d-q}$. Note that the Gauss constraint does not exist for $q=0$.

\subsubsection{Temporal-pseudo-spontaneous symmetry breaking}
\label{sec:TPSSB}

The effects of temporal-spontaneous symmetry breaking are further manifest when the symmetry is also approximately broken. In this case, the equilibrium action of the theory can additionally depend on the background phase field $\Phi$ introduced in \cref{sec:appx_higher_form} that shifts under U(1)$_q$ transformations as \cref{eq:Phi-trans}. The effective theory will now also possess a spatial U(1)$^\ell_{q-1}$ global symmetry associated with the conservation of defects
\begin{align}
    \Phi &\to \Phi
    + \begin{cases}
        \df\Lambda_\ell, &\text{for } q > 0, \\
        a_\ell, & \text{for } q=0,
    \end{cases}
    \quad\text{s.t.}\quad 
    \lie_\beta\Lambda_\ell = 0,
    \label{eq:U1defect-trans-eqb}
\end{align}
which is the equilibrium version of \cref{eq:U1defect-trans}. 
For \tightmath{$q>1$}, we run into the same issue concerning thermal equilibrium for the U(1)$^\ell_{q-1}$ global symmetry as we did for the U(1)$_q$ global symmetry previously. To circumvent this, we will require that the the U(1)$^\ell_{q-1}$ symmetry is at least temporally-spontaneously broken in the thermal state, giving rise to a \tightmath{$(q-2)$}-form temporal-Goldstone $\varphi_\ell$ with transformation $\varphi_\ell \to \varphi_\ell - \iota_\beta\Lambda_\ell$. The equilibrium action can now depend on the new \emph{defect chemical potential} $\mu_\ell$, defined as
\begin{equation}
    \frac{\mu_\ell}{T_0}
    = \ell \begin{cases}
        \iota_\beta \Phi - \varphi - \df\varphi_\ell, 
        &\quad \text{for } q>1, \\
        \iota_\beta \Phi - \varphi + \mu^\ell_0/T_0
        &\quad \text{for } q=1, \\
        \varnothing 
        &\quad \text{for } q=0,
    \end{cases} 
    \label{eq:qform-mu-ell}
\end{equation}
which is invariant under both U(1)$_q$ and U(1)$^\ell_{q-1}$ symmetry transformations. It is easy to see that $\mu_\ell$ is purely spatial as well like $\mu$, i.e. $\iota_\beta\mu_\ell = 0$. Note that $\mu_\ell$ does not exist for $q=0$, as there is no concept of temporal-spontaneous breaking symmetry breaking.
This is to be expected because the scalar source term $L$ associated with an approximate U(1)$_0$ symmetry, i.e. $\dow_\mu J^\mu = -\ell L$, does not feature any conservation of its own.

Depending on the application in mind, we can also completely-spontaneously break the U(1)$^\ell_{q-1}$ symmetry, giving rise to the full Goldstone field $\phi_\ell$ introduced in \cref{sec:Higgs-zeroT}. However, as long as the U(1)$_q$ symmetry is only spontaneously broken in the time direction, the effective theory would still be the same because there are no possible symmetry-invariants that depend on the spatial components of $\phi_\ell$.

Using $\mu_\ell$, we can write down a new term in the equilibrium action 
\eqref{eq:freeE-string} that was previously disallowed by symmetries
\begin{align}
  S_\eqb 
  &= \int_{\Sigma_\beta} \frac{\chi}{2} \mu \wedge {*\mu}
    + \frac{\chi_\ell}{2} \mu_\ell \wedge {*\mu_\ell}.
    \label{eq:freeE-varphi-Phi}
\end{align}
This results in a defect current obtained by varying the equilibrium action with respect to $\Phi$ as
\begin{equation}
    L_\eqb = u\wedge n_\ell,
    \label{eq:Leqb-TPSSB}
\end{equation}
where $n_\ell = \chi\mu_\ell$,
while the U(1)$_q$ current is still given by \cref{eq:Jeqb-TSSB}.
We can see that $\mu_\ell$ acts as the chemical potential for defects, with $\chi_\ell$ being the associated susceptibility. We emphasise that the temporal-spontaneous breaking of U(1)$_q$ and U(1)$^\ell_{q-1}$ symmetries, for $q>0$ and $q>1$ respectively, is crucial to be able to generate a thermodynamic density of defects. 
The configuration equation for $\varphi$ now gives rise to the Gauss-constraint \eqref{eq:cons-gauss-broken} including defects, i.e.
\begin{subequations}
\begin{gather}
    \df{*n}
    = (-)^{q+1}\ell {*n_\ell}.
    % \nn\\
    % \quad\overset{A=0,\Phi=0}{\implies}\quad 
    % \df{*\df\varphi}
    % + (-)^{q} k_0^2\, {*\varphi}
    % = 0,     
    \label{eq:gauss-broken-eqb}
\end{gather}
 The Bianchi identity \eqref{eq:bianchi-eqb} remains the same, but now we also have another such identity associated with $\mu_\ell$, i.e.
\begin{equation}
    \df\mu_\ell = \ell\lb \mu - \iota_u\Xi\rb.
    % \quad\Longrightarrow\quad 
    % \df\df\varphi_\ell = 0.
    \label{eq:bianchi-ell-eqb}
\end{equation}
\end{subequations}
For nonzero $\ell$, the identity \eqref{eq:bianchi-eqb} follows as a differential of \cref{eq:bianchi-ell-eqb}. 

Using this identity back into \cref{eq:gauss-broken-eqb}, we can see that due to the presence of defects, both $q$-form charge and defects admit a finite inverse correlation length given by $k_0 = \ell\sqrt{\chi_\ell/\chi}$. Indeed, switching off the background fields, we find 
% \begin{align}
%     \lb \dow^2 - k_0^2\rb n
%     &= 0, \nn\\
%     %
%     \lb \dow^2 - k_0^2 \rb n_\ell
%     &= 0.
%     \label{eq:mu-relax}
% \end{align}
\begin{align}
    \Big( \df{*\df}
    + (-)^{p} k_0^2{*} \Big) {*n}
    &= 0, \nn\\
    \Big( \df{*\df} 
    + (-)^{q} k_0^2{*} \Big) n_\ell
    &= 0.
    \label{eq:mu-relax}
\end{align}
Note that the transverse components of $n$ and the longitudinal components of $n_\ell$ are identically zero in the absence of background fields, i.e. $\df n = \df{*n_\ell} = 0$.

% \begin{align}
%     \Big( \df{*\df}
%     + (-)^{p} k_0^2{*} \Big) {*n}
%     &= (-)^{pq+p+q} \ell^2\chi_\ell \iota_u\Xi, \nn\\
%     %
%     \Big( \df{*\df} 
%     + (-)^{q} k_0^2{*} \Big) n_\ell
%     &= 
%     - \ell\chi_\ell\df{*\iota_u\Xi}.
%     \label{eq:mu-relax}
% \end{align}

% To understand the physical implications of these equations, we can go to momentum space and decompose the U(1)$_q$ density into longitudinal and transverse components as $n = \hat k\wedge n_\| + n_\perp$, where $\hat k = k/|k|$ denotes the direction of the wavevector $k$, and $\iota_k n_\| = \iota_k n_\perp = 0$. Due to the Bianchi identity, we have that $\ell n_\ell = i|k| n_\|$. Using \cref{eq:mu-relax}, one finds that $n_\|$ (and hence $n_\ell$) has a finite correlation length $k_0$, while $n_\perp$ has infinite correlation length. To wit
% \begin{align}
%     G_{n_\| n_\|}
%     = \frac{-\ell^2}{k^2} G_{n_\ell n_\ell}
%     &= \frac{\chi k_0^2}{k^2 + k_0^2}, \nn\\
%     G_{n_\perp n_\perp}
%     &= \chi.
% \end{align}

% The U(1)$_q$ charge defined in \cref{eq:Q-eqb} is no longer conserved under deformations of the defining surface $\Sigma_{d-q}$. However, the U(1)$^\ell_{q-1}$ topological charge of defects defined over a surface $\Sigma_{d-q+1}$ is conserved
% \begin{align}
%     \ell Q^\eqb_\ell[\Sigma_{d-q+1}]
%     &= \ell\chi_\ell\int_{\Sigma_{d-q+1}}{*\mu_\ell},
% \end{align}
% and is given by just the value of the non-conserved U(1)$_q$ charge at the boundary of $\Sigma_{d-q+1}$.

\subsubsection{Applications}
\label{sec:applications-fluid}

\noindent
\emph{Relaxed fluids:} The equilibrium currents \eqref{eq:Jeqb-TSSB} and \eqref{eq:Leqb-TPSSB} describe the equilibrium state of a  U(1)$_q$ fluid relaxed due to the presence of U(1)$^\ell_{q-1}$ defects. This is also the theory we will be left with after phase transition from a U(1)$_q$ superfluid with U(1)$^\ell_{q-1}$ defects, mediated by the proliferation of vortices. We will discuss more about vortices in the next subsection. 

\noindent
\emph{Superfluids with vortices:}
The symmetry breaking pattern outlined above can also be used to describe a U(1)$_p$ superfluid with vortices, in a phase where the fluctuations of the U(1)$_p$ charge itself have been relaxed due to the presence of impurities or can be ignored at the time-scales under consideration. This leaves only the superfluid velocity as the low-energy degree of freedom. In this interpretation, 
the $q$-form $\chi\mu$ represents the spatial Hodge dual of the superfluid velocity, which is conserved in the absence of vortices due to the Bianchi identity, while the \tightmath{$(q-1)$}-form $\ell\chi_\ell\mu_\ell$ represents the vortex density.\footnote{Conventional 0-form superfluids with vortices were discussed in \cite{Davison:2016hno} but the vortex susceptibility was not accounted for.} The parameters $1/\chi$ and $\ell\chi_\ell$ can be seen as superfluid density and vortex susceptibility respectively. In the presence of vortices, the correlation length of the superfluid velocity becomes finite and is given by $1/k_0$.
We will revisit this system in more details in the subsequent subsections and see how this arises from a consistent limit of the full superfluid including U(1)$_p$ charge fluctuations. 

\noindent
\emph{Electrohydrodynamics with free electric charges:}
The same setup can also be applied to the theory of ``electrohydrodynamics''. This is seen as a limit of electromagnetism near the mass scale of charged matter fields, so that electric fields are not strongly screened, but looking at small enough time-scales that the fluctuations of magnetic fields can be effectively neglected. For a U(1)$_{q-1}^\loc$ gauge theory, in the absence of free charges, this theory furnishes a U(1)$_q$ global symmetry associated with conserved electric field lines, with the $q$-form $\mu$ playing the role of electric fields and $\chi$ that of electric permittivity. In the presence of free electric charges, however, the U(1)$_q$ symmetry is violated by the \tightmath{$(q-1)$}-form free charge density $\ell\chi_\ell\mu_\ell$ with susceptibility $\ell\chi_\ell$. In this interpretation, $1/k_0$ represents the Debye screening length, which decreases with the increasing susceptibility of free charges.

\noindent
\emph{Magnetohydrodynamics with free magnetic charges:}
Perhaps the best explored example of temporal-spontaneous symmetry breaking is magnetohydrodynamics~\cite{Armas:2018atq, Armas:2018zbe}. This applies to electromagnetism with charged matter at energy scales well above the mass scale of the matter fields. In this regime, electric fields and charged particles have mutually screened each other, leaving a theory of just the conserved magnetic fields. For a U(1)$_{p-1}^\ell$ gauge theory, magnetic fields realise a U(1)$_{q}$ global symmetry. In this interpretation, the $q$-form $\chi\mu$ plays the role of magnetic fields that are conserved due to the Bianchi identity, with $\chi$ being the magnetic permeability.
Explicit breaking of the U(1)$_{q}$ symmetry amounts to the introduction of the magnetic monopoles. We can identify the \tightmath{$(q-1)$}-form $\ell\chi_\ell\mu_\ell$ as the magnetic monopole density and $\ell\chi_\ell$ as the susceptibility of magnetic charge. The static correlation length $1/k_0$, in this scenario, is the magnetic equivalent of the Debye screening length. 
While fundamental magnetic monopoles have not been observed in nature, this effective theory can also be applied to ``emergent magnetic monopoles'' observed in various condensed matter systems such as spin ice~\cite{Castelnovo:2007qi} and anomalous hall effect~\cite{Fang:2003ir}. Later in our discussion, we will set up a theory of magnetic monopoles more formally as topological defects of the U(1)$^\loc_{p-1}$ gauge field, and show how we reduce to the description above when electric fields are screened.

\subsection{Pseudo-spontaneous symmetry breaking and relaxed pseudo-superfluids}
\label{sec:appx-SF-Coulomb-eqb}

The discussion of the previous section can be easily extended to the superfluid phase where the U(1)$_q$ global symmetry is completely-spontaneously broken, giving rise to the $q$-form Goldstone field $\phi$ introduced in \cref{sec:SSB-zeroT}. When the symmetry is further explicitly broken, the pseudo-superfluid can exist in one of two phases depending on the status of the U(1)$^\ell_{q-1}$ global symmetry in \cref{eq:U1defect-trans-eqb}. The first of these is the relaxed phase, where the U(1)$^\ell_{q-1}$ global symmetry is only spontaneously broken in the time-direction, or spontaneously unbroken for $q=1$ or simply unbroken for $q=0$. This results in the spatial components of the Goldstone $\phi$, or just $\phi$ as a whole for $q=0$, remaining massless. This phase will describe a relaxed superfluid or the relaxed phase of electromagnetism at finite temperature. It is also in this phase that we will be able to consistently introduce vortices (or magnetic monopoles) as topological defects in the configurations of the spatial components of $\phi$, or $\phi$ as a whole for $q=0$. 

The other phase of a pseudo-superfluid is the pinned phase, where the U(1)$^\ell_{q-1}$ global symmetry is completely-spontaneously broken, or simply broken for $q=0$. This results in the Goldstone $\phi$ becoming massive and applies to the theory of pinned superfluids or the Higgs phase of electromagnetism (superconductivity) at finite temperature. In this subsection, we devote our attention to the relaxed phase of a pseudo-superfluid and return to the pinned phase in \cref{sec:appx-SF-Higgs-eqb}.

\subsubsection{Spontaneous symmetry breaking}
\label{sec:CSSB-thermal}

Let us start with the higher-form superfluid where the U(1)$_q$ global symmetry is not yet explicitly broken. Noting the definition of superfluid velocity from \cref{eq:xi-defn}, and upon identifying the temporal-Goldstone field as $\varphi = \iota_\beta\phi/c_\phi$, one can check that $\iota_u\xi = c_\phi\mu$ in equilibrium.
However, the equilibrium action \eqref{eq:freeE-string} can now contain a new term containing the spatial components of the superfluid velocity
\begin{align}
    S_\eqb
    &= \int_{\Sigma_\beta} \frac{\chi}{2} \mu \wedge {*\mu}
    - \frac{f_s}{2} \xi\wedge{*\xi} \nn\\
    &\qquad\qquad\qquad
    - \iota_u(\xi\wedge \tilde A),
    \label{eq:freeE-string-sf}
\end{align}
where $f_s$ is the superfluid density we met in \cref{sec:SSB-zeroT}. We have also introduced the background coupling term for the U(1)$_p$ topological symmetry coming from \cref{eq:LG_goldstone}.
Varying the equilibrium action with respect to $A$, accounting for the bulk action in \cref{eq:bulk-lagrangian}, we can read off the associated U(1)$_q$ and U(1)$_p$ currents in equilibrium
\begin{align}
    J_\eqb
    &= u\wedge n - c_\phi f_s\,\rmP_u\xi \nn\\
    &= \lb 1 - c_\phi^2f_s/\chi\rb u\wedge n
    - c_\phi f_s\,\xi, \nn\\
    \tilde J_\eqb
    &= (-)^q u\wedge{*\xi}
    - c_\phi\, {*\mu} \nn\\
    &= \star\xi,
    \label{eq:Jeqb-SSB}
\end{align}
where $\rmP_u = (1 + u\wedge \iota_u)$ denotes the projection operator transverse to $u^\mu$. We still find a nonzero $q$-form density $J^{ti\ldots}_\eqb = n^{i\ldots}$ as before, but now we also have a nonzero flux $J^{i\ldots}_\eqb = - c_\phi f_s\xi^{i\ldots}$ characteristic of superfluids. 
Comparing this expression to the zero temperature version of the current in \cref{eq:J-electric}, we can identify the term proportional to $ 1-c_\phi^2 f_s/\chi$ in $J_\eqb$ as the contribution coming from the matter polarisation tensor ${\cal M}$. Whereas, the expression for $\tilde J_\eqb$  is the same as in \cref{eq:tildeJ}.

\begin{subequations}
The configuration equations for the temporal-components of $\phi$ lead to the same Gauss constraint from \cref{eq:gauss-eqb}, but modified due to $\tilde A$, i.e.
\begin{equation}
    \df{*n} 
    = c_\phi\rmP_u\tilde F.
    % \quad\overset{A,\tilde A=0}{\Longrightarrow}\quad 
    % \df{*\df\varphi} = 0.
\end{equation}
In addition, we now also have the configuration equations coming from the spatial components of $\phi$, leading to
\begin{gather}
    \df {*\xi} = (-)^{q+1} \frac{1}{f_s} \iota_u\tilde F.
    % \quad\overset{A,\tilde A=0}{\Longrightarrow}\quad 
    % \df{*\df\phi} = 0,
    \label{eq:gauss-eqb-spatial}
\end{gather}
Because the fluxes in the superfluid phase are nonzero, these equations are necessary to ensure that the dynamical components of the U(1)$_q$ conservation equations \eqref{eq:cons-dyn-broken} are satisfied in thermal equilibrium. The Bianchi identity generalises from \cref{eq:max-2} to \cref{eq:bianchi-eqb}.
\end{subequations}

% \begin{subequations}
% The associated conserved U(1)$_q$ and U(1)$_p$ charges are now given as
% \begin{align}
%     Q_\eqb[\Sigma_{d-q}] 
%     &= \int_{\Sigma_{d-q}}
%     \chi\,{*\mu}
%     - c_\phi f_s (-)^q\, u\wedge {*\xi} \nn\\
%     &\hspace{10em}
%     - c_\phi \tilde A
%     \label{eq:Qeqb-SSB} \\
%     \tilde Q_\eqb[\Sigma_{d-p}] 
%     &= 
%     \int_{\Sigma_{d-p}} 
%     (-)^{pq+p+q}\, \rmP_u\xi 
%     - \tilde c_\phi\,u\wedge\mu \nn\\
%     &\hspace{10em}
%     - \tilde c_\phi A.
%     \label{eq:tQeqb-SSB}
% \end{align}
% Note that the flux terms do not contribute to these integrals if the surfaces $\Sigma_{d-q}$ and $\Sigma_{d-p}$ are stationary in the thermal rest frame.
% \label{eq:Qeqb-SSB-both}
% \end{subequations}

\subsubsection{Pseudo-spontaneous symmetry breaking}
\label{sec:Coulomb-eqb}

Let us introduce a weak explicit breaking of the U(1)$_q$ global symmetry. We will focus on the relaxed phase of the pseudo-superfluid, where the U(1)$^\ell_{q-1}$ symmetry is only spontaneously broken in the time-direction, allowing us to define the defect chemical potential $\mu_\ell$ as in \cref{eq:qform-mu-ell}. However, there are no symmetry-invariants in the theory that contain the spatial components of $\phi$ without acted upon by derivatives. We can try to define a misalignment tensor as
\begin{equation}
    \psi = \ell
    \begin{cases}
    \phi - c_\phi \Phi - c_\phi u\wedge \df\varphi_\ell,
    &\quad \text{for } q>1, \nn\\
    \phi - c_\phi \Phi + c_\phi u\, \mu^\ell_0/T_0,
    &\quad \text{for } q=1, \nn\\
    \phi - c_\phi \Phi,
    &\quad \text{for } q=0.
    \end{cases}
    \label{eq:psi-Coulomb}
\end{equation}
Taking $\Lambda = \df\lambda$, $\Lambda_\ell = \lambda$ (or $\Lambda = a$, $a_\ell = a$ for $q=0$), we can see that $\psi$ transforms under the spatial version of the U(1)$^\loc_{q-1}$ symmetry in \cref{eq:psi-trans}, i.e.
\begin{equation}
    \psi = \psi - \ell c_\phi 
    \begin{cases}
    \rmP_u\df\lambda,
    &~~ \text{for } q>0, \\
    a, &~~ \text{for } q=0.
    \end{cases}
    \quad\text{s.t.}\quad 
    \lie_\beta\lambda = 0.
    \label{eq:psi-trans-eqb}
\end{equation}
For $q=0$, $\psi$ has a constant shift symmetry. For $q>0$, the time-component of $\psi$ is gauge-invariant and is precisely given by $\mu_\ell$; to wit $\iota_u\psi = -c_\phi\mu_\ell$ in equilibrium. However, the spatial components behave like a gauge field.

All in all, we are allowed to add the superfluid density term from \cref{eq:freeE-string-sf} to the equilibrium action \eqref{eq:freeE-varphi-Phi}, i.e.
\begin{align}
    S_\eqb
    &= \int_{\Sigma_\beta} \frac{\chi}{2} \mu \wedge {*\mu}
    - \frac{f_s}{2} \xi\wedge{*\xi}
    + \frac{\chi_{\ell}}{2} \mu_\ell \wedge {*\mu_\ell} \nn\\
    &\qquad\qquad\qquad
    - \iota_u(\xi\wedge \tilde A).
    \label{eq:freeE-SSB-Coulomb}
\end{align}
The equilibrium versions of U(1)$_q$, U(1)$_p$ currents and the U(1)$^\ell_{q-1}$ defect current are the same as \cref{eq:Jeqb-SSB} and \eqref{eq:Leqb-TPSSB} respectively. The classical configuration equations for the temporal-components of $\phi$ are given by \cref{eq:gauss-broken-eqb} modified with the $\tilde A$ term
\begin{gather}
    \df{*n}
    = c_\phi\rmP_u\tilde F
    + (-)^{q+1}\ell {*n_\ell}.
    % \nn\\
    % \quad\overset{A,\tilde A,\Phi=0}{\implies}\quad 
    % \df{*\df\varphi}
    % = (-)^{q+1} k_0^2\, {*\varphi}.
    \label{eq:gauss-broken-anom-eqb}
\end{gather}
The configuration equations for the spatial components of $\phi$ are still given by \cref{eq:gauss-eqb-spatial}. The Bianchi identities are the same as \cref{eq:max-2} and \eqref{eq:bianchi-ell-eqb}. The presence of defects screens the $q$-form charges and defects same as \cref{eq:mu-relax}, but the spatial components of $\xi$ remain unscreened.

% \begin{align}
%     \Big(\df{*\df}
%     + (-)^{p} k_0^2 {*} \Big) {*n}
%     &= c_\phi \df{* \tilde F}
%     + (-)^{pq+p+q} \ell^2\chi_\ell \iota_u\Xi, \nn\\
%     %
%     \Big(\df{*\df}
%     + (-)^{q} k_0^2 {*} \Big) n_\ell
%     &= 
%     \ell\, \frac{c_\phi\chi_\ell}{\chi}\rmP_u\tilde F
%     - \ell\chi_\ell \df{*\iota_u\Xi}.
%     \label{eq:mu-relax-Coulomb}
% \end{align}

% Due to the presence of defects, the U(1)$_q$ charge given in \cref{eq:Qeqb-SSB} is now only approximately conserved. However, the associated defects are topologically conserved
% \begin{align}
%     \ell Q^\eqb_\ell[\Sigma_{d-q+1}] 
%     &= \ell\int_{\Sigma_{d-q+1}}
%     \chi_\ell\,{*\mu_\ell}.
%     \label{eq:Qell-PSSB}
% \end{align}
% The U(1)$_p$ charge given in \cref{eq:tQeqb-SSB} is still conserved.

\subsubsection{Vortices and magnetic monopoles}

We can also explicitly break the U(1)$_p$ magnetic global symmetry of a U(1)$_q$ superfluid. This results in the introduction of vortices, or magnetic monopoles in the context of U(1)$^\loc_{q-1}$ gauge theory. 
An interesting consequence of finite temperature is that we can do this even when the U(1)$_q$ symmetry is explicitly broken, provided that we are in the relaxed phase of the U(1)$_q$ pseudo-superfluid, as we now explain. 
In the presence of vortices, the U(1)$_q$ Goldstone field $\phi$ need not be well-behaved and we can decompose its gradient into a vortex-free and vortex-induced part as in \cref{eq:V-intro}. Of course, any such splitting is inherently redundant and we must supplement it with a new local U(1)$^\loc_q$ gauge symmetry
\begin{align}
    \bar\phi \to \bar\phi - \tilde \ell a, \qquad
    V \to V + \df a.
\end{align}
This should not be confused with the local U(1)$^\loc_{q-1}$ gauge symmetry we have previously introduced. Without loss of generality, we can work in a gauge where $\iota_\beta V = 0$ and $\iota_\beta\bar\phi = \iota_\beta\phi = c_\phi\varphi$. This leaves us with purely spatial residual U(1)$^\loc_q$ gauge transformations
\begin{equation}
    \iota_\beta a = \lie_\beta a = 0.
    \label{eq:V-gauge}
\end{equation}
In fixing this gauge, we have essentially chosen to shift all the singular ``vortex information'' into the spatial components of $\phi$ and picked out the temporal components $\varphi$ to be smooth. 

As long as the U(1)$^\loc_{q-1}$ gauge symmetry in \cref{eq:psi-trans-eqb} is respected, all the dependence on $\phi$ in the effective theory can only appear via $\varphi$ or $\xi$, both of which are smooth in the presence of vortices. Therefore, the relaxed phase of U(1)$^\loc_{q-1}$ gauge theories is compatible with vortices (magnetic monopoles) at finite temperature. Note that for ordinary U(1)$_0$ superfluids, there is no notion of the temporal-Goldstone $\varphi$, so the restriction is still the same as it was at zero temperature, i.e. all the dependence on the U(1)$_0$ phase $\phi$ must arise via $\xi$  to allow for vortices in the configurations of $\phi$.

Due to the presence of vortices, the equilibrium action in \cref{eq:freeE-SSB-Coulomb} can now contain a new term
\begin{align}
    S_\eqb
    &= \int_{\Sigma_\beta} \frac{\chi}{2} \mu \wedge {*\mu}
    - \frac{f_s}{2} \xi\wedge{*\xi} \nn\\
    &\qquad\qquad
    + \frac{\chi_{\ell}}{2} \mu_\ell \wedge {*\mu_\ell} 
    - \frac{1/\tilde\chi_\ell}{2} \df V \wedge {*\df V} \nn\\
    &\qquad\qquad
    - \iota_u(\xi\wedge \tilde A) 
    - \tilde\ell\,(-1)^{q}{\iota_u(\df V\wedge \tilde\Phi)}.
    \label{eq:freeE-defect}
\end{align}
We have also introduced a source term coupled to $\df V$ in the equilibrium action. The equilibrium versions of U(1)$_q$, U(1)$_p$ currents and the U(1)$^\ell_{q-1}$ defect current are the same as \cref{eq:Jeqb-SSB} and \eqref{eq:Leqb-TPSSB} respectively. However, we now have a U(1)$^\ell_{p-1}$ defect current
\begin{equation}
    \tilde L_\eqb 
    = - u \wedge {*\df V}
    = (-)^{q} {\star\df V}.
\end{equation}
Varying this with respect to the time-components of $\bar\phi$, we recover the Gauss constraint in \cref{eq:gauss-broken-anom-eqb}. However, varying with respect to the spatial components of $\bar\phi$ and $V$, we now find
\begin{subequations}
\begin{align}
    \df {*\xi} 
    &= (-)^{q+1} \frac{1}{f_s} \iota_u\tilde F, 
    \label{eq:dyn-broken-eqb-mon-xi} \\
    \df {*(\df \xi - c_\phi F)}
    &= (-)^{q+1} \tilde\ell^2\tilde\chi_\ell
    \lb f_s{*\xi}
    - (-)^q \iota_u\tilde\Xi
    \rb.
    \label{eq:dyn-broken-eqb-mon-V}
\end{align}
For any nonzero $\tilde\ell$, the first equation is just the derivative of the second. This is to be expected on account of the U(1)$^\loc_{q}$ spatial gauge symmetry remaining after the gauge-fixing in \cref{eq:V-gauge}. On the other hand, for $\tilde\ell=0$, the second equation becomes trivial, while the first reverts to the original vortex-free version of the configuration equations in \cref{eq:gauss-eqb-spatial}. The Bianchi identity in the presence of vortices  becomes \cref{eq:max-2-broken},
\end{subequations}
while the Bianchi identity \eqref{eq:bianchi-ell-eqb} remains the same. 

The $q$-form charges and defects are screened as in \cref{eq:mu-relax}. In the presence of vortices, the spatial components of $\xi$ are also relaxed as we can see from \cref{eq:dyn-broken-eqb-mon-V}. In the absence of background fields, this yields
\begin{align}
    \Big(\df {*\df} + (-)^q \tilde k_0^2 {*} \Big)\xi
    &= 0,
    \label{eq:xi-relax-vor}
\end{align}
% \begin{align}
%     \Big(\df {*\df} + (-)^q \tilde k_0^2 {*} \Big)\xi
%     &= c_\phi \df{*F} 
%     + \frac{\tilde k_0^2}{f_s} \iota_u\tilde\Xi,
%     \label{eq:mu-relax-vor}
% \end{align}
where $\tilde k_0 = \tilde\ell\sqrt{\tilde\chi_\ell f_s}$ denotes the new inverse correlation length associated with the spatial components of $\xi$. 

% \hl{What's the correlation length doing below?}
% \begin{align}
%     G_{\xi_\perp \xi_\perp}
%     = \frac{-\tilde \ell^2}{k^2} G_{\tilde n_\ell \tilde n_\ell}
%     &= \frac{1}{f_s} \frac{\tilde k_0^2}{k^2 + \tilde k_0^2}, \nn\\
%     G_{\xi_\|\xi_\|}
%     &= \frac{1}{f_s}.
% \end{align}

% Since the U(1)$_q$ and U(1)$_p$ symmetries are both explicitly broken, the respective charges in \cref{eq:Qeqb-SSB-both} are only approximately conserved. 
% The U(1)$^\ell_{q-1}$ conserved topological charge is given by \cref{eq:Qell-PSSB}, while the U(1)$^\ell_{p-1}$ conserved topological charge is 
% \begin{align}
%     \tilde\ell \tilde Q^\eqb_\ell[\Sigma_{d-p+1}] 
%     &= (-)^{pq+q+1} \tilde\ell \int_{\Sigma_{d-p+1}}
%     \df V.
%     \label{eq:tQell-PSSB}
% \end{align}

\subsubsection{Duality transformations}

As we have discussed at length in \cref{sec:higher-form-general}, when the U(1)$_q$ global symmetry is not explicitly broken, we can employ the duality procedure in \cref{sec:SSB-zeroT} to dress the theory of U(1)$_q$ superfluids with vortices equivalently as a U(1)$_p$ pseudo-superfluid in the relaxed phase. The ingredients we need are the dual-Goldstone terms similar to \cref{eq:lagrange-mult} to impose the Bianchi identities
\begin{align}
    S_\eqb
    &\sim \int_{\Sigma_\beta} \iota_u\!
    \lb \frac{1}{c_\phi}
    \df\tilde\phi \wedge (\xi - c_\phi A - \tilde\ell V) \rb \nn\\
    &\qquad\qquad
    + \iota_u \!\lb \frac{\tilde\ell}{c_\phi} (-)^{p+1} \df\tilde\phi_\ell \wedge \df V \rb.
    \label{eq:lagrange-eqb}
\end{align}
We can now go ahead and integrate out $\xi$ and $\df V$ to give
\begin{subequations}
\begin{align}
    \xi 
    % &= - \frac{1/f_s}{c_\phi} \lb {\star\tilde\xi}
    % + \frac{\chi_n}{\chi} u\wedge \iota_u {\star\tilde\xi} \rb \nn\\
    &=
    (-)^{pq+p+q} \frac{1}{f_s} {*\tilde\mu}
    + (-)^{p+1}\frac{c_\phi}{\chi} u\wedge {*\tilde\xi}, \\
    \df V 
    &= -(-)^{pq+q} \tilde\chi_\ell\, {*\tilde\mu_\ell},
\end{align}
\end{subequations}
where $\tilde\mu$, $\tilde\mu_\ell$ are defined using $\tilde\varphi = \iota_\beta\tilde\phi/\tilde c_\phi$, $\tilde\varphi_\ell = \iota_\beta\tilde\phi/\tilde c_\phi$ similar to \cref{eq:qform-mu,eq:qform-mu-ell}.
Substituting these back into the equilibrium action together with the Lagrange multiplier terms in \cref{eq:lagrange-eqb}, we recover the relaxed phase of a U(1)$_p$ pseudo-superfluid without vortices in \cref{eq:freeE-SSB-Coulomb}, with the same substitutions outlined in \cref{eq:EM-mappings,eq:EM-mappings-vor} together with 
\begin{gather}
    \mu\leftrightarrow\tilde\mu, \qquad 
    \mu_\ell\leftrightarrow\tilde\mu_\ell, \nn\\
    \chi\leftrightarrow \tilde\chi, \qquad 
    \chi_\ell\leftrightarrow\tilde\chi_\ell, 
\end{gather}
where we have identified $\tilde\chi = 1/f_s$.

% dual representation of the theory
% \begin{align}
%     S_\eqb 
%     &= \int_{\Sigma_\beta}\frac{\tilde\chi}{2} 
%     \tilde\mu \wedge {*\tilde\mu}
%     - \frac{\tilde f_s}{2} 
%     \tilde\xi \wedge {*\tilde\xi}
%     + \frac{\tilde\chi_\ell}{2} 
%     \tilde\mu_\ell \wedge {*\tilde\mu_\ell}
%     \nn\\
%     &\qquad\qquad
%     - \iota_u(\tilde\xi \wedge A)
%     - c_\phi\iota_u(A \wedge \tilde A).
% \end{align}
% Comparing this to , we can see that we have arrived at the relaxed phase of a U(1)$_p$ pseudo-superfluid without vortices. The susceptibility and superfluid density play the roles inverse of each other 
% \begin{equation}
%     \tilde\chi = \frac{1}{f_s}, \qquad 
%     \tilde f_s = \frac{1}{\chi},
% \end{equation}
% whereas $\tilde\chi_\ell$ becomes the susceptibility of the conserved U(1)$_p$ defects. The last term in the first line allows us to probe the U(1)$_q$ current $J = \star\tilde\xi$. On the other hand, the gauge-non-invariant term in the last line is due to the $\rmU(1)_q\times\rmU(1)_p$ anomaly in the theory and can be removed by re-expressing the bulk Lagrangian as in \cref{eq:bulk-lagrangian-dual}.

Normally, at zero temperature, we cannot perform a duality transformation when the U(1)$_q$ symmetry is explicitly broken. This is because the description now explicitly depends on the Goldstone field $\phi$ and not on its derivatives $\xi$ alone. However, when describing the relaxed phase of a U(1)$_q$ superfluid in thermal equilibrium, the transformation \eqref{eq:psi-trans-eqb} implies that equilibrium action only explicitly depends on the time-component $\varphi = \iota_\beta\phi/c_\phi$ of the Goldstone field $\phi$ and not on its spatial components. Therefore, we still have access to the partial duality transformation, where we only integrate out the spatial components of $\xi$. To this end, instead of \cref{eq:lagrange-eqb}, we only introduce a \tightmath{$(p-1)$}-form Lagrange multiplier $\tilde\varphi$ and a \tightmath{$(p-2)$}-form Lagrange multiplier $\tilde\varphi_\ell$ for the spatial components of the Bianchi identities, i.e.
\begin{align}
    S_\eqb
    & \sim - (-)^{pq+p+q} T_0\int_{\Sigma_\beta}  
     \df\tilde\varphi \wedge (\xi - c_\phi A - \tilde\ell V) \nn\\
    &\qquad\qquad\qquad\qquad\qquad
    -  (-)^{p} \tilde\ell\,\df\tilde\varphi_\ell \wedge \df V.
 \label{eq:lagrange-eqb-spatial}
\end{align}
The configuration equations for the spatial components of $\xi$ and $\df V$ read
\begin{align}
    \rmP_u\xi 
    &= \frac{1}{f_s} (-)^{pq+p+q} 
    {*\tilde\mu}, \nn\\
    \rmP_u\df V 
    &= -(-)^{pq+q} \tilde\chi_\ell\, {*\tilde\mu_\ell}.
\end{align}
We can use this, together with the time-components 
of the two objects $\iota_u\xi=c_\phi\mu$, $\iota_u\df V = 0$, to find an equivalent form of the equilibrium action
\begin{align}
    S_\eqb
    &= \int_{\Sigma_\beta}
    \frac{\chi}{2} \mu\wedge{*\mu}
    + \frac{\tilde\chi}{2} \tilde\mu \wedge{*\tilde\mu} \nn\\
    &\qquad\qquad
    + \frac{\chi_\ell}{2} \mu_\ell\wedge{*\mu_\ell}
    + \frac{\tilde\chi_\ell}{2} 
    \tilde\mu_\ell\wedge{*\tilde\mu_\ell} \nn\\
    &\qquad\qquad
    + c_\phi \lb TF \wedge \tilde\varphi - \mu\wedge\tilde A\rb.
    \label{eq:freeE-coulomb-partial}
\end{align}
Interestingly, we observe that this representation corresponds to a system with a temporally-spontaneously broken approximate 
$\rmU(1)_q\times\rmU(1)_p$ symmetry, with a mixed anomaly between the two symmetry groups. 
The terms in the last line are precisely the anomalous contribution to the equilibrium action that can be treated by using the bulk anomaly-inflow action; see \cref{eq:Sbulk-eqb}.

% expressing the bulk Lagrangian in \cref{eq:bulk-lagrangian} as
% \begin{align}
%     \iota_u\!\lb c_\phi\, \df A \wedge \tilde A \rb 
%     &= - c_\phi\, \df \iota_u A \wedge \tilde A
%     + (-)^{q} c_\phi\,\df A \wedge \iota_u\tilde A \nn\\
%     &= - c_\phi \df\mu \wedge \tilde A
%     + (-)^{q} c_\phi\,\df A \wedge \tilde\mu \nn\\
%     &\qquad 
%     + (-)^{q} c_\phi\, T \df A \wedge \df\tilde\varphi \nn\\
%     &= (-)^{q} c_\phi\,\mu \wedge \tilde F
%     + (-)^{p} \tilde c_\phi\,\tilde\mu \wedge F \nn\\
%     &\qquad
%     + c_\phi\,
%     \df\lb T_0 F \wedge \tilde\varphi
%     - \mu \wedge \tilde A \rb.
%     \label{eq:anomaly-free-eqb}
% \end{align}
% The bulk terms in the second-last line are entirely gauge-invariant, while the boundary terms in the last line precisely cancel the gauge non-invariant terms in \cref{eq:freeE-coulomb-partial}.

Varying the full equilibrium action with respect to the background sources, we can read out the respective currents 
\begin{align}
    J_\eqb 
    &= u\wedge n - \tilde c_\phi\, {*\tilde\mu}, \nn\\
    \tilde J_\eqb 
    &= u\wedge\tilde n
    - c_\phi\,{*\mu}, \nn\\
    L_\eqb 
    &= u\wedge n_\ell, \nn\\
    \tilde L_\eqb 
    &= u\wedge\tilde n_\ell,
    \label{eq:curr-SSB-dual}
\end{align}
where we have further identified $\tilde n = \tilde\chi\tilde\mu$ and $\tilde n_\ell = \tilde\chi_\ell\tilde\mu_\ell$.
We can extremise the equilibrium action with respect to $\varphi$ and $\tilde\varphi$ to recover the anomalous Gauss constraints
\begin{subequations}
\begin{align}
    \df{*n}
    &= c_\phi \rmP_u \tilde F
    + (-)^{q+1}\ell {*n_\ell}, 
    \label{eq:config-Coulomb-dual-1}
    \\
    \df{*\tilde n}
    &= \tilde c_\phi \rmP_u F
    + (-)^{p+1}\tilde\ell {*\tilde n_\ell}~,
    \label{eq:config-Coulomb-dual-2}
\end{align}
together with the Bianchi identities
\begin{align}
    \df\mu &= -\iota_u F \label{eq:bianchi-Coulomb-dual-1}, \\
    \df\tilde\mu &= - \iota_u \tilde F
    \label{eq:bianchi-Coulomb-dual-2} \\
    \df\mu_\ell &= \ell\lb \mu - \iota_u\Xi\rb, 
    \label{eq:bianchi-Coulomb-dual-3}
    \\
    \df\tilde\mu_\ell &= \tilde\ell\lb \tilde\mu 
    - \iota_u\tilde\Xi\rb.
    \label{eq:bianchi-Coulomb-dual-4}
\end{align}
\label{eq:config-Coulomb}%
\end{subequations}
The first and fifth of these equations \eqref{eq:config-Coulomb-dual-1} and \eqref{eq:bianchi-Coulomb-dual-3} are just the original U(1)$_q$ Gauss constraint in \cref{eq:gauss-broken-anom-eqb} and the Bianchi identity in \cref{eq:bianchi-ell-eqb}. The second and third equations \eqref{eq:config-Coulomb-dual-2} and \eqref{eq:bianchi-Coulomb-dual-1} implement the Bianchi identities \eqref{eq:max-2-broken}. Finally, the forth and sixth equations \eqref{eq:bianchi-Coulomb-dual-2} and \eqref{eq:bianchi-Coulomb-dual-4} implement \cref{eq:dyn-broken-eqb-mon-xi} and \eqref{eq:dyn-broken-eqb-mon-V} respectively. 

Due to the final two Bianchi identities, all charges are screened as 
\begin{align}
    \Big(\df{*\df}
    + (-)^{p} k_0^2{*} \Big) {*n}
    &= 0, \nn\\
    \Big(\df{*\df}
    + (-)^{q} k_0^2 {*} \Big) n_\ell
    &= 0, \nn\\
    \Big(\df {*\df} 
    + (-)^{q} \tilde k_0^2 {*}\Big) {*\tilde n}
    &= 0, \nn\\
    \Big(\df{*\df}
    + (-)^{p} \tilde k_0^2 {*} \Big)\tilde n_\ell
    &= 0.
    \label{eq:relax-Coulomb-dual}
\end{align}
% \begin{align}
%     \Big(\df{*\df}
%     + (-)^{p} k_0^2{*} \Big) {*n}
%     &= c_\phi \df{* \tilde F}
%     + (-)^{pq+p+q} \ell^2\chi_\ell \iota_u\Xi, \nn\\
%     %
%     \Big(\df{*\df}
%     + (-)^{q} k_0^2 {*} \Big) n_\ell
%     &= 
%     \ell\, \frac{c_\phi\chi_\ell}{\chi} \rmP_u \tilde F
%     - \ell\chi_\ell \df{*\iota_u\Xi}, \nn\\
%     %
%     \Big(\df {*\df} 
%     + (-)^{q} \tilde k_0^2 {*}\Big) {*\tilde n}
%     &= \tilde c_\phi \df{*F} 
%     + (-)^{pq+p+q} \tilde\ell^2\tilde\chi_\ell \iota_u\tilde\Xi, \nn\\
%     %
%     \Big(\df{*\df}
%     + (-)^{p} \tilde k_0^2 {*} \Big)\tilde n_\ell
%     &= 
%     \tilde \ell\, \frac{\tilde c_\phi\tilde\chi_\ell}{\tilde\chi}
%     \rmP_u F
%     - \tilde\ell\tilde\chi_\ell \df{*\iota_u\tilde\Xi}.
%     \label{eq:relax-Coulomb-dual}
% \end{align}
The first two equations are precisely \cref{eq:mu-relax}, whereas the last two are a consequence of \cref{eq:xi-relax-vor}.

% The approximately conserved U(1)$_p$ and U(1)$_q$ operators are given as
% \begin{align}
%     Q_\eqb[\Sigma_{d-q}] 
%     = \int_{\Sigma_{d-q}}
%     \chi\,{*\mu}
%     - c_\phi\, u\wedge {\tilde\mu}, \nn\\
%     \tilde Q_\eqb[\Sigma_{d-p}] 
%     = \int_{\Sigma_{d-p}}
%     \tilde\chi\,{*\tilde\mu}
%     - \tilde c_\phi\, u\wedge {\mu}.
%     \label{eq:dual-charges-undefected}
% \end{align}
% On the other hand, the topologically conserved U(1)$^\ell_{q-1}$ and U(1)$^\ell_{p-1}$ defect operators are given as
% \begin{align}
%     \ell Q^\eqb_\ell[\Sigma_{d-q+1}] 
%     &= \ell\int_{\Sigma_{d-q+1}}
%     \chi_\ell\,{*\mu_\ell}, \nn\\
%     %
%     \tilde\ell \tilde Q^\eqb_\ell[\Sigma_{d-p+1}] 
%     &= \tilde\ell\int_{\Sigma_{d-p+1}}
%     \tilde\chi_\ell\,{*\tilde\mu_\ell}.
% \end{align}

This representation of the theory is neat because it is symmetric in the U(1)$_q$ and U(1)$_p$ sectors. It also allows us to infer that a the relaxed phase of a U(1)$_q$ pseudo-superfluid with vortices is dual to the relaxed phase of a U(1)$_p$ pseudo-superfluid with vortices. Under this duality, all ``tilde'' quantities get exchanged with the respective ``untilde'' quantities.

\subsubsection{Applications}

\noindent
\emph{Superfluids with vortices:} 
To summarise, the model above describes a theory of relaxed U(1)$_q$ or U(1)$_p$ superfluids in the presence of vortices. If we increase the susceptibility of vortices by increasing $\tilde\ell$, thereby decreasing the correlation length $1/\tilde k_0$, we can integrate out the spatial components of the superfluid velocity $\tilde\chi\tilde\mu = (-)^q {*\xi}$ from the setup altogether using \cref{eq:relax-Coulomb-dual}. This leads us to the theory of relaxed fluids with temporally-spontaneously broken U(1)$_q$ global symmetry discussed in \cref{sec:fluids-eqb}. In this sense, the proliferation of vortices controls the phase transition from superfluids to fluids.
Analogously, by increasing the susceptibility of U(1)$^\ell_{q-1}$ defects by increasing $\ell$, thereby decreasing the correlation length $1/k_0$, we can integrate out the $q$-form chemical potential $\chi\mu = (-)^p {*\tilde\xi}$ from the setup using \cref{eq:relax-Coulomb-dual}. This results in a theory of just dynamical superfluid velocity, furnishing a temporally-spontaneously broken U(1)$_p$ global symmetry in \cref{sec:fluids-eqb}; see \cref{sec:applications-fluid}. These notions of phase transitions will be made more precise in the subsequent hydrodynamic discussion.

\noindent
\emph{Electromagnetism with free electric and magnetic charges:}
The setup outlined above can also be understood as describing the relaxed phase of a U(1)$^\loc_{q-1}$ or U(1)$^\loc_{p-1}$ gauge theory in thermal equilibrium, in the presence of free electric and possibly magnetic charges. Taking the U(1)$^\loc_{q-1}$ interpretation, we can identify $-c_\phi\mu$ as electric field, $\tilde\chi\tilde\mu$ as magnetic fields, $\chi/c_\phi^2$ as the electric permittivity, $\tilde\chi$ as the magnetic permeability. Furthermore, we can identify $-\ell\chi_\ell\mu_\ell/c_\phi$ as the electric charge density and $\tilde\ell\tilde\chi_\ell\tilde\mu_\ell$ as the magnetic charge density, with $\ell^2\chi_\ell/c_\phi^2$ and $\tilde\ell^2\tilde\chi_\ell$ their respective susceptibilities.

Therefore, in the presence of free electric charges, electric fields and electric charges themselves are screened to a finite Debye length $1/k_0$. If we proliferate the free electric charges by increasing $\ell$, thereby decreasing the electric Debye screening length $1/k_0$, we can integrate out electric fields and free electric charges from the description using the equations above. This leaves us with the theory of magnetohydrodynamics with magnetic monopoles, with an approximate U(1)$_p$ symmetry detailed in \cref{sec:fluids-eqb}; see \cref{sec:applications-fluid}. Similarly, in the presence of free magnetic charges/monopoles, magnetic field and magnetic charges are screened to a finite magnetic Debye length $1/\tilde k_0$. If we wish, we can identically switch off the effects of magnetic monopoles by setting $\tilde\ell$ to zero, which takes $\tilde k_0\to 0$ and unscreens the magnetic fields. On the other hand, if we proliferate the magnetic monopoles by increasing $\tilde\ell$, thereby decreasing the magnetic Debye screening length $1/\tilde k_0$, we can integrate out magnetic fields and free magnetic charges from the description. In this case, we are left with the theory of electrohydrodynamics with free electric charges, with an approximate U(1)$_q$ symmetry detailed in \cref{sec:fluids-eqb}.

% On the other hand in the dual U(1)$^\loc_{q-1}$ interpretation, we instead have the identifications for the electromagnetic fields and densities
% \begin{alignat}{2}
%     E &\equiv -\iota_u\tilde\xi = - \tilde c_\phi\tilde\mu, \qquad 
%     %
%     &\rho_e 
%     &\equiv - \frac{\tilde\ell}{\tilde c_\phi}
%     \tilde\chi_\ell\tilde\mu_\ell, \nn\\
%     %
%     B &\equiv -\iota_u{\star\tilde\xi} = \chi\mu, \qquad
%     %
%     &\rho_m
%     &\equiv - \frac{\ell}{c_\phi}
%     \chi_\ell\mu_\ell,
% \end{alignat}
% susceptibilities 
% \begin{alignat}{2}
%     \chi_E &= \frac{1}{\tilde c_\phi^2}\tilde\chi, \qquad 
%     &\chi_e &= 
%     \frac{\tilde\ell^2}{\tilde c_\phi^2}\tilde\chi_\ell, \nn\\
%     \chi_B &= \chi, \qquad 
%     &\chi_m &= \frac{\ell^2}{c_\phi^2}\chi_\ell.
% \end{alignat}
% and the external densities and currents
% \begin{alignat}{2}
%     \rho_e^\ext &= - \iota_u \tilde K_\ext, \qquad 
%     &j^\ext_e &= \rmP_u \tilde K_\ext, \nn\\
%     \rho_m^\ext &= - \iota_u K_\ext, \qquad 
%     &j^\ext_m &= \rmP_u K_\ext.
% \end{alignat}
% The resultant Maxwell's equations are precisely the same as before, with $\tilde c_\phi$ and $c_\phi$ interchanged.

\subsection{Pseudo-spontaneous symmetry breaking and pinned pseudo-superfluids}
\label{sec:appx-SF-Higgs-eqb}

Let us switch gears and consider the pinned phase of a U(1)$_q$ pseudo-superfluid, where the topological U(1)$^\ell_{q-1}$ global symmetry is completely spontaneously broken, giving mass to the $q$-form pseudo-Goldstone field $\phi$. The resultant model also describes a U(1)$^\loc_{q-1}$ gauge theory in the Higgs phase, where the U(1)$^\loc_{q-1}$ gauge symmetry is spontaneously broken and the dynamical $q$-form gauge field $\phi$ acquires a mass. For $q=0$, this model describes a pinned U(1)$_0$ superfluid with a massive 0-form pseudo-Goldstone field $\phi$, while for $q=1$ it describes a U(1)$_0$ superconductor with a massive dynamical 1-form gauge field $\phi$. We will talk more about these applications after setting up the model.

\subsubsection{Pseudo-spontaneous symmetry breaking}

Since both the U(1)$_q$ and U(1)$^\ell_{q-1}$ global symmetries are completely spontaneously broken, we have access to the full Goldstone fields $\phi$ and $\phi_\ell$. Using these, we can construct the completely gauge-invariant misalignment tensor $\psi$ defined in \cref{eq:psi-Higgs}. One can check that we still have $\iota_u\psi = - c_\phi\mu_\ell$, so this definition of $\psi$ only differs from its relaxed phase definition in \cref{eq:psi-Coulomb} by spatial terms. The new gauge-invariant information in $\psi$ allows us to introduce a new mass term for $\psi$ in the equilibrium action \eqref{eq:freeE-SSB-Coulomb} as 
\begin{align}
    S_\eqb
    &= \int_{\Sigma_\beta} \frac{\chi}{2} \mu \wedge {*\mu}
    - \frac{f_s}{2} \xi\wedge{*\xi}
    + \frac{\chi_{\ell}}{2} \mu_\ell \wedge {*\mu_\ell} \nn\\
    &\qquad\qquad
    - \frac{m^2}{2} \psi\wedge{*\psi}
     \nn\\
    &\qquad\qquad
    - \iota_u(\xi\wedge\tilde A)
    + (-)^q \iota_u(\psi \wedge \tilde A_\psi)~.
    \label{eq:freeE-SSB}
\end{align}
Here $m$ is the pinning mass that we also encountered previously in \cref{sec:SSB-zeroT-broken}. We have also included the $\tilde A_\psi$ coupling term for $\psi$ from \cref{eq:sf-action-Higgs}.
The U(1)$_q$ charge current is still given by \cref{eq:Jeqb-SSB}, while the defect current can be obtained by varying the equilibrium action, together with the bulk action in \cref{eq:bulk-S-higgs}, with respect to $\Phi$, to yield
\begin{align}
    L_\eqb 
    &= \chi_\ell\, u\wedge\mu_\ell + c_\phi m^2\,\rmP_u\psi 
    \nn\\
    &= \lb \chi_\ell - c_\phi^2 m^2 \rb
    u\wedge\mu_\ell + c_\phi m^2\,\psi.
\end{align}

The configuration equations for the time-component of $\phi$ give rise to the same Gauss constraint in \cref{eq:gauss-broken-anom-eqb}, but with the renewed definition of $\tilde F$ in \cref{eq:new-tildeF}. However, the configuration equations coming from the spatial components of $\phi$ modify from \cref{eq:gauss-eqb-spatial} to
\begin{gather}
    \df {*\xi} 
    = (-)^{q} \frac{\ell m^2}{f_s} {*\psi}
    - (-)^{q} \frac{1}{f_s} \iota_u\tilde F.
    % \nn\\
    % \quad\overset{A=0,\Phi=0}{\implies}\quad 
    % \df{*\df\phi}
    % = (-)^{q+1} (k^\phi_0)^2\, {*\phi},
\end{gather}
The Bianchi identities associated with $\psi$ is given by \cref{eq:Bianchi-psi} and that associated with $\xi$ is given in \cref{eq:max-2}.
Since $\phi$ acquires a mass, the temporal as well as spatial components of the superfluid velocity $\xi$ and $\psi$ are screened. 

The screening of the temporal components $c_\phi\mu = \iota_u\xi$ and $c_\phi\mu_\ell = -\iota_u\psi$ is still given by \cref{eq:mu-relax}, however the screening of the spatial components of $\xi$ and $\psi$ takes the form
\begin{align}
    \lb \df{*\df} - (-)^p (k_0^\phi)^2 * \rb  {*\xi}
    &= 0, \nn\\
    \lb \df{*\df} - (-)^{q} (k_0^\phi)^2 * \rb \psi
    &= 0, 
\end{align}
% \begin{align}
%     \df{*\df {*\xi}} - (-)^{pq} &(k_0^\phi)^2 \rmP_u\xi \nn\\
%     &= -\tilde c_\phi (k_0^\phi)^2 \rmP_u\Xi
%     - (-)^{q} \frac{1}{f_s} \df{*\iota_u\tilde F}, \nn\\
%     %
%     \df{*\df\psi} - (-)^{q} (k_0^\phi)^2 {*\psi}
%     &=
%     (-)^{q+1} \frac{\ell}{f_s} \iota_u\tilde F
%     - c_\phi\ell \df{*\Xi}, 
% \end{align}
where $k_0^\phi = \ell m/\sqrt{f_s}$ denotes the finite inverse correlation length of $\phi$ due to pinning. Recall that $\xi$ was also screened in the presence of vortices in \cref{eq:xi-relax-vor}. However, the precise form of the screening equations are different in the two cases.

\subsubsection{Duality transformations}

As we discussed in \cref{sec:Higgs-zeroT}, the pinned phase of a U(1)$_q$ pseudo-superfluid is self-dual under $q\leftrightarrow p+1$. To see the finite temperature realisation of this duality, we need to introduce the Lagrange multiplier $\tilde\phi$ and $\tilde\phi_\psi$ to impose the Bianchi identities
\begin{align}
    S_\eqb
    &\sim \int_{\Sigma_\beta} 
    \iota_u \lb \frac{1}{c_\phi} \df\tilde\phi \wedge 
    (\xi - c_\phi A) \rb
    \nn\\
    &\qquad\qquad
    + \iota_u \lb \frac{1}{c_\phi} 
    \tilde\phi_\psi \wedge \lb \df\psi - \ell\xi + \ell c_\phi\Xi \rb \rb~.
\end{align}
Together with the equilibrium action \cref{eq:freeE-SSB}, this can be used to integrate out $\xi$ and $\psi$ from the theory to give 
\begin{subequations}
\begin{align}
    \xi 
    &=
    (-)^{pq+p+q} \frac{1}{f_s} {*\tilde\mu}
    + (-)^{p+1}\frac{c_\phi}{\chi} u\wedge {*\tilde\xi}, \\
    \psi 
    &= \frac{(-)^{pq+q}}{m^2} {*\tilde\mu_\psi}
    + \frac{c_\phi}{\chi_\ell} u \wedge {*\tilde\xi_\psi},
\end{align}
\end{subequations}
where the new definition of $\tilde\xi$ is being used from \cref{eq:newdefs-xi} in the pinned phase. The dual chemical potentials, on the other hand, are defined as
\begin{subequations}
\begin{align}
    \frac{\tilde\mu_\psi}{T_0}
    &= \iota_\beta \tilde A_\psi
    - \df\tilde\varphi_\psi, \label{eq:mu-psi-def} \\
    \frac{\tilde\mu}{T_0}
    &= 
    \iota_\beta \tilde A
    - \df\tilde\varphi
    - \ell\tilde\varphi_\psi,
\end{align}
\end{subequations}
where $\tilde\varphi = \iota_\beta\tilde\phi/\tilde c_\phi$ and $\tilde\varphi_\psi = \iota_\beta\tilde\varphi_\psi/\tilde c_\phi$. Substituting these back, we obtain back the same functional form of the equilibrium action with the mappings in \cref{eq:Higgs-mappings} together with
\begin{gather}
    \mu\leftrightarrow\tilde\mu_\psi, \qquad
    \mu_\ell \leftrightarrow \tilde\mu \nn\\
    \chi\leftrightarrow\frac{1}{m^2}, \qquad
    \chi_\ell \leftrightarrow \tilde\chi.
\end{gather}

We can also derive an intermediate picture purely in terms of temporally-spontaneously broken higher-form symmetry by performing a partial duality transformation. To this end, we only introduce the temporal components of the Lagrange multipliers to impose the spatial components of the Bianchi identities, i.e.
\begin{align}
    S_\eqb
    &\sim -(-)^{pq+p+q} T_0
    \int_{\Sigma_\beta} 
    \df\tilde\varphi \wedge (\xi - c_\phi A)
    \nn\\
    &\qquad\qquad\qquad
    - \tilde\varphi_\psi \wedge \lb \df\psi - \ell\xi + \ell c_\phi\Xi \rb.
\end{align}
This still allows us to integrate out the spatial components of $\xi$ and $\psi$ as
\begin{subequations}
\begin{align}
    \rmP_u\xi 
    &=
    (-)^{pq+p+q} \frac{1}{f_s} {*\tilde\mu}, \\
    \rmP_u\psi 
    &= \frac{(-)^{pq+q}}{m^2} {*\tilde\mu_\psi}.
\end{align}
\end{subequations}
Further using $\iota_u\xi = c_\phi\mu$ and $\iota_u\psi = - \chi_\phi\mu_\ell$, we can substitute these expressions into the onshell action to obtain a dual form
\begin{align}
    S_\eqb
    &= \int_{\Sigma_\beta} \frac{\chi}{2} \mu \wedge {*\mu}
    + \frac{1}{2f_s} \tilde\mu\wedge{*\tilde\mu}
    + \frac{\chi_{\ell}}{2} \mu_\ell \wedge {*\mu_\ell} \nn\\
    &\qquad
    + \frac{1}{2m^2} \tilde\mu_\psi\wedge{*\tilde\mu_\psi}
     \nn\\
    &\qquad
    + c_\phi \!\lb T_0 F \wedge \tilde\varphi
    - \mu\wedge\tilde A \rb \nn\\
    &\qquad
    + c_\phi (-)^q \!\lb 
    T_0 \ell\Xi \wedge \tilde\varphi_\psi
    - \mu_\ell \wedge \tilde A_\psi \rb.
    \label{eq:lagrange-eqb-spatial-Higgs}
\end{align}
The anomalous terms in the last two lines of this expression can be removed by using the bulk anomaly-inflow action; see \cref{eq:Sbulk-eqb-Higgs}. The remaining terms in the dual form manifest the $q\leftrightarrow p+1$ symmetry of the pinned phase of a U(1)$_q$ pseudo-superfluid.

Varying the full action with respect to the background sources, we can read off the respective currents 
\begin{align}
    J_\eqb
     &= u \wedge n
     - \tilde c_\phi\,{*\tilde\mu}, \nn\\
     \tilde J_\eqb
     &= u \wedge \tilde n -
     c_\phi\,{*\mu}, \nn\\
     L_\eqb
     &= u \wedge n_\ell 
     + (-)^{p} \tilde c_\phi {*\tilde\mu_\psi}  
     , \nn\\
     \tilde J_{\psi,\eqb}
     &= u \wedge \tilde n_\psi
     - (-)^{q} c_\phi{*\mu_\ell}.
\end{align}
The anomalous Gauss constraints take the similar form
\begin{subequations}
\begin{align}
    \df{*n}
    &= c_\phi \rmP_u \tilde F
    + (-)^{q+1}\ell {*n_\ell}, \nn\\
    \df{*\tilde n_\psi}
    &= (-)^{p+1} \tilde c_\phi \ell \rmP_u\Xi
    + (-)^{p}\ell {*\tilde n},
\end{align}
together with the Bianchi identities
\begin{align}
    \df\mu &= -\iota_u F \label{eq:bianchi-Higgs-dual-1}, \\
    \df\tilde\mu_\psi &= - \iota_u \tilde F_\psi, 
    \label{eq:bianchi-Higgs-dual-2} \\
    \df\mu_\ell &= \ell \mu - \ell\iota_u\Xi, 
    \label{eq:bianchi-Higgs-dual-3}
    \\
    \df\tilde\mu &= \ell\tilde\mu_\psi
    - \iota_u\tilde F.
    \label{eq:bianchi-Higgs-dual-4}
\end{align}
\end{subequations}
All charged objects are also screened in the pinned phase. But the actual form of the relaxation is different and is given as
\begin{align}
    \Big(\df{*\df}
    + (-)^{p} k_0^2{*} \Big) {*n}
    &= 0, \nn\\
    \Big(\df{*\df}
    + (-)^{q} k_0^2 {*} \Big) n_\ell
    &= 0, \nn\\
    \Big(\df {*\df} 
    - (-)^{q} (k^\phi_0)^2 {*}\Big) {*\tilde n_\psi}
    &= 0, \nn\\
    \Big(\df{*\df}
    - (-)^{p} (k^\phi_0)^2 {*} \Big)\tilde n
    &= 0.
\end{align}
These are just rewriting of the equations found the respective equations found previously.

% \begin{align}
%     &\iota_u\lb \df A \wedge \tilde A
%     + (-)^q \ell \Xi \wedge \tilde A_\psi \rb \nn\\
%     &= (-)^q \mu \wedge \tilde F
%     + (-)^{pq+q} \tilde\mu \wedge F
%     \nn\\
%     &\qquad 
%     - \mu_\ell \wedge \tilde F_\psi
%     + (-)^{pq+p+q}\tilde\mu_\psi \wedge \ell\Xi \nn\\
%     &\qquad 
%     + \df\lb T_0 F \wedge\tilde\varphi 
%     - \mu \wedge \tilde A 
%     \rb \nn\\
%     &\qquad 
%     + (-)^q  \df\lb T_0 \ell\Xi \wedge\tilde\phi_\psi
%     - \mu_\ell \wedge \tilde A_\psi
%     \rb.
% \end{align}

% \begin{align}
%     S_\eqb
%     &= \int_{\Sigma_\beta} \frac{\chi}{2} \mu \wedge {*\mu}
%     - \frac{f_s}{2} \xi\wedge{*\xi}
%     + \frac{\chi_{\ell}}{2} \mu_\ell \wedge {*\mu_\ell} \nn\\
%     &\qquad\qquad
%     - \frac{m^2}{2} \psi\wedge{*\psi}
%      \nn\\
%     &\qquad\qquad
%     - \frac{c_\phi}{\tilde c_\phi}\mu\wedge\tilde\xi
%     - (-1)^{q+1}\xi\wedge\tilde\mu \nn\\
%     &\qquad
%     - \frac{(-)^q}{\tilde c_\phi} c_\phi
%     \mu_\ell \wedge \tilde\xi_\psi
%     + \psi \wedge \tilde\mu_\psi
%     \nn\\
%     &\qquad 
%     - \iota_u (\tilde\xi \wedge A)
%     + (-)^p \iota_u (\tilde\xi_\psi \wedge \ell\Phi) \nn\\
%     &\qquad 
%     - c_\phi \iota_u \lb 
%      A \wedge \tilde A 
%     + (-)^{q} \ell\Phi \wedge \tilde A_\psi \rb
% \end{align}

\subsubsection{Applications}

\noindent
\emph{Superconductivity:}
The above construction applies to the spontaneously broken phase of a U(1)$^\loc_{q-1}$ gauge theory, with massive gauge fields $\psi$. For $q=1$, this describes the theory of superconductivity with a spontaneously-broken U(1)$_0$ gauge symmetry. The massive gauge fields $\xi$ can only propagate over a finite correlation length $1/k_0^\phi$, identified as the London penetration depth in a superconductor.

\section{Higher-form pseudo-hydrodynamics}
\label{sec:higher-form-hydro}

In this section, we discuss the theory of hydrodynamics with an approximate higher-form symmetry in the temporally-spontaneously broken phase. The setup we employ here is a direct generalisation of our recent work on approximate 0-form symmetries~\cite{Armas:2021vku} and utilises the hydrodynamic framework for higher-form symmetries from~\cite{Armas:2018ibg,Armas:2018zbe}. We will assume relativistic symmetry through this discussion. The generalisation to Galilean systems or systems without any boost symmetry is involved but conceptually straight-forward, and we will consider it in a future publication.

\subsection{Hydrodynamic fields and equations}

When we leave thermal equilibrium, we no longer have the luxury to base our effective description on an equilibrium action like in \cref{sec:fluids-eqb}. We must instead rely on the framework of hydrodynamics, where the starting point are symmetries and the respective conservation laws. The rationale being that out-of-equilibrium processes are generically dissipative and the only long-lived modes that are relevant for the low-energy effective description are those corresponding to conserved quantities. As we have discussed in detail in the previous subsections, for a system respecting an approximate U(1)$_q$ global symmetry, the relevant conservation law is
\begin{subequations}
\begin{align}
  \df{\star J} &= (-1)^{q+1}\ell\,L~~.
  \label{eq:EE-1}
\end{align}
Furthermore, we assume the system to be thermodynamically isolated and invariant under spacetime translations, in which case we also have an energy-momentum conservation equation 
\begin{align}
    \nabla_\mu T^{\mu\nu}
    &= (F\cdot J)^\nu + (\ell\,\Xi \cdot L)^\nu \label{eq:EE-3},
\end{align}
where $T^{\mu\nu}$ is the symmetric energy-momentum tensor and $\nabla_\mu$ denotes the covariant derivative operator associated with the background spacetime metric $g_{\mu\nu}$.  
We have used the notation $(F\cdot J)^\nu = 1/(q+1)!\,F^{\nu\lambda_1\ldots}J_{\lambda_1\ldots}$ and similarly for the second term on the right.
The force terms sourcing the energy-momentum conservation are a direct generalisation of the Lorentz force for 0-form symmetries. Depending on the particular application in mind, one might also need to consider additional conservation equations associated with particle number, etc. We will ignore these for now for simplicity.
\label{eq:hydro-eqs-TSSB}
\end{subequations}

To solve these hydrodynamic equations, we need the appropriate degrees of freedom. Following our previous work on 1-form hydrodynamics~\cite{Armas:2018zbe}, we note that the hydrodynamic fields can be identified as a set of symmetry parameters corresponding to the global symmetries of the system under consideration. For our case of interest, these are a $q$-form parameter $\Lambda_\beta$ and a diffeomorphism parameter $\beta^\mu = u^\mu/T$, with $u^\mu$ being the normalised timelike fluid velocity (with $u^\mu u_\mu = -1$) and $T$ the fluid temperature. These fields transform under infinitesimal diffeomorphisms $\chi^\mu$ and U(1)$_q$ transformations $\Lambda$ as
\begin{subequations}
\begin{align}
    \beta^\mu &\to \beta^\mu + \lie_\chi\beta^\mu, \\
    \Lambda_\beta
    &\to \Lambda_\beta + \lie_\chi \Lambda_\beta
    - \lie_\beta \Lambda.
\end{align}
When the U(1)$_q$ global symmetry is explicitly broken, for $q>0$, we also have a new global U(1)$^\ell_{q-1}$ symmetry associated with the conservation of defects. Correspondingly, we introduce a \tightmath{$(q-1)$}-form parameter $\Lambda_\beta^\ell$ transforming under diffeomorphisms $\chi^\mu$ and \tightmath{$(q-1)$}-form gauge transformations $\Lambda_\ell$ as
\begin{align}
    \Lambda^\ell_\beta
    &\to \Lambda_\beta^\ell + \lie_\chi \Lambda^\ell_\beta
    - \lie_\beta \Lambda_\ell.
\end{align}
\label{eq:hydro-trans}%
\end{subequations}
In equilibrium, the hydrodynamic fields take trivial values $\beta^\mu = \delta^\mu_t/T_0$ and $\Lambda_\beta = \Lambda_\beta^\ell = 0$. Using \cref{eq:hydro-trans}, we can see that these remain unchanged under time-independent diffeomorphisms and gauge transformations.  

\begin{subequations}
As we discussed in \cref{sec:fluids-eqb}, we need to temporally-spontaneously break the U(1)$_q$ and U(1)$^\ell_{q-1}$ global symmetries of the theory, for \tightmath{$q>0$} and \tightmath{$q>1$} respectively, to allow for non-trivial chemical potentials. This gives rise to temporal-Goldstone fields $\varphi$ and $\varphi_\ell$, satisfying $\iota_\beta\varphi=\iota_\beta\varphi_\ell=0$, and transforming as
\begin{align}
    \varphi
    &\to \varphi
    + \lie_\chi \varphi
    - \iota_\beta \Lambda, \\
    \varphi_\ell
    &\to \varphi_\ell
    + \lie_\chi \varphi_\ell
    - \iota_\beta \Lambda_\ell.
\end{align}
\label{eq:varphi-trans-TSSB}%
\end{subequations}
We will also need Josephson equations to govern the time-evolution of these temporal-Goldstones, which take the usual form
\begin{subequations}
\begin{align}
    \lie_\beta\varphi
    &= \iota_\beta\Lambda_\beta, \label{eq:Joseph-TSSB-1} \\
    \lie_\beta\varphi_\ell
    &= \iota_\beta\Lambda^\ell_\beta. \label{eq:Joseph-TSSB-2}
\end{align}
\label{eq:Joseph-TSSB}%
\end{subequations}
The $\Lambda_\beta$ and $\Lambda_\beta^\ell$ terms on the right are necessitated by symmetries. In principle, these equations can admit derivative corrections.
However, by judiciously redefining the time-components of $\Lambda_\beta$ and $\Lambda_\beta^\ell$, we can absorb all such corrections and impose the above equations exactly. 

An astute reader will notice the mismatch between the equations of motion and the degrees of freedom. The two Josephson equations determine the evolution of $\varphi$ and $\varphi_\ell$, while the two conservation equations determine the dynamics of $\beta^\mu$ and $\Lambda_\beta$. However, there is no independent equation of motion for the field $\Lambda_\beta^\ell$. This can be pinned down to a secret gauge symmetry in the theory. Consider a U(1)$_q$ transformation with parameter $\Lambda = \df\lambda$, together with a U(1)$^\ell_{q-1}$ transformation with parameter $\Lambda_\ell = \lambda$. This leaves both the background fields $A$ and $\Phi$ invariant, but instead acts as a U(1)$^\loc_{q-1}$ local gauge transformation on the dynamical fields of the theory
\begin{alignat}{2}
    \Lambda_\beta 
    &\to \Lambda_\beta - \lie_\beta \df\lambda,  \qquad
    & \Lambda_\beta^\ell
    &\to \Lambda_\beta^\ell - \lie_\beta\lambda, \nn\\
    \varphi &\to \varphi - \iota_\beta\df\lambda, \qquad
    &\varphi_\ell 
    &\to \varphi_\ell - \iota_\beta\lambda.
    \label{eq:gauge-freedom-TSSB}
\end{alignat}
This gauge symmetry accounts for the additional unphysical degrees of freedom in the description. We could fix a gauge to try and rid ourselves of the unphysical degrees of freedom, but it is cleaner to play along and keep the gauge symmetry manifest for the moment. 

We can use the dynamical field content of the theory to define the gauge-invariant U(1)$_q$ chemical potential $\mu$ and the U(1)$^\ell_{q-1}$ defect chemical potential $\mu_\ell$ via
\begin{subequations}
\begin{align}
    \frac{\mu}{T}
    &= \Lambda_\beta + \iota_\beta A
    - \df\varphi, \label{eq:mu-def-hydro} \\
    \frac{\mu_\ell}{T}
    &= \ell \lb 
    \Lambda_\beta^\ell 
    + \iota_\beta\Phi
    - \df\varphi_\ell
    - \varphi 
    \rb.
    \label{eq:muell-def-hydro}
\end{align}
\label{eq:mu-defs-hydro}%
\end{subequations}
The Josephson equations \eqref{eq:Joseph-TSSB} imply that both of these chemical potentials are purely spatial
\begin{equation}
    \iota_\beta\mu = \iota_\beta\mu_\ell = 0,
    \label{eq:mu-spatiality}
\end{equation}
as we obtained in equilibrium. Together, $\beta^\mu$, $\mu$, and $\mu_\ell$ make up the right amount of degrees of freedom to solve for using the hydrodynamic equations \eqref{eq:hydro-eqs-TSSB}.

Before we proceed, it is important to set a derivative counting scheme. To keep the effects of explicit symmetry breaking small in the effective theory, we take the symmetry breaking parameter to be of $\ell\sim{\cal O}(\dow)$. We choose the usual derivative-scaling for the gauge field $A\sim{\cal O}(\dow^0)$, while for the dynamical and background phase fields  we choose $\varphi,\Phi\sim{\cal O}(\dow^{-1})$, so that $\mu$ and $\mu_\ell$ are ${\cal O}(\dow^0)$. This is the same derivative counting we employed in our recent work on approximate 0-form symmetries~\cite{Armas:2021vku}.

\subsection{Second law of thermodynamics and constitutive relations}
\label{sec:secondlaw}

An important aspect of the hydrodynamic framework is the local second law of thermodynamics. As such, all we need to complete the hydrodynamic equations is a set of constitutive relations for the currents $J$, $L$, $T^{\mu\nu}$ in terms of the hydrodynamic variables $u^\mu$, $T$, $\mu$, and $\mu_\ell$ (for $q>0$), along with the background field strengths $F$ and $\Xi$. However, the second law of thermodynamics imposes constraints on the constitutive relations as it postulates the existence of an entropy current $S^\mu$ whose divergence is locally positive semi-definite
\begin{equation}
    \nabla_\mu S^\mu \geq 0,
\end{equation}
for all possible solutions of the hydrodynamic equations. In practise, it is useful to work with the free energy current $N^\mu$ defined as
\begin{align}
    TN^\mu 
    &= TS^\mu +
    T^{\mu\nu} u_\nu 
    + (J\cdot\mu)^\mu
    + (L\cdot\mu_\ell)^\mu.
    \label{eq:N-def}
\end{align}
In terms of this, the statement of the second law turns into the so-called \emph{adiabaticity equation}
\begin{align}
    \nabla_\mu N^\mu 
    &= \half T^{\mu\nu} \delta_\scB g_{\mu\nu}
    + J\cdot\delta_\scB A 
    + \ell L\cdot\delta_\scB \Phi
    + \Delta,
    \label{eq:adiabaticity}
\end{align}
for some positive-semi-definite quadratic form $\Delta$ representing entropy production. To compactify the notation, we have utilised the variations of the background fields along the hydrodynamic data
\begin{alignat}{2}
    \delta_\scB g_{\mu\nu}
    &= \lie_\beta g_{\mu\nu}
    &&= 2\nabla_{(\mu}\beta_{\nu)}, \nn\\
    \delta_\scB A
    &= \lie_\beta A + \df\Lambda_\beta 
    &&= \df\frac{\mu}{T} + \iota_\beta F, \nn\\
    \delta_\scB\Phi
    &= \lie_\beta\Phi - \ell\Lambda_\beta
    + \df\Lambda^\ell_\beta
    &&= \df\frac{\mu_\ell}{\ell T} 
    + \iota_\beta\Xi - \frac{\mu}{T}.
    \label{eq:deltaB-defs}
\end{alignat}
The primary benefit of the adiabaticity equation over the statement of the second law is that it is satisfied even off-shell, i.e. even when the hydrodynamic conservation equations are not necessarily imposed.

\subsubsection{Constitutive relations}

The adiabaticity equation can be used to classify the hydrodynamic constitutive relations consistent with the second law of thermodynamics to arbitrary orders in derivatives~\cite{Haehl:2014zda, Haehl:2015pja, Jain:2016rlz}. 
We will not perform this exercise here.
Instead, for illustrative purposes, we focus just on ``canonical transport'', characterised by a simple free energy current
\begin{equation}
    N^\mu = P\,\beta^\mu
    + {\cal N}^\mu~.
\end{equation}
Here $P$ denotes the thermodynamic pressure that generically depends on all the zero-derivative order Lorentz scalars in the theory constructed out $T$, $\mu$, and $\mu_\ell$. The vector ${\cal N}^\mu$ denotes possible non-canonical corrections to the free energy current.
Since the higher-form chemical potentials carry Lorentz indices, the number of such scalars crucially depends on the rank $q$ of the higher-form symmetry and the number of spatial dimensions $d$.\footnote{For $q=0$ and any $d$, we only have $T$ and $\mu$ as we expect from ordinary 0-form hydrodynamics. For $q=1$ and any $d$, we have $T$ and $\mu_\mu\mu^\mu$, but also $\mu_v$ because the defects themselves furnish a $0$-form symmetry. For higher-$q$, the counting becomes dimension dependent. For instance for $q=2$, $d=3$, we have $T$, $\mu_{\mu\nu}\mu^{\mu\nu}$, $\mu^v_\mu \mu_v^\mu$, and the product $\mu_{\mu\nu}\mu^{\nu\rho}\mu^v_{\rho}\mu_v^{\mu}$. But for $q=2$, $d\geq 4$, we also have independent chemical potential chains like $\mu_{\mu\nu}\mu^{\nu\rho}\mu_{\rho\sigma}\mu^{\sigma\mu}$.} We can sweep this complication under the rug for now by pretending that $P$ is a function of the thermodynamic parameters directly and expressing its derivatives as
\begin{subequations}
\begin{align}
    \df P 
    &= s\delta T + n\cdot\delta\mu
    + n_\ell\cdot\delta\mu_\ell 
    - \half r^{\mu\nu}\delta g_{\mu\nu}. 
\end{align}
The cost we pay is that the differential of $P$ now also explicitly depends on the differential of the spacetime metric used for the contraction of indices. We can identify $s$ as the thermodynamic entropy density, $n$ as the $q$-form charge density, $n_\ell$ as the $(q-1)$-form defect density, and $r^{\mu\nu}$ the anisotropic stress tensor. We can also define the thermodynamic energy density 
\begin{equation}
    \epsilon = -P + Ts + \mu\cdot n
    + \mu_\ell\cdot n_\ell.
\end{equation}
\end{subequations}
Since $P$ is a scalar, it cannot arbitrarily depend on the metric and hence $r^{\mu\nu}$ is not independent. In fact, requiring the Lie derivative of $P$ to be given by the Lie derivatives of its constituents, we can find that 
\begin{align}
    r^{\mu\nu}
    &= \frac{1}{(q-1)!} n^\mu{}_{\rho_1\ldots}
    \mu^{\nu\rho_1\ldots} 
    + \frac{1}{(q-2)!} 
    (n_\ell)^\mu{}_{\rho_1\ldots}
    \mu_\ell^{\nu\rho_1\ldots}, \nn\\
    r^{[\mu\nu]} &= 0.
\end{align}
The condition in the last line can be understood as a constraint on the densities due to the Lorentz invariance.
In the simplified case when the pressure only depends on $T$, $\mu^2$, and $\mu_\ell^2$, we find that
\begin{equation}
    n = \chi\mu, \qquad 
    n_\ell = \chi_\ell\mu_\ell,    
    \label{eq:simple-EOS}
\end{equation}
where $\chi$ and $\chi_\ell$ are susceptibilities of $q$-form charge and $(q-1)$-form defects respectively. One can check that the Lorentz-invariance constraint is trivially satisfied. 

Let us now move on to the constitutive relations. Using the fact that $\nabla_\mu (P\,\beta^\mu) = \half P\, g^{\mu\nu}\delta_\scB g_{\mu\nu} + \delta_{\scB} P$ into the adiabaticity equation \eqref{eq:adiabaticity}, and using
\begin{align}
    \delta_\scB T 
    &= \frac{T}{2} u^\mu u^\nu \delta_\scB g_{\mu\nu}, \nn\\
    \delta_\scB \mu 
    &= \frac{\mu}{2} u^\mu u^\nu \delta_\scB g_{\mu\nu}
    + \iota_u \delta_\scB A, \nn\\
    \delta_\scB \mu_\ell 
    &= \frac{\mu_\ell}{2} u^\mu u^\nu \delta_\scB g_{\mu\nu}
    + \ell\iota_u \delta_\scB \Phi~,
\end{align}
we can read off the constitutive relations 
\begin{align}
    T^{\mu\nu}
    &= (\epsilon+P) u^\mu u^\nu + P\,g^{\mu\nu}
    - r^{\mu\nu}
    + {\cal T}^{\mu\nu}
    , \nn\\
    J
    &= u\wedge n + {\cal J}, \nn\\
    L
    &= u \wedge n_\ell + {\cal L}~.
    \label{eq:consti-ansatz}
\end{align}
Here $\cal T^{\mu\nu}$, ${\cal J}$, ${\cal L}$ denote the respective dissipative corrections, which we will return to below. These constitutive relations satisfy the adiabaticity equation in the absence of dissipative corrections.

We can use the second law of thermodynamics to constrain the derivative corrections that can appear in the hydrodynamic constitutive relations. Plugging the constitutive relations \eqref{eq:consti-ansatz} into the adiabaticity equation \eqref{eq:adiabaticity}, we can derive a constraint equation to be satisfied by the derivative corrections
\begin{align}
    - \half{\cal T}^{\mu\nu} \delta_\scB g_{\mu\nu}
    &- {\cal J}\cdot\delta_\scB A \nn\\
    &- \ell{\cal L}\cdot\delta_\scB \Phi
    + \nabla_\mu{\cal N}^\mu
    = \Delta \geq 0.
    \label{eq:constraint-adiabaticity}
\end{align}
We already fixed the redefinition freedom in the time-component of $\Lambda_\beta$ to set $\iota_\beta\mu = 0$. Further, using the redefinition freedom in the spatial components of $\Lambda_\beta$, along with those in $\beta^\mu$ and $\varphi$, we can choose these dissipative corrections to satisfy the Landau frame condition
\begin{equation}
    {\cal T}^{\mu\nu}u_\nu 
    = \iota_u {\cal J} = \iota_u {\cal L} = 0.
\end{equation}

\subsubsection{Dissipative corrections}

For simplicity, we shall ignore any non-canonical transport characterised by ${\cal N}^\mu$ and further focus only on fluids enjoying parity and charge-conjugation invariance; see \cref{tab:CPT} for discrete transformation properties of various fields. The most general such derivative corrections up to first order in derivatives are given as
\begin{align}
    {\cal T}^{\mu\nu} 
    &= - T\eta P^{\mu\rho}P^{\nu\sigma}
    \delta_\scB g_{\rho\sigma}
    - \frac{T}{2}\!\lb \zeta - {\textstyle\frac{2}{d}}\eta \rb
    P^{\mu\nu}P^{\rho\sigma}
    \delta_\scB g_{\rho\sigma} \nn\\
    &= - 2\eta P^{\lambda(\mu} \nabla_\rho u^{\nu)}
    - \lb \zeta - {\textstyle\frac{2}{d}}\eta \rb
    P^{\mu\nu}\nabla_\lambda u^\lambda, \nn\\[0.5em]
    {\cal J}
    &= -T\sigma\,\rmP_u\delta_\scB A \nn\\
    &= -\sigma
    \lb T\,\rmP_u\df\frac{\mu}{T} 
    + \iota_u F
    \rb, \nn\\[0.5em]
    {\cal L}
    &= -\ell T\sigma_\ell\,\rmP_u\delta_\scB \Phi \nn\\
    &= - \sigma_\ell
    \lb T\,\rmP_u\df\frac{\mu_\ell}{T} 
    + \ell\lb \iota_u\Xi - \mu \rb
    \rb,
    \label{eq:diss-consti-normal}
\end{align}
where $\Delta^{\mu\nu} = g^{\mu\nu} + u^\mu u^\nu$ is the projector transverse to the fluid velocity.
The shear viscosity $\eta$, bulk viscosity $\zeta$, conductivity $\sigma$, and the defect conductivity $\sigma_\ell$ are all non-negative dissipative transport coefficients. 

In writing these constitutive relations, for simplicity, we have ignored possible anisotropic transport in $\mu$. When expanded around the isotropic equilibrium state $\mu=0$, such terms only contribute to the hydrodynamic equations non-linearly. More generally, we might be interested in anisotropic equilibrium states, such as magnetohydrodynamics with a constant density of magnetic fields in equilibrium, where such anisotropic contributions to the constitutive relations will become important; see~\cite{Armas:2018zbe}.

Let us do a quick sanity check. As we said before, in equilibrium, the hydrodynamic fields take trivial values $\beta^\mu = \delta^\mu_t/T_0$ and $\Lambda_\beta = 0$. Furthermore, all background sources must be time-independent in equilibrium, implying that
\begin{equation}
    \delta_\scB g_{\mu\nu} = \delta_\scB A = \delta_\scB\Phi = 0.
\end{equation}
This sets all the dissipative corrections in the constitutive relations to be identically zero and we can verify that we get the same expressions for the currents as we did using the equilibrium action in \cref{sec:fluids-eqb}.

% where we have used
% \begin{align}
%     \frac1T \delta_\scB T 
%     &= \frac{1}{2} u^\mu u^\nu 
%     \delta_{\scB} g_{\mu\nu}, \nn\\
%     T\delta_{\scB}\! \lb \frac1T\mu_{\mu_1\ldots\mu_q} \rb
%     &= u^\mu \delta_\scB A_{\mu\mu_1\ldots\mu_q}, \nn\\
%     T\delta_\scB\!\lb \frac1T\mu^v_{\mu_1\ldots\mu_{q-1}} \rb
%     &= u^\mu \delta_\scB\Phi_{\mu\mu_1\ldots\mu_{q-1}}.
% \end{align}

\subsection{Linearised analysis and mode spectrum}
\label{sec:modes-hydro}

To understand the physical effects of these coefficients, let us switch off the background sources and take the simplified equation of state \eqref{eq:simple-EOS}. The hydrodynamic equations are trivially satisfied by the equilibrium configuration $T=T_0$, $u^\mu = \delta^\mu_t$, together with $\mu=\mu_\ell = 0$. In principle, we can also consider equilibrium states with $\mu\neq 0$, but for $q>0$, these states are anisotropic and would need a more careful treatment that we leave for future work. We can look at small perturbations around the said equilibrium state, which will be described by linearised hydrodynamics.

When expanding around the isotropic state with $\mu=0$, energy and momentum fluctuations decouple from the charge fluctuations and are simply given by 
\begin{align}
    \dow_t\epsilon
    &= - \dow_i\pi^i, \nn\\
    \dow_t \pi^i
    &= - v_\|^2 \dow^i \epsilon
    + D_\pi^\perp\, \dow^2\pi^i
    +  \lb D_\pi^\| - D_\pi^\perp \rb
    \dow^i\dow_k\pi^k,
\end{align}
where $\pi^i = (\epsilon+p) u^i$ denote the momentum fluctuations and we have identified
\begin{equation}
    v_\|^2 = \frac{\dow P}{\dow\epsilon}, \quad
    D^\perp_\pi = \frac{\eta}{\epsilon+P}, \quad 
    D_\pi^\| 
    = \frac{\zeta + 2{\textstyle\frac{d-1}{d}}\eta}{\epsilon+P}.
\end{equation}
They give rise to the familiar longitudinal fluid sound and transverse shear diffusion modes
\begin{align}
    \omega &= \pm v_\| k - \frac{i}{2} D_\pi^\| k^2 + \ldots, \nn\\
    \omega &= - iD_\pi^\perp k^2.
    \label{eq:hydro-sound}
\end{align}
Around a state with $\mu\neq 0$, the energy-momentum sector generically non-trivially couples with the charge sector. We shall not consider this case in this work.

Coming back to the charge fluctuations, the two higher-form currents take the form
\begin{align}
    J
    &= u\wedge n -\sigma \rmP_u\df\mu , \nn\\
    L
    &= u\wedge n_\ell + \ell\sigma_\ell \mu
    - \sigma_\ell \rmP_u\df\mu_\ell.
\end{align}
We can see that $\sigma$ and $\sigma_\ell$ are conductivities for U(1)$_q$ charge $n$ and U(1)$^\ell_{q-1}$ defects $n_\ell$ respectively. On the other hand, defect current also has a term proportional to $\mu$ also controlled by $\sigma_\ell$, which tends to relax $n$. The conservation equations become
\begin{align}
    (\dow_t + \Gamma) n
    &= D_n \dow^2 n
    - (-)^{pq+p+q} (D_\ell - D_n) \df{*\df{*n}}, \nn\\
    *\df{*n} 
    &= (-)^{pq+p+q+1} \ell n_\ell,
    \label{eq:linear-eqs}
\end{align}
where we have defined the diffusion constants $D_n$, $D_\ell$, and the relaxation rate $\Gamma$ as
\begin{gather}
    D_n = \frac{\sigma}{\chi}, \qquad 
    D_\ell = \frac{\sigma_\ell}{\chi_\ell}, \qquad 
    \Gamma = \frac{\ell^2\sigma_\ell}{\chi}.
    \label{eq:damping-attenuation-coeff}
\end{gather}
The first equation in \cref{eq:linear-eqs} determines the dynamics of U(1)$_q$ charge density $n$, whereas the second equation defines the defect density $n_\ell$.
Let us go to the momentum space with frequency $\omega$ and wavevector $k$, and
decompose the U(1)$_q$ charge density as $n = k/|k|\wedge n_\| + n_\perp$, such that $\iota_k n_\| = \iota_k n_\perp = 0$. We can see that transverse components of the U(1)$_q$ charge $n_\perp$ evolves independently of the defects and give rise to the usual charge diffusion mode 
\begin{equation}
    \omega = -iD_n k^2 - i\Gamma.
    \label{eq:perp-diff}
\end{equation}
However, this mode is now damped due to the presence of explicit symmetry breaking.  There is another damped diffusion mode associated with the fluctuations of the longitudinal components of the U(1)$_q$ charge $n_\|$. This mode ise absent in the absence of defects due to the Gauss constraint $n_\| = -\ell/k^2\,n_\ell$. It reads
\begin{equation}
    \omega = -i D_\ell k^2 - i\Gamma,
\end{equation}
Such a mode does not exist for $q=0$.
Note that defects are exactly conserved, but since they are topological and cannot propagate locally, the associated diffusion mode is still relaxed.
Due to their topological nature, the damping coefficient $\Gamma$ and the attenuation coefficient $D_\|$ in this mode are related via the damping-attenuation relation
\begin{equation}
    \Gamma = D_\ell k_0^2,
    \label{eq:damping-attenuation}
\end{equation}
where $k_0 = \ell\sqrt{\chi_\ell/\chi}$ is the inverse hydrostatic correlation length of defects introduced in \cref{sec:fluids-eqb}. Such relations are a generic hallmark of systems with a spontaneously broken approximate symmetry, in this case the temporal part of the higher-form symmetry.

If we proliferate $q$-form defects by increasing the magnitude of $\ell$, the relaxation rate $\Gamma$ becomes finite and both the higher-form charged modes drop out of the spectrum. The resultant theory is that of a neutral fluid without any higher-form symmetry and implements the phase transition from (c) to (h) in \cref{fig:chart}. 

We can also use the hydrodynamic equations to obtain the retarded correlation functions of the U(1)$_q$ charge. These have been summarised in \cref{app:correlators}.

\section{Relaxed higher-form pseudo-superfluid dynamics}
\label{sec:higher-form-Coulomb}

In this section, we will discuss the hydrodynamic theory of a U(1)$_q$ pseudo-superfluid in the relaxed phase. For $q=0$, this amounts to an ordinary U(1)$_0$ superfluid with charge relaxation effects. For $q>1$, this describes non-equilibrium effects in a U(1)$_{q-1}^\loc$ gauge theory in the presence of free electric charges. We will also study the effects of vortices in the hydrodynamic theory, interpreted as magnetic monopoles in the context of gauge theories. 

\subsection{Hydrodynamic fields and equations}

We will start with the conventional formulation of (pseudo-)superfluid dynamics in terms of the Goldstone phase field $\phi$ and superfluid velocity $\xi$. In the later half of this subsection, we will explicitly show that this system can equivalently be posed as hydrodynamics with a temporally-spontaneously broken approximate anomalous $\rmU(1)_q\times\rmU(1)_p$ symmetry. This generalises the work of~\cite{Delacretaz:2019brr, Armas:2018zbe}, where the authors discussed explicitly-unbroken U(1)$_0$ and U(1)$_1$ superfluids. 

\subsubsection{Conventional formulation}

The hydrodynamic description of a system is based on the associated conservation equations. As we discussed at length in \cref{sec:SSB-zeroT-broken}, the conservation equations of a U(1)$_q$ superfluid in the relaxed phase (with vortices) take the form
\begin{subequations}
\begin{align}
    \df{\star J}
    &= c_\phi\tilde F + (-)^{q+1} \ell\, {\star L}, 
    \label{eq:hydro-eqs-Coulomb-dual-1} \\
    \df{\star\tilde J} 
    &= \tilde c_\phi F + (-)^{p+1} \tilde\ell\, {\star\tilde L},
    \label{eq:hydro-eqs-Coulomb-dual-2} \\
    \df{\star L}
    &= 0, \label{eq:hydro-eqs-Coulomb-dual-1-L} \\
    \df{\star\tilde L}
    &= 0, \label{eq:hydro-eqs-Coulomb-dual-2-L} \\
    \nabla_\mu T^{\mu\nu} 
    &= (F\cdot J)^\nu 
    + (\tilde F\cdot\tilde J)^\nu \nn\\
    &\qquad
    + (\ell\, \Xi\cdot L)^\nu
    + (\tilde\ell\, \tilde\Xi\cdot \tilde L)^\nu.
    \label{eq:hydro-eqs-Coulomb-dual-3}
\end{align}
\label{eq:hydro-eqs-Coulomb-dual}%
\end{subequations}
We have included additional Lorentz force terms in the energy-momentum conservation corresponding to the new conserved currents.
In the conventional formulation of superfluid dynamics, the U(1)$_p$ conservation equation \eqref{eq:hydro-eqs-Coulomb-dual-2} and the associated U(1)$^\ell_{p-1}$ defect conservation equation \eqref{eq:hydro-eqs-Coulomb-dual-2-L} are trivially satisfied due to the Bianchi identity \eqref{eq:max-2-broken}. The respective U(1)$_p$ current $\tilde J$ is defined in terms of the superfluid velocity $\xi$ in \cref{eq:tildeJ}, while the U(1)$^\ell_{p-1}$ defect current $\tilde L$ is given in terms of the vortex-induced part $V$ of the superfluid phase in \cref{eq:tildeL}. To solve the remaining hydrodynamic equations, we introduce the set of hydrodynamic fields $\beta^\mu$, $\Lambda_\beta$, and $\Lambda_\beta^\ell$ from \cref{sec:higher-form-hydro}, with the transformation properties given in \cref{eq:hydro-trans}. We can define the U(1)$_q$ chemical potential $\mu$ and the U(1)$^\ell_{q-1}$ defect chemical potential $\mu_\ell$ same as in \cref{eq:mu-defs-hydro}.

\begin{subequations}
Out of equilibrium, the configuration equation for $\phi$ is replaced by the respective Josephson equation determining its time-derivatives, i.e.
\begin{equation}
    \lie_\beta\phi
    = c_\phi\Lambda_\beta + \frac1T {\cal K}
    \qquad\text{where}\quad 
    \iota_\beta{\cal K} = 0,
    \label{eq:Joseph1}
\end{equation}
where ${\cal K}$ denotes arbitrary corrections to the Josephson equation that need to be determined as part of the constitutive relations. The $\Lambda_\beta$ term here is necessitated by U(1)$_q$-invariance, rendering the derivative corrections ${\cal K}$ to be gauge-invariant. We have used the redefinition freedom in the time-component of $\Lambda_\beta$ to set $\iota_\beta{\cal K} = 0$, keeping with our previous convention in \cref{eq:Joseph-TSSB-1} upon identifying $\varphi = \iota_\beta\phi/c_\phi$. We can also use the redefinition freedom in $\Lambda_\beta$ as a whole to rid ourselves of ${\cal K}$ entirely, but we have other plans for this freedom later in our discussion. 
Since the U(1)$^\ell_{q-1}$ symmetry is temporally-spontaneously broken in the relaxed phase, we also have a similar Josephson equation for pseudo-Goldstone $\varphi_\ell$, taking the same from as in \cref{eq:Joseph-TSSB-2}. Analogously, in the presence of vortices, we have another Josephson equation for the vortex-induced part $V$ of the superfluid velocity. This takes the form
\begin{align}
    \lie_\beta V
    &= \frac1T {\cal V}, \qquad 
    \text{where}\quad 
    \iota_\beta {\cal V} = 0,
    \label{eq:Joseph-V-Coulomb}
\end{align}
\label{eq:joseph-Coulomb}
\end{subequations}
with ${\cal V}$ denoting the admissible derivative corrections. In line with our gauge choice $\iota_\beta V = 0$ in \cref{eq:V-gauge}, we have also taken $\iota_\beta{\cal V} = 0$. The consequence of having vortices is that the evolution of $\xi$ is now dependent on both $\phi$ and $V$. The can be seen by expressing the two Josephson equations as
\begin{subequations}
\begin{align}
    \iota_\beta\xi 
    &= c_\phi \frac{\mu}{T}
    + \frac1T {\cal K},
    \label{eq:Joseph-xi-Coulomb} \\
    \lie_\beta\xi 
    &= 
    c_\phi\! \lb \df\frac{\mu}{T} + \iota_\beta F \rb
    + \df{\cal K}
    + \frac1T \tilde\ell\,{\cal V}.
    \label{eq:xi-EOM}
\end{align}
\end{subequations}

% For a U(1)$_q$ pseudo-superfluid, we have the approximate U(1)$_q$ charge and energy-momentum conservation equations
% \begin{subequations}
% \begin{align}
%     \df{\star J}
%     &= c_\phi(-)^q\,{\star K_\ext}
%     + (-)^{q+1} \ell\, {\star L}, 
%     \label{eq:hydro-eqs-Coulomb-1} \\
%     %
%     % \df{\star L}
%     % &= 0, \\
%     %
%     \nabla_\mu T^{\mu\nu}
%     &= (F\cdot J)^\nu 
%     + (\ell\, \Xi\cdot L)^\nu
%     - (\xi \cdot {K_\ext})^\nu.
%     \label{eq:hydro-eqs-Coulomb-2}
% \end{align}
% \label{eq:hydro-eqs-Coulomb}%
% \end{subequations}
% We have introduced a $q$-form background field $K_\ext$ into the conservation equations coupled to the U(1)$_q$ Goldstone phase field $\phi$, satisfying $\df{\star K_\ext} = 0$. In the context of gauge theories, $K_\ext$ can be understood as ``external conserved current'' coupled to the dynamical gauge fields $\phi$; see our discussion around \cref{eq:max-2-broken}.

Similar to our discussion around \cref{eq:gauge-freedom-TSSB}, we have a U(1)$^\loc_{q-1}$ gauge freedom in this theory given by 
\begin{alignat}{2}
    \Lambda_\beta 
    &\to \Lambda_\beta - \lie_\beta \df\lambda,  \qquad
    & \Lambda_\beta^\ell
    &\to \Lambda_\beta^\ell - \lie_\beta\lambda, \nn\\
    \phi &\to \phi - c_\phi\,\df\lambda, \qquad
    &\varphi_\ell 
    &\to \varphi_\ell - \iota_\beta\lambda.
\end{alignat}
We can see that $\phi$ acts as dynamical gauge field for this U(1)$^\loc_{q-1}$ symmetry.

% However, we need a new 

% Since $\phi$ is singular, it does not make sense to introduce a source for it in the equations of motion directly. We instead introduce a new \tightmath{$(q+1)$}-form source $U_\ext$ coupled to $\xi$ directly. The source for $\phi$ can formally be defined from here as $\star K_\ext = (-)^q\df{\star U_\ext}$ and shows up in the U(1)$_q$ conservation equation \eqref{eq:hydro-eqs-Coulomb-1}. 
% However, the energy-momentum conservation \eqref{eq:hydro-eqs-Coulomb-2} now needs to be modified to include a new source term
% \begin{align}
%     \nabla_\mu T^{\mu\nu}
%     &= (F\cdot J)^\nu 
%     + (\ell\, \Xi\cdot L)^\nu \nn\\
%     &\qquad
%     - (\xi \cdot {K_\ext})^\nu
%     + (\tilde\ell\, \df V \cdot U_\ext)^\nu.
%     \label{eq:hydro-eqs-Coulomb-2-p}
% \end{align}

\subsubsection{Dual formulation}

The set of hydrodynamic equations outlined above can also be described in terms of a temporally-spontaneously broken approximate anomalous $\rmU(1)_q\times\rmU(1)_p$ symmetry. In this formalism, we do not interpret the U(1)$_p$ conservation equation in \eqref{eq:hydro-eqs-Coulomb-dual-2} as a Bianchi identity, but rather treat both U(1)$_q$ and U(1)$_p$ symmetries democratically.
The hydrodynamic fields are taken to be a set of symmetry parameters $\Lambda_\beta$, $\Lambda^\ell_\beta$, $\tilde\Lambda_\beta$, $\tilde\Lambda^\ell_\beta$, and $\beta^\mu$, and their transformation properties are given analogous to \cref{eq:hydro-trans}. We will take the U(1)$_q$, U(1)$_p$ higher-form symmetries and the emergent U(1)$^\ell_{q-1}$, U(1)$^\ell_{p-1}$ topological symmetries to be temporally spontaneously broken, for \tightmath{$q>0$}, \tightmath{$q>1$}, \tightmath{$p>0$}, \tightmath{$p>1$} respectively, giving rise to the respective temporal-Goldstones $\varphi$, $\tilde\varphi$, $\varphi_\ell$, $\tilde\varphi_\ell$, with the same transformation properties analogous to \cref{eq:varphi-trans-TSSB}. Their dynamics is governed by the Josephson equations that take the same form as \eqref{eq:Joseph-TSSB}.

This representation of the theory has two copies of gauge symmetries U(1)$^\loc_{q-1}$, U(1)$^\loc_{p-1}$ analogous to \cref{eq:gauge-freedom-TSSB}. However, we do not need to worry about these as long as we work with the gauge-invariant higher-form chemical potentials $\mu$, $\tilde\mu$, and the associated defect chemical potentials $\mu_\ell$, $\tilde\mu_\ell$ defined similar to \cref{eq:mu-defs-hydro}. Together with $\beta^\mu$, these make up the right amount of degrees of freedom to solve for using the hydrodynamic equations \eqref{eq:hydro-eqs-Coulomb-dual}.

% Let us consider a superfluid, where a $q$-form U(1) generalised global symmetry of the effective theory is spontaneously broken in the ground state. This gives rise to a Goldstone phase field $\phi$ that transforms under the U(1)$_q$ symmetry as $\phi \to \phi - c_\phi\Lambda$, where $c_\phi$ denotes the constant charge of $\phi$. Let us also introduce a background  U(1)$_q$ gauge field $A$ transforming as $A \to A + \df\Lambda$, so that we can define the ``superfluid velocity'' as the gauge-covariant derivative of the Goldstone, i.e. $\xi = \df\phi+c_\phi A$. The above setup is a straight-forward generalisation of ordinary 0-form superfluids for $q=0$; see e.g.~\cite{landau1959fluid, Bhattacharya:2011tra}.
% Furthermore, if we take $c_\phi = 0$ together with $q=0$, the same setup can also apply to uni-directional charge density wave states, where the uncharged scalar field $\phi$ being interpreted as the Goldstone of spontaneously broken translations in one spatial direction.

Let us see how this description is related to the original conventional formulation of a U(1)$_q$ (pseudo-)superfluid. Essentially, we can use the constitutive relations for the temporal parts of the currents $\iota_u\tilde J$, $\iota_u\tilde L$ to relate $\tilde\mu_\ell$, $\tilde\mu_\ell$ to the spatial components of the superfluid velocity $\xi$ and defect density $\df V$ respectively via 
\begin{align}
    \iota_u\tilde J &= (-)^{q+1}{*\xi}, \nn\\
    \iota_u {\tilde L} 
    &= {*\df V}.
\end{align}
On the other hand, the physical information in the two Josephson equations \eqref{eq:joseph-Coulomb} can be identified with the constitutive relations for the spatial parts of the currents; to wit
\begin{align}
    \rmP_u\tilde J 
    &= - {*\iota_u\xi} 
    = - c_\phi {*\mu}
    + {\cal\tilde J}, \nn\\
    \rmP_u{\tilde L} 
    &= (-)^{q+1} {*\iota_u\df V}
    = {\cal\tilde L},
    \label{eq:current-identifications}
\end{align}
where we have defined 
\begin{align}
    {\cal\tilde J} = - {*\cal K}, \qquad 
    {\cal\tilde L} = (-)^{q+1}{*{\cal V}}.
    \label{eq:K-J-reln}
\end{align}
This ensures that the two formulations describe the same physical system.

% The time-component of this
% field is usually fixed in terms of the U(1)$_0$ chemical potential $\mu$ via the
% Josephson's relation like $\xi_t = c_\phi\mu + \ldots$, with the ellipses suppressing
% possible derivative corrections; we shall return to this relation in the due
% course. 

% The effective theory describing a superfluid is also be invariant under the
% global shift of the Goldstone phase $\phi \to \phi + \lambda$ that takes us
% around the circle of equivalent ground states arising from spontaneous symmetry breaking. It embodies the fact that the true degrees of freedom are the changes in Goldstone phase, not the phase itself. Note that, despite the similar action on $\phi$, the global U(1) symmetry and this shift symmetry are actually distinct; the former is a true global
% symmetry of the theory and generically acts on all the charged matter in the
% system, while the latter is exclusive to the Goldstone. In particular, while the
% global U(1) symmetry is gauged by the background $A_\mu$, we shall leave the
% shift symmetry un-gauged. 

% \begin{equation}
%     K^{\lambda_1\ldots\lambda_{q+1}}_{\ext(q+1)}
%     = \frac{(-1)^q}{(p+1)!} \epsilon^{\lambda_1\ldots\lambda_{q+1}
%     \mu_1\ldots\mu_{p+1}}
%     \tilde A^{(p+1)}_{\mu_1\ldots \mu_{p+1}}
% \end{equation}

% \begin{equation}
%     \frac{(-1)^{q}}{(p+1)!}\epsilon_{\nu_1\ldots\nu_{q+1}
%     \mu_1\ldots\mu_{p+1}}
%     \tilde J^{\mu_1\ldots\mu_{p+1}}_{(p+1)} 
%     = 
%     \xi_{\nu_1\ldots\nu_{q+1}}^{(q+1)}.
% \end{equation}

\subsection{Second law of thermodynamics and constitutive relations}

Having set up the hydrodynamic model, let us construct the respective constitutive relations. Using the free energy current defined in \cref{eq:N-def}, we find that the adiabaticity equation \eqref{eq:adiabaticity} for a superfluid modifies to
\begin{align}
    \nabla_\mu N^\mu 
    &= \half T^{\mu\nu} \delta_\scB g_{\mu\nu}
    + J\cdot\delta_\scB A 
    + \ell L\cdot\delta_\scB \Phi
  \nn\\
    &\qquad 
    - K_\ext \cdot \delta_\scB \phi
    + \ell U_\ext \cdot\delta_\scB V
    + \Delta,
\end{align}
where we have defined $\star K_\ext = (-)^{q}{\tilde F}$ and $\star U_\ext = {\tilde\Xi}$.
In addition to the variational definitions in \cref{eq:deltaB-defs}, we have further defined
\begin{alignat}{2}
    \delta_\scB \phi 
    &= \lie_\beta\phi - \Lambda 
    &&= \iota_\beta\xi - \frac{\mu}{T}, \nn\\
    \delta_\scB V
    &= \lie_\beta V 
    &&= \iota_\beta \df V.
    \label{eq:phi-variation}
\end{alignat}
We need to solve the adiabaticity equation for the constitutive relations for $T^{\mu\nu}$, $J$, $L$, together with corrections to the Josephson equations ${\cal K}$, ${\cal V}$, in terms of $u^\mu$, $T$, $\mu$, $\mu_\ell$, $\xi$, $\df V$, and the background fields, given some $N^\mu$ and $\Delta\geq 0$, arranged order-by-order in derivatives.

\subsubsection{Constitutive relations}

Following our discussion in \cref{sec:secondlaw}, we can set up the thermodynamics for a U(1)$_q$ superfluid. Unlike \cref{sec:secondlaw}, however, the thermodynamic pressure $P$ can now also depend on the spatial components of the superfluid velocity $\tilde n = (-)^q {*\xi}$ and the vortex density $\tilde n_\ell = -{*\df V}$. We can parameterise its variation as
\begin{align}
    \delta P 
    &= s\delta T 
    + n\cdot\delta\mu
    + n_\ell\cdot\delta\mu_\ell 
    - {*\tilde\mu} \cdot \delta{*\tilde n}
    - {*\tilde\mu_\ell} \cdot\delta {*\tilde n_\ell}
    \nn\\
    &\qquad 
    - \half r^{\mu\nu}\df g_{\mu\nu}, \nn\\
    \epsilon
    &= -P + Ts 
    + \mu\cdot n
    + \mu_\ell\cdot n_\ell.
    \label{eq:thermo-Coulomb}
\end{align}
This equation can be seen as defining the conjugate chemical potentials $\tilde\mu$ and $\tilde\mu_\ell$. By using $\xi$ as the independent degree of freedom in the conventional formulation, we are naturally led to work in the canonical ensemble with respect to $\tilde n$ and $\tilde n_\ell$. We can find the associated $r^{\mu\nu}$ using Lorentz invariance
\begin{align}
    r^{\mu\nu}
    &= \frac{1}{(q-1)!} n^\mu{}_{\rho_1\ldots}
    \mu^{\nu\rho_1\ldots} 
    + \frac{1}{(q-2)!} 
    (n_\ell)^\mu{}_{\rho_1\ldots}
    \mu_\ell^{\nu\rho_1\ldots} \nn\\
    &\quad 
    - \frac{1}{q!} {*\tilde n}^\mu{}_{\rho_1\ldots}
    {*\tilde\mu}^{\nu\rho_1\ldots} 
    - \frac{1}{(q+1)!} 
    ({*\tilde n_\ell})^\mu{}_{\rho_1\ldots}
    {*\tilde\mu_\ell}^{\nu\rho_1\ldots}, \nn\\
    r^{[\mu\nu]} &= 0.
\end{align}
We can take a simplified equation of state where $P$ only depends on the squares of the higher-form objects, leading to the thermodynamic relations
\begin{subequations}
\begin{equation}
    \delta P = s\df T 
    + \half\chi  \delta\mu^2
    + \half\chi_\ell \delta\mu_\ell^2
    - \frac{1}{2\tilde\chi} \delta \tilde n^2
    - \frac{1}{2\tilde\chi_\ell} \delta \tilde n_\ell^2.
\end{equation}
In this case, various densities are related to the chemical potentials as
\begin{equation}
    n = \chi\mu, \quad 
    n_\ell = \chi_\ell\mu_\ell, \quad
    \tilde n = \tilde\chi\tilde\mu, \quad 
    \tilde n_\ell = \tilde\chi_\ell\tilde\mu_\ell.
\end{equation}
Additionally, the anisotropic stress tensor $r^{\mu\nu}$ is automatically symmetric.
\end{subequations}

To analyse the adiabaticity equation, let us perform a canonical decomposition of the free energy current. We take the parametrisation
\begin{equation}
    N^\mu 
    = P\,\beta^\mu 
    + \frac1T ({*{\cal K}} \cdot \tilde\mu)^\mu
    + \frac{(-)^{q}}{T} (*{\cal V} \cdot {\tilde\mu_\ell})^\mu
     + {\cal N}^\mu,
     \label{eq:N-current-SF}
\end{equation}
where ${\cal N}^\mu$ denotes non-canonical contributions. The additional terms proportional to ${\cal K}$ and ${\cal V}$ have been included for later convenience.
We are now ready to derive the constitutive relations of an ideal superfluid. Following a procedure similar to the one used for \cref{eq:consti-ansatz}, together with the variational formulae
\begin{align}
    {*\tilde\mu}\cdot\delta_\scB {*\tilde n}
    &= T(-)^{pq+q+1} 
    {*\df\frac{\tilde\mu}{T}} \cdot \delta_\scB\phi \nn\\
    &\qquad 
    + (-)^{pq+p+q} 
    {*\tilde\mu}\cdot \lb 
    c_\phi\delta_\scB A
    + \tilde\ell\delta_\scB V
    \rb \nn\\
    &\qquad 
    + c_\phi (-)^{p} 
    {*(\mu\wedge\tilde\mu)}^{\mu} u^{\nu}\delta_\scB g_{\mu\nu} \nn\\
    &\qquad 
    - (-)^{pq+q} 
    \star\!\df( u\wedge \tilde\mu \wedge \delta_\scB\phi)
    , \nn\\[0.3em]
    {*\tilde\mu_\ell}\cdot\delta_\scB{*\tilde n_\ell}
    &= T(-)^{pq+q+p+1}
    {*\df\frac{\tilde\mu_\ell}{T}}
    \cdot\delta_\scB V
    \nn\\
    &\qquad
    - (-)^{pq+p+q} \star\!\df(u\wedge{\tilde\mu_\ell}
    \wedge\delta_\scB V),
    \label{eq:delta-n}
\end{align}
we can find the constitutive relations
\begin{align}
    T^{\mu\nu}
    &= (\epsilon+P) u^\mu u^\nu + P\,g^{\mu\nu}
    - r^{\mu\nu} \nn\\
    &\qquad 
    - 2c_\phi (-)^{p} 
    {*(\mu\wedge\tilde\mu)}^{(\mu} u^{\nu)}
    + {\cal T}^{\mu\nu}
    , \nn\\
    J
    &= u\wedge n - \tilde c_\phi\,{*\tilde\mu} + {\cal J}, \nn\\
    L
    &= u \wedge n_\ell + {\cal L}.
    \label{eq:consti-ansatz-SF}
\end{align}
These satisfy the adiabaticity equation with $\Delta=0$ when all the derivative corrections have been switched off.
The total derivative terms in \cref{eq:delta-n} precisely account for the two additional terms in the free energy current decomposition in \cref{eq:N-current-SF}.

Plugging the constitutive relations \eqref{eq:consti-ansatz-SF} into the adiabaticity equation together with the Josephson equations \cref{eq:joseph-Coulomb}, we can read of the constraint equation to be satisfied by the dissipative corrections 
\begin{align}
    &- \half {\cal T}^{\mu\nu} \delta_\scB g_{\mu\nu}
    - {\cal J}\cdot\delta_\scB A 
    - \ell {\cal L}\cdot\delta_\scB \Phi \nn\\
    &\qquad
    - {\cal\tilde J}\cdot \delta_\scB\tilde A
    - \tilde\ell  {\cal\tilde L}\cdot \delta_\scB\tilde\Phi
    + \nabla_\mu {\cal N^\mu}
    = \Delta  \geq 0,
    \label{eq:constraint-adiabaticity-sf}
\end{align}
where the variations $\delta_\scB\tilde A$ and $\delta_\scB\tilde\Phi$  are defined similar to \cref{eq:deltaB-defs}.
The Landau frame condition is given as
\begin{equation}
    {\cal T}^{\mu\nu}u_\nu 
    = \iota_u {\cal J} = \iota_u {\cal L} 
    = \iota_u{\cal\tilde J}
    = \iota_u{\cal\tilde L}
    = 0.
\end{equation}

\subsubsection{Dual formulation}

The same constitutive relations can also be derived using the dual formulation in terms of anomalous higher-form symmetries. The relevant free energy current is given by a canonical transformation of \cref{eq:N-def} as
\begin{align}
    TN'^\mu
    &= TN^\mu   
    + (\tilde J \cdot \tilde\mu)^\mu
    + (\tilde L \cdot \tilde\mu_\ell )^\mu  \nn\\
    &= TS^\mu 
    + T^{\mu\nu} u_\nu 
    + (J\cdot\mu)^\mu 
    + (\tilde J \cdot \tilde\mu)^\mu \nn\\
    &\qquad 
    + (L\cdot\mu_\ell)^\mu
   + (\tilde L \cdot \tilde\mu_\ell )^\mu.
   \label{eq:freeE-dual-Coulomb}
\end{align}
One can check that the respective adiabaticity equation takes a similar form to \cref{eq:adiabaticity} but generalised to two higher-form symmetries, i.e.
\begin{align}
    \nabla_\mu N'^\mu 
    &= \rmN_\rmH 
    + \half T^{\mu\nu} \delta_\scB g_{\mu\nu}
    + J\cdot\delta_\scB A 
    + \tilde J\cdot\delta_\scB \tilde A  \nn\\
    &\qquad 
    + \ell L\cdot\delta_\scB \Phi
    + \tilde\ell \tilde L\cdot\delta_\scB \tilde\Phi
    + \Delta.
    \label{eq:adiabaticity_sf}
\end{align}
Here $\delta_\scB\tilde A$ and $\delta_\scB\tilde\Phi$ are defined similarly to \cref{eq:deltaB-defs}.
We also get a new anomaly-induced ``Hall free energy'' contribution in the adiabaticity equation given by
\begin{align}
    \rmN_\rmH 
    &= (-)^{q+1}\frac{c_\phi}{T}
    \lb (-)^{pq}
    \mu\cdot{\star\tilde F}
    + \tilde\mu\cdot{\star F} \rb.
\end{align}

Similarly, the canonical decomposition of the free energy current in \cref{eq:N-current-SF} becomes
\begin{equation}
    N'^\mu 
    = P'\beta^\mu 
    - (-)^{p}\frac{c_\phi}{T}
    *\!\lb\mu\wedge\tilde\mu \rb^\mu 
     + {\cal N}^\mu,
     \label{eq:Nmu-SF}
\end{equation}
where the conjugate pressure is given as
\begin{align}
    P' 
    &= P + \tilde n \cdot \tilde\mu
    + \tilde n_\ell \cdot \tilde\mu_\ell.
\end{align}
The differential of the new pressure yields
\begin{align}
    \delta P'
    &= s\delta T 
    + n\cdot\delta\mu
    + \tilde n\cdot\delta\tilde\mu
    + n_\ell\cdot\delta\mu_\ell 
    + \tilde n_\ell\cdot\delta\tilde\mu_\ell \nn\\
    &\qquad 
    - \half r'^{\mu\nu}\delta g_{\mu\nu}, \nn\\
    \epsilon
    &= -P' + Ts 
    + \mu\cdot n
    + \tilde\mu\cdot \tilde n \nn\\
    &\qquad 
    + \mu_\ell\cdot n_\ell
    + \tilde \mu_\ell\cdot \tilde n_\ell,
\end{align}
together with
\begin{align}
    r'^{\mu\nu}
    &= r^{\mu\nu}
    + P^{\mu\nu} \lb \tilde n \cdot \tilde\mu
    + \tilde n_\ell \cdot \tilde\mu_\ell \rb \nn\\
    &= \frac{1}{(q-1)!} n^\mu{}_{\rho_1\ldots}
    \mu^{\nu\rho_1\ldots} 
    + \frac{1}{(q-2)!} 
    (n_\ell)^\mu{}_{\rho_1\ldots}
    \mu_\ell^{\nu\rho_1\ldots} \nn\\
    &\quad 
    + \frac{1}{(p-1)!} \tilde n^\mu{}_{\rho_1\ldots}
    \tilde\mu^{\nu\rho_1\ldots} 
    + \frac{1}{(p-2)!} 
    (\tilde n_\ell)^\mu{}_{\rho_1\ldots}
    \tilde\mu_\ell^{\nu\rho_1\ldots}, \nn\\
    r'^{[\mu\nu]} &= 0.
\end{align}
Using these, together with the identifications in \cref{eq:current-identifications},  we can read off the constitutive relations in the dual formulation 
\begin{align}
    T^{\mu\nu}
     &= (\epsilon+P')u^{\mu}u^\nu + P'\, g^{\mu\nu}
     - r'^{\mu\nu} \nn\\
     &\qquad 
     - 2c_\phi(-)^{p}
     *\!\lb\mu\wedge\tilde\mu \rb^{(\mu} 
     u^{\nu)}
     + {\cal T}^{\mu\nu}, \nn\\
     J
     &= u \wedge n
     - \tilde c_\phi\,{*\tilde\mu}
     + {\cal J}, \nn\\
     \tilde J
     &= u \wedge \tilde n -
     c_\phi\,{*\mu}
     + {\cal\tilde J}, \nn\\
     L
     &= u \wedge n_\ell 
     + {\cal L}
     , \nn\\
     \tilde L
     &= u \wedge \tilde n_\ell
     + \tilde{\cal L}.
\end{align}
Note that the definitions of the dissipative corrections $\cal\tilde J$ and $\cal\tilde L$ coincide with \cref{eq:K-J-reln}. Ignoring all derivative corrections, we can see that we have arrived at exactly the same form of the constitutive relations as derived using an equilibrium action in \cref{sec:appx-SF-Coulomb-eqb}.
The constraint equation for dissipative corrections in the higher-form formulation takes the same form given in \cref{eq:constraint-adiabaticity-sf}.

\subsubsection{Dissipative corrections}
\label{sec:dissipation-Coulomb}

\begin{table}[t]
\def\arraystretch{1.5}
    \centering
    \begin{tabular*}{0.7\columnwidth}{@{\extracolsep{\fill}} c|ccc}
        & $\rmC$ & $\rmP$ & $\rmT$ \\
        \hline\hline 

        $T^{tt}$, $g_{tt}$, $T$ 
        & $+$ & $+$ & $+$  \\
        $T^{ti}$, $g_{ti}$, $u^i$ 
        & $+$ & $-$ & $-$  \\
        $T^{ij}$, $g_{ij}$
        & $+$ & $+$ & $+$  \\

        \hline
        
        $J^{ti\ldots}$, $A_{ti\ldots}$, $\mu_{i\ldots}$  
        & $-$ & $(-)^{q+1}$ & $+$ \\
        $J^{ij\ldots}$, $A_{ij\ldots}$  
        & $-$ & $(-)^{q}$ & $-$ \\

        $L^{ti\ldots}$, $\Phi_{ti\ldots}$, $\mu^\ell_{i\ldots}$  
        & $-$ & $(-)^{q}$ & $+$ \\
        $L^{ij\ldots}$, $\Phi_{ij\ldots}$  & $-$ & $(-)^{q+1}$ & $-$ \\

        \hline 
        
        $\tilde J^{ti\ldots}$, $\tilde A_{ti\ldots}$, $\tilde\mu_{i\ldots}$ 
        & $-$ & $(-)^{p}$ & $-$ \\
        $\tilde J^{ij\ldots}$, $\tilde A_{ij\ldots}$  
        & $-$ & $(-)^{p+1}$ & $+$ \\
        $\tilde L^{ti\ldots}$, $\tilde\Phi_{ti\ldots}$, $\tilde\mu^\ell_{i\ldots}$  
        & $-$ & $(-)^{p+1}$ & $-$ \\
        $\tilde L^{ij\ldots}$, $\tilde\Phi_{ij\ldots}$  
        & $-$ & $(-)^{p}$ & $+$ \\
        
        \hline 

        $\epsilon^{tij\ldots}$ 
        & $+$ & $(-)^{d-1}$ & $+$
        
    \end{tabular*}
    \caption{Discrete symmetries of various objects. Under the duality transformation $p\leftrightarrow q$, the spacetime discrete symmetries will change as $\rmP \leftrightarrow \rmC\rmP$, $\rmT \leftrightarrow \rmC\rmT$. }
    \label{tab:CPT}
\end{table}

For simplicity, let us assume the constitutive relations to be parity and charge-conjugation invariant. The discrete transformation properties of various objects are given in \cref{tab:CPT}. We also ignore any all anisotropic dissipative transport. With these in mind, the constitutive relations for the dissipative energy-momentum tensor ${\cal T}^{\mu\nu}$ take precisely the same form as before in \cref{eq:diss-consti-normal}. Next, we have transport in the higher-form currents. It turns out that the dissipative $(q+1)$-form flux ${\cal J}$ and the dissipative $p$-form defect flux ${\cal\tilde L}$ mutually couple, give rise to 
\begin{align}
    {\cal J}
    &= - T\sigma \,\rmP_u \delta_\scB A
    - (-)^{pq+p+q}T\tilde\ell\tilde\gamma_\times
    {*\delta_\scB\tilde\Phi} \nn\\
    &= -\sigma
    \lb T\,\rmP_u\df\frac{\mu}{T} 
    + \iota_u F
    \rb \nn\\
    &\qquad 
    - (-)^{pq+p+q} \tilde\gamma_\times
    {*\!\lb 
    T\df\frac{\tilde\mu_\ell}{T} 
    + \tilde\ell\lb \iota_u\tilde\Xi - \tilde\mu \rb 
    \rb } , \nn\\[0.5em]
    \tilde{\cal L}
    &= - T\tilde\ell\tilde\sigma_\ell\,\rmP_u\delta_\scB\tilde\Phi
    + (-)^q T\tilde\gamma'_\times\, 
    {*\delta_\scB A} \nn\\
    &= 
    - \tilde\sigma_\ell
    \lb T\,\rmP_u\df\frac{\tilde\mu_\ell}{T} 
    + \tilde\ell\lb \iota_u\tilde\Xi - \tilde\mu \rb
    \rb \nn\\
    &\qquad 
    + (-)^q \tilde\gamma'_\times\, {*\!\lb 
    T\df\frac{\mu}{T} 
    + \iota_u F\rb}.
\end{align}
We see the same coupling structure in the dissipative $(p+1)$-form flux ${\cal\tilde J}$ and the dissipative $q$-form defect flux ${\cal L}$, taking the form
\begin{align}
    {\cal\tilde J}
    &= - T\tilde\sigma \,\rmP_u \delta_\scB \tilde A
    - (-)^{pq+p+q}
    T\ell\gamma_\times{*\delta_\scB\Phi} \nn\\
    &= -\tilde\sigma
    \lb T\,\rmP_u\df\frac{\tilde\mu}{T} 
    + \iota_u \tilde F
    \rb \nn\\
    &\qquad 
    - (-)^{pq+p+q} \gamma_\times
    {*\!\lb 
    T\df\frac{\mu_\ell}{T} + \ell\lb \iota_u\Xi - \mu \rb 
    \rb } , \nn\\[0.5em]
    {\cal L}
    &= - T\ell\sigma_\ell\,\rmP_u\delta_\scB\Phi
    + (-)^p T\gamma'_\times
    {*\delta_\scB \tilde A} \nn\\
    &= 
    - \sigma_\ell
    \lb T\,\rmP_u\df\frac{\mu_\ell}{T} 
    + \ell\lb \iota_u\Xi - \mu \rb \rb \nn\\
    &\qquad 
    + (-)^p \gamma'_\times\, 
    {*\!\lb T\df\frac{\tilde\mu}{T} 
    + \iota_u \tilde F\rb}.
    \label{eq:tildeJ-L-diss}
\end{align}
The coefficient $\sigma$, $\tilde\sigma$ are the conductivities of U(1)$_q$, U(1)$_p$ charges respectively, which we met already in \cref{sec:higher-form-hydro}. Similarly, $\sigma_\ell$, $\tilde\sigma_\ell$ are conductivities of the respective defects. They are also responsible for the relaxation of U(1)$_q$, U(1)$_p$ charges, through the linear term in the chemical potentials, eventually giving rise to the damping-attenuation relations like \cref{eq:damping-attenuation}. 
However, interestingly, we now also have cross coefficients $\gamma_\times$, $\gamma'_\times$, $\tilde\gamma_\times$, $\tilde\gamma'_\times$ coupling the U(1)$_q$ and U(1)$_p$ symmetries. The physical significance of these will be clear below. In the anisotropic case, we expect more intricate couplings between various currents, leading to many more transport coefficients.

% \begin{align}
%      \nn\\[0.5em]
%     %
%     %
%     \tilde{\cal J}
%     &= - T\tilde\sigma \Big(
%     \delta_\scB \tilde A 
%     + u\wedge\iota_u \delta_\scB \tilde A \Big)
%     - T\ell\gamma_\times {\star(u\wedge \delta_\scB\Phi)}, \nn\\[0.5em]
%     %
%     {\cal L}
%     &= - T\ell\sigma_\Phi \Big( \delta_\scB \Phi
%     + u\wedge\iota_u\delta_\scB \Phi \Big)
%     - T\gamma'_\times {\star(u\wedge \delta_\scB\tilde A)}, \nn\\[0.5em]
% \end{align}

Requiring the microscopic theory underlying the hydrodynamic description to be invariant under time-reversal transformations, the primed and unprimed cross coefficients are related via the Onsager's relations
\begin{equation}
    \gamma'_\times
    = \gamma_\times, \qquad
    \tilde\gamma'_\times 
    = \tilde\gamma_\times,
\end{equation}
leaving only two of the cross coefficients to be independently physical.  
With the Onsager's relations in place, we find that the cross coefficients do now contribute to the entropy production quadratic form $\Delta$ obtained through \cref{eq:constraint-adiabaticity-sf}. Demanding entropy production to be positive semi-definite, we find the inequality constraints on the remaining six coefficients
\begin{gather}
    \eta,\zeta,\sigma,\tilde\sigma,
    \sigma_\ell,\tilde\sigma_\ell \geq 0.
\end{gather}
We will discuss the physical implications of these coefficients in the next subsection.

% The $\gamma_\times$ coefficient is related to explicitly broken U(1)$_q$ symmetry and shows up in the constitutive relations for the U(1)$_q$ violation operator $L$ and the U(1)$_p$ flux $\tilde J$ (which, in the U(1)$_q$ superfluid interpretation, is dual to the Josephson relation for the U(1)$_q$ Goldstone). Physically, this gives rise to the $q$-form analogue of the $\lambda$ coefficient discussed in our recent paper~\cite{Armas:2021vku}.

For completeness, let us also record the corrections to the $\phi$ and $V$ Josephson equations obtained using \cref{eq:K-J-reln}, leading to
\begin{align}
    T {\cal K} 
    & = (-)^{pq+q} \tilde\sigma
    *\!\lb T\df\frac{\tilde\mu}{T} 
    + \iota_u \tilde F
    \rb \nn\\
    &\qquad 
    + (-)^{pq+p+q} \gamma_\times
    \lb 
    T\,\rmP_u\df\frac{\mu_\ell}{T} + \ell\lb \iota_u\Xi - \mu \rb 
    \rb , \nn\\[0.5em]
    T{\cal V}
    &= 
    (-)^{pq+p+q} \tilde\sigma_\ell
    *\!\lb T\df\frac{\tilde\mu_\ell}{T} 
    + \tilde\ell\lb \iota_u\tilde\Xi - \tilde\mu \rb
    \rb \nn\\
    &\qquad 
    - \tilde\gamma'_\times \lb 
    T\,\rmP_u\df\frac{\mu}{T} 
    + \iota_u F\rb.
\end{align}
In particular, notice that there is a term proportional to $\mu$ in the constitutive relations for $\cal K$ coming with the coefficient $\gamma_\times$. This will modify the leading order $c_\phi\mu$ term in the Josephson equation \eqref{eq:Joseph-xi-Coulomb} to $\lambda c_\phi\mu$, where $\lambda = 1 - \ell\gamma_\times/\tilde c_\phi$. Similarly, there is a term proportional to $\tilde\mu = (-)^q {*\xi}/\tilde\chi$ in the constitutive relations for ${\cal V}$ coming with the coefficient $\tilde\sigma_\ell$. Plugging this into the equation of motion \eqref{eq:xi-EOM}, this term will relax the spatial components of $\xi$ with relaxation rate $\tilde\Gamma = \tilde\ell^2\tilde\sigma_\ell/\tilde\chi$. More on this in a moment.

We have carefully parametrised the constitutive relations, so that the theory is self-dual under the exchange of tilde and untilde quantities. This implements the $\rmU(1)_q \leftrightarrow \rmU(1)_p$ duality of the relaxed phase of a U(1)$_q$ pseudo-superfluid with vortices.

\subsection{Linearised analysis and mode spectrum}

To understand the physical meaning of the hydrodynamic equations and various transport coefficients, let us perform the same simplifications as in \cref{sec:modes-hydro}. In particular, we focus on fluctuations around an equilibrium state with zero densities, so the energy-momentum fluctuations decouple and lead to the same longitudinal sound and shear diffusion modes in \cref{eq:hydro-sound}. The higher-form charge currents, on the other hand, take the form
\begin{align}
    J
    &= u \wedge n
    - \tilde\lambda\tilde c_\phi
    {*\tilde\mu} 
    -\sigma \rmP_u\df\mu 
    - (-)^{pq+p+q}\tilde\gamma_\times
    {*\df\tilde\mu_\ell}, \nn\\[0.5em]
    \tilde J
    &= u \wedge \tilde n
    - \lambda c_\phi {*\mu} 
    - \tilde\sigma \rmP_u\df\tilde\mu 
    - (-)^{pq+p+q}\gamma_\times
    {*\df\mu_\ell}, \nn\\[0.5em]
    L
    &= u \wedge n_\ell 
    + \ell\sigma_\ell \mu
    - \sigma_\ell \rmP_u\df\mu_\ell
    + (-)^p \gamma_\times {*\df\tilde\mu }, \nn\\[0.5em]
    \tilde L
    &= u \wedge \tilde n_\ell 
    + \tilde\ell\tilde\sigma_\ell \tilde\mu
    - \tilde\sigma_\ell \rmP_u\df\tilde\mu_\ell
    + (-)^q \tilde\gamma_\times {*\df\mu },
\end{align}
where 
\begin{equation}
    \lambda = 1
    - \frac{\ell\gamma_\times}{\tilde c_\phi}, \qquad 
    \tilde\lambda 
    = 1
    - \frac{\tilde\ell\tilde\gamma_\times}{c_\phi},
\end{equation}
renormalise $c_\phi$ and $\tilde c_\phi$ respectively. In the conventional formulation, the constitutive relations for $\tilde J$ and $\tilde L$ give rise to the Josephson equations 
\begin{align}
    \iota_u\xi
    &= \lambda c_\phi \mu
    + (-)^{pq} \frac{\tilde\sigma}{\tilde\chi}
    {*\df{*\xi}}
    + (-)^{pq+p+q} \gamma_\times\rmP_u\df\mu_\ell, \nn\\[0.5em]
    \iota_u\df V
    &= - \frac{\tilde\ell\tilde\sigma_\ell}{\tilde\chi} 
    \rmP_u\xi \nn\\
    &\qquad 
    + (-)^{pq+p+q} \tilde\sigma_\ell
    {* \df\tilde\mu_\ell} 
    - \tilde\gamma_\times\rmP_u\df\mu.
    \label{eq:joseph-linear-coulomb}
\end{align}
We see that the $\lambda$ coefficient modifies the leading order Josephson equation and screens the effective chemical potential as seen by the superfluid phase.
The same result was also found for 0-form superfluids in~\cite{Armas:2021vku}, with the coefficient $\gamma_\times/\tilde c_\phi$ being identified as $-\sigma_\times/\sigma_\phi$ from that paper. For $q>0$, this coefficient also controls the coupling of the Josephson relation to the U(1)$_q$ defect chemical potential. The coefficient $\tilde\sigma$, on the other hand, causes the Goldstone to diffuse and can be identified with $1/\sigma_\phi$ from~\cite{Armas:2021vku}.

Using the constitutive relations, we can see that the two higher-form conservation equations become
\begin{align}
    (\dow_t + \Gamma) n
    &= (-)^{p}\frac{\lambda_s}{\tilde\chi} \tilde c_\phi {*\df\tilde n} \nn\\
    &\qquad 
    + D_n \dow^2 n
    - (-)^{pq+p+q} (D_\ell - D_n) {\df{*\df n}}, \nn\\
    (\dow_t + \tilde\Gamma) \tilde n
    &= (-)^{q}\frac{\lambda_s}{\chi} c_\phi {*\df n} \nn\\
    &\qquad
    + \tilde D_n \dow^2 \tilde n
    - (-)^{pq+p+q} (\tilde D_\ell - \tilde D_n) {\df{*\df \tilde n}}, \nn\\
    *\df{*n} 
    &= (-)^{pq+p+q+1} \ell n_\ell, \nn\\
    *\df{*\tilde n} 
    &= (-)^{pq+p+q+1} \tilde\ell \tilde n_\ell,
    \label{eq:linear-SF}
\end{align}
where we have defined two copies of damping and attenuation coefficients as in \cref{eq:damping-attenuation-coeff}. We have also defined a combined renormalisation factor 
\begin{align}
    \lambda_s 
    &= 1
    - \frac{\ell\gamma_\times}{\tilde c_\phi} 
    - \frac{\tilde\ell\tilde\gamma'_\times}{c_\phi}
    = \lambda + \tilde\lambda - 1.
\end{align}
For completeness, let us also write down the conservation equations in terms of the superfluid velocity $\xi$, leading to
\begin{align}
    (\dow_t + \Gamma) {n}
    &= (-)^{pq} \lambda_s 
    c_\phi f_s {*\df{*\xi}}
    + D_n \dow^2 n
    \nn\\
    &\qquad
    - (-)^{pq+p+q} (D_\ell - D_n) {\df{*\df n}}, \nn\\
    (\dow_t + \tilde\Gamma)\rmP_u\xi
    &= \frac{\lambda_s}{\chi} c_\phi \rmP_u\df n
    + \tilde D_n \dow^2\rmP_u\xi
    \nn\\
    &\qquad 
    - (-)^{pq+p+q}
    (\tilde D_\ell - \tilde D_n ) {*\df{*\df\xi}}, \nn\\
    *\df{*n} 
    &= (-)^{pq+p+q+1} \ell n_\ell, \nn\\
    {*\df\xi}
    &= - \tilde\ell\tilde n_\ell.
    \label{eq:linear-SF-conv}
\end{align}

Let us decompose the U(1)$_q$ and U(1)$_p$ charge densities into longitudinal and transverse components according to $n = k/|k|\wedge n_\| + n_\perp$ and $\tilde n = k/|k|\wedge \tilde n_\| + \tilde n_\perp$. In the same spirit, decomposing the spatial components of the superfluid velocity, $\rmP_u\xi = k/|k|\wedge \xi_\| + \xi_\perp$, we can see that 
\begin{equation}
    \tilde n_\perp = \frac{(-)^q}{|k|} {*(k\wedge \xi_\|)}, \qquad 
    \tilde n_\| = \frac{-1}{|k|} {*(k \wedge \xi_\perp)}.
\end{equation}
If it were not for the $\lambda_s$ coupling terms in \cref{eq:linear-SF}, the transverse components of higher-form densities $n_\perp$ and $\tilde n_\perp$ (or the longitudinal components of the superfluid velocity $\xi_\|$), would lead to two independent damped diffusion modes like \cref{eq:perp-diff}. However, because of the coupling terms, these combine into the dispersion relations
\begin{equation}
    \big( i\omega - D_n k^2 - \Gamma \big)
    \big( i\omega - \tilde D_n k^2 - \tilde\Gamma \big)
    + v_\perp^2 k^2
    = 0,
\end{equation}
where we have defined the transverse sound speed $v_\perp$ as
\begin{equation}
    v_\perp^2 = \lambda^2_s \frac{c_\phi^2}{\chi\tilde\chi}.
\end{equation}
If we take $\ell,\tilde\ell\sim k$, i.e. assume both the defect-induced relaxation $\Gamma$ and the vortex-induced relaxation $\tilde\Gamma$ to be ${\cal O}(k^2)$, this yields a second sound mode characteristic of superfluids
\begin{equation}
    \omega = \pm v_\perp k - \frac{i}{2}
    \lb \Gamma + \tilde\Gamma + (D_n+\tilde D_n) k^2 \rb
    + \ldots.
\end{equation}
However, the sound mode is damped due to presence of defects and vortices. 

If we proliferate vortices by taking $\tilde\ell\sim\mathcal{O}( 1)$, thereby making $\tilde\Gamma\sim{\cal O}(k^0)$, the dispersion relations instead give rise to a charge diffusion mode weakly damped due to defects and a strongly damped mode 
\begin{align}
    \omega
    &= - i\Gamma
    -i\lb D_n + \frac{v_\perp^2}{\tilde\Gamma} \rb k^2
    + \ldots, \nn\\
    \omega
    &= - i\tilde\Gamma
    % - i \lb \tilde D_\perp - \frac{v_\perp^2}{\tilde\Gamma} \rb
    % k^2 
    + \ldots.
\end{align}
The resultant mode spectrum is precisely that of a U(1)$_q$ (pseudo)-fluid discussed in \cref{sec:modes-hydro}. Depending on the presence of explicit breaking of the U(1)$_q$ symmetry, this implements the phase transition from (f) to (a) or (g) to (c) in \cref{fig:chart}.
Instead, if we proliferate U(1)$_{q}$ defects by taking $\ell\sim\mathcal{O}( 1)$, thereby making $\Gamma\sim{\cal O}(k^0)$, the charge diffusion mode gets strongly relaxed and we instead get a weakly damped vortex diffusion mode
\begin{align}
    \omega
    &= - i\Gamma
    % - i \lb \tilde D_\perp - \frac{v_\perp^2}{\tilde\Gamma} \rb
    % k^2 
    + \ldots, \nn\\
    \omega
    &= - i\tilde\Gamma
    -i\lb \tilde D_n + \frac{v_\perp^2}{\Gamma} \rb k^2
    + \ldots.
\end{align}
This is a signature of a U(1)$_p$ (pseudo-)fluid and implements the phase transition from (d)$=\!\text{(f)}_{q\leftrightarrow p}$ to (a)$_{q\leftrightarrow p}$ or (g)$_{q\leftrightarrow p}$ to (c)$_{q\leftrightarrow p}$ depending on the presence of vorticess. 

Finally, fluctuations in the longitudinal components of the higher-form densities $n_\|$ and $\tilde n_\|$ (or the transverse components of the superfluid velocity $\xi_\perp$), we find two defect diffusion modes mediated by $n_\ell$ and $\tilde n_\ell$ respectively 
\begin{align}
    \omega = - i\Gamma - i D_\ell k^2, \qquad 
    \omega = - i\tilde\Gamma - i\tilde D_\ell k^2.
\end{align}
These modes would be absent in the absence of defects and vortices respectively. These modes will also be absent for $q=0$ and $p=0$ respectively.

\section{Pinned higher-form pseudo-superfluid dynamics}
\label{sec:higher-form-Higgs}

In this section, we discuss the hydrodynamics of a U(1)$_q$ pseudo-superfluid in the pinned phase. For $q=0$, this describes a pinned U(1)$_0$ superfluid.\footnote{Pinned conventional 0-form superfluids were considered in \cite{Delacretaz:2021qqu, Ammon:2020znv} a complete discussion in \cite{Armas:2021vku}.}

\subsection{Hydrodynamic equations}

Following our discussion in \cref{sec:Higgs-zeroT}, the conservation equations for a U(1)$_q$ pseudo-superfluid in the pinned phase are given by 
\begin{subequations}
\begin{align}
    \df{\star J}
    &= c_\phi \tilde F + (-)^{q+1}\ell\,{\star L}, 
    \label{eq:hydro-Higgs-1} \\
    \df{\star\tilde J_\psi} 
    &= 
    (-)^{p+1} \tilde c_\phi \ell\Xi
    + (-)^{p}\ell\,{\star\tilde J}, 
    \label{eq:hydro-Higgs-2} \\
    \df{\star L} 
    &= (-)^q c_\phi \tilde F_\psi, 
    \label{eq:hydro-Higgs-3} \\
    \df{\star \tilde J} 
    &= \tilde c_\phi F,
    \label{eq:hydro-Higgs-4} \\
    \nabla_\mu T^{\mu\nu} 
    &= (F\cdot J)^\nu 
    + (\tilde F\cdot\tilde J)^\nu \nn\\
    &\qquad
    + (\ell\, \Xi\cdot L)^\nu
    + (\tilde F_\psi \cdot \tilde J_\psi)^\nu.
    \label{eq:hydro-Higgs-5}
\end{align}
\label{eq:hydro-eqs-Higgs}%
\end{subequations}
In the conventional formulation of a pseudo-superfluid (gauge theory), the U(1)$^\psi_{p+1}$ conservation equation \eqref{eq:hydro-Higgs-2} and the U(1)$_p$ conservation equation \eqref{eq:hydro-Higgs-4} are identically satisfied due to the Bianchi identities associated with the misalignment tensor $\psi$ and the superfluid velocity $\xi$. The definitions of the respective currents are given in \cref{eq:tildeJ,eq:tildeJpsi}. For the remaining hydrodynamic equations, we have the set of hydrodynamic fields $\beta^\mu$, $\Lambda_\beta$, and $\Lambda_\beta^\ell$ from \cref{sec:higher-form-hydro}, with the transformation properties given in \cref{eq:hydro-trans}. These can be used to define the U(1)$_q$ chemical potential $\mu$ and the U(1)$^\ell_{q-1}$ defect chemical potential $\mu_\ell$ same as in \cref{eq:mu-defs-hydro}.

The Josephson equation for the U(1)$_q$ Goldstone $\phi$ also takes the same for as in \cref{eq:Joseph1}. In the pinned phase, there is a similar Josephson equation associated with the U(1)$^\ell_{q-1}$ Goldstone $\phi_\ell$, taking the form
\begin{equation}
    \lie_\beta\phi_\ell = c_\phi\Lambda_\beta^\ell
    + \frac1T{\cal K}_\ell
    \qquad\text{where}\quad 
    \iota_\beta{\cal K}_\ell = 0.
    \label{eq:Joseph-ell-Higgs}
\end{equation}
Similar to our discussion around \cref{eq:Joseph1}, we have used the redefinition freedom in the temporal components of $\Lambda_\beta^\ell$ to absorb derivative corrections in the temporal direction. The Josephson equation \eqref{eq:Joseph-TSSB-2} for the temporal-Goldstone field $\varphi_\ell = \iota_\beta\phi_\ell/c_\phi$ follows as the time-component of the equation above. Recast in terms of gauge-invariant variables, the two Josephson equations lead to the relations
\begin{subequations}
\begin{align}
    \iota_\beta\xi 
    &= c_\phi\frac{\mu}{T} + \frac1T {\cal K}, \nn\\
    \iota_\beta\psi 
    &= - c_\phi \frac{\mu_\ell}{T}
    - \frac{\ell}{T}{\cal K}_\ell.
\end{align}
\label{eq:joseph-Higgs}
\end{subequations}

Similar to our discussion around \cref{eq:gauge-freedom-TSSB}, we also have a U(1)$^\loc_{q-1}$ gauge freedom in this theory
\begin{alignat}{2}
    \Lambda_\beta 
    &\to \Lambda_\beta - \lie_\beta \df\lambda,  \qquad
    & \Lambda_\beta^\ell
    &\to \Lambda_\beta^\ell - \lie_\beta\lambda, \nn\\
    \phi &\to \phi - c_\phi\,\df\lambda, \qquad
    &\phi_\ell 
    &\to \phi_\ell - c_\phi\lambda.
\end{alignat}
The field $\phi$ acts as the dynamical gauge field associated with the U(1)$^\loc_{q-1}$ gauge symmetry, while the field $\phi_\ell$ can be thought of as the Goldstone arising from its spontaneous breaking.
Following usual lore of Higgs mechanism, the dynamical gauge field $\phi$ eats the Goldstone $\phi_\ell$ to acquire a mass, precisely giving rise to the gauge-invariant misalignment tensor $\psi$. 

\subsubsection{Dual formulation}

We can also view the pinned phase of a U(1)$_q$ pseudo-superfluid from the perspective of hydrodynamics with a temporally-spontaneously broken approximate anomalous $\rmU(1)_q\times\rmU(1)^\psi_{p+1}$ global symmetry. The relevant hydrodynamic variables are a set of symmetry parameters $\Lambda_\beta$, $\Lambda^\ell_\beta$, $\tilde\Lambda^\psi_\beta$, $\tilde\Lambda_\beta$, and $\beta^\mu$, and their transformation properties are given analogous to \cref{eq:hydro-trans}. We will take all the higher-form symmetries to be temporally-spontaneously broken, giving rise to the respective temporal Goldstone fields $\varphi$, $\varphi_\ell$, $\tilde\varphi$, and $\tilde\varphi_\psi$, with transformation properties similar to \eqref{eq:varphi-trans-TSSB}.
We can use these to define U(1)$_q$ and U(1)$^\ell_{q-1}$ chemical potentials same as \cref{eq:muell-def-hydro}. On the other hand, we have two more chemical potentials for U(1)$_p$ and U(1)$^\psi_{p+1}$ symmetries, given as
\begin{subequations}
\begin{align}
    \frac{\tilde\mu_\psi}{T}
    &= \tilde\Lambda_\beta^\psi + \iota_\beta \tilde A_\psi
    - \df\tilde\varphi_\psi, \label{eq:mu-psi-def-hydro} \\
    \frac{\tilde\mu}{T}
    &= 
    \tilde\Lambda_\beta
    + \iota_\beta \tilde A
    - \df\tilde\varphi
    - \ell\tilde\varphi_\psi.
\end{align}
\end{subequations}
Note that the definition of $\tilde\mu$ needs to be modified to make it invariant under the new U(1)$^\psi_{p+1}$ transformations exclusive to the pinned phase. The temporal Goldstone fields satisfy Josephson equations similar to \cref{eq:Joseph-TSSB}. These imply that all the chemical potentials are purely spatial, satisfying constraints similar to \cref{eq:mu-spatiality}. 

To relate this formalism to the conventional formulation, we can notice that the constitutive relations for the temporal parts of the currents $\iota_u\tilde J$, $\iota_u\tilde J_\psi$ to relate $\tilde\mu_\ell$, $\tilde\mu_\psi$ to the spatial components of the superfluid velocity $\xi$ and misalignment tensor $\psi$ respectively via 
\begin{align}
    \iota_u \tilde J 
    &= (-)^{q+1} {*\xi}, \nn\\
    \iota_u \tilde J_\psi 
    &= -{*\psi}.
\end{align}
On the other hand, the two Josephson equations \eqref{eq:joseph-Higgs} can be seen as determining the constitutive relations for the spatial parts of the currents; to wit
\begin{align}
    \rmP_u\tilde J 
    &= - {*\iota_u\xi} 
    = - c_\phi {*\mu}
    + {\cal\tilde J}, \nn\\
    \rmP_u{\tilde J_\psi} 
    &= (-)^{q} {*\iota_u\psi}
    = (-)^{q+1} c_\phi{*\mu_\ell}
    + {\cal\tilde J}_\psi,
    \label{eq:current-identifications-higgs}
\end{align}
where we have defined 
\begin{align}
    {\cal\tilde J} = - {*\cal K}, \qquad 
    {\cal\tilde J}_\psi = (-)^{q+1} \ell {*{\cal K}_\ell}.
    \label{eq:K-J-reln-Higgs}
\end{align}
This ensures that both the formulations describe the same pinned phase of a U(1)$_q$ pseudo-superfluid.

\subsection{Second law of thermodynamics and constitutive relations}

Using the free energy current defined in \cref{eq:N-def}, we find that the adiabaticity equation \eqref{eq:adiabaticity} for a superfluid modifies to
\begin{align}
    \nabla_\mu N^\mu 
    &= \half T^{\mu\nu} \delta_\scB g_{\mu\nu}
    + J\cdot\delta_\scB A 
    + \ell L\cdot\delta_\scB \Phi
  \nn\\
    &\qquad 
    - K_\ext \cdot \delta_\scB \phi
    - \ell Q_\ext \cdot \delta_\scB \phi_\ell
    + \Delta.
\end{align}
where we have defined $\tilde F_\psi = -{\star Q_\ext}$. The variations of the phase fields are defined similar to \cref{eq:phi-variation}.

\subsubsection{Conventional formulation}

Thermodynamics in the pinned phase can be defined similar to \cref{eq:thermo-Coulomb}. However, the thermodynamic pressure $P$ does not depend on the $(p-1)$-form vortex chemical potential $\tilde\mu_\ell$ anymore and instead depends on the misalignment tensor $\tilde n_\psi = {*\psi}$. We find
\begin{align}
    \delta P 
    &= s\delta T 
    + n\cdot\delta\mu
    + n_\ell\cdot\delta\mu_\ell 
    - {*\tilde\mu} \cdot \delta{*\tilde n} 
    - {*\tilde\mu_\psi} \cdot \delta {*\tilde n_\psi}
    \nn\\
    &\qquad 
    - \half r^{\mu\nu}\df g_{\mu\nu}, \nn\\
    \epsilon
    &= -P + Ts 
    + \mu\cdot n
    + \mu_\ell\cdot n_\ell.
\end{align}
This equation can be seen as defining the conjugate chemical potentials $\tilde\mu$ and $\tilde\mu_\ell$. We can find the associated anisotropic stress tensor $r^{\mu\nu}$ using Lorentz invariance
\begin{align}
    r^{\mu\nu}
    &= \frac{1}{(q-1)!} n^\mu{}_{\rho_1\ldots}
    \mu^{\nu\rho_1\ldots} 
    + \frac{1}{(q-2)!} 
    (n_\ell)^\mu{}_{\rho_1\ldots}
    \mu_\ell^{\nu\rho_1\ldots} \nn\\
    &\quad 
    - \frac{1}{q!} {*\tilde n}^\mu{}_{\rho_1\ldots}
    {*\tilde\mu}^{\nu\rho_1\ldots} 
    - \frac{m^2}{(q-1)!} 
    {(*\tilde n_\psi)}^\mu{}_{\rho_1\ldots} 
    {*\tilde\mu}_\psi^{\nu\rho_1\ldots}, \nn\\
    r^{[\mu\nu]} &= 0.
\end{align}
If we take a simplified equation of state where $P$ only depends on the squares of the higher-form objects, the thermodynamic relations become
\begin{subequations}
\begin{equation}
    \delta P = s\df T 
    + \half\chi  \delta\mu^2
    + \half\chi_\ell \delta\mu_\ell^2
    - \frac{1}{2\tilde\chi} \delta \tilde n^2
    - \frac{m^2}{2} \delta \tilde n_\psi^2.
\end{equation}
In this case, various densities are related to the chemical potentials as
\begin{equation}
    n = \chi\mu, \quad 
    n_\ell = \chi_\ell\mu_\ell, \quad
    \tilde n = \tilde\chi\tilde\mu, \quad 
    \tilde n_\psi = \frac{1}{m^2}\tilde\mu_\psi.
\end{equation}
As before, the anisotropic stress tensor $r^{\mu\nu}$ is automatically symmetric.
\end{subequations}

The free energy current can be decomposed in the canonical parametrisation similar to \cref{eq:N-current-SF} as 
\begin{equation}
    N^\mu 
    = P\,\beta^\mu 
    + \frac1T ({*{\cal K}} \cdot \tilde\mu)^\mu
    + \frac{(-)^{q}}{T}\ell (*{\cal K}_\ell \cdot {\tilde\mu_\psi})^\mu
     + {\cal N}^\mu,
     \label{eq:N-current-higgs}
\end{equation}
where ${\cal N}^\mu$ denotes non-canonical contributions.
Using the variational formulae
\begin{align}
    {*\tilde\mu_\psi}\cdot\delta_\scB {*\tilde n_\psi}
    &= 
    - \ell (-)^{s}T{*\df\frac{\tilde\mu_\psi}{T}}
    \cdot\delta_\scB \phi_\ell \nn\\
    &\qquad 
    + \ell (-)^{pq+q}{*\tilde\mu_\psi}\cdot
    \lb \delta_\scB \phi - c_\phi\delta_\scB \Phi \rb \nn\\
    &\qquad 
    - c_\phi (-)^{p+q}{*(\mu_\ell\wedge\tilde\mu_\psi)}^{(\mu} u^{\nu)} \delta_\scB g_{\mu\nu} \nn\\
    &\qquad 
    - \ell (-)^{s}{\star \df(u \wedge \tilde\mu_\psi\wedge\delta_\scB \phi_\ell)},
\end{align}
we can derive the constitutive relations for an ideal U(1)$_q$ pseudo-superfluid in the pinned phase
\begin{align}
    T^{\mu\nu}
    &= (\epsilon+P) u^\mu u^\nu + P\,g^{\mu\nu}
    - r^{\mu\nu} \nn\\
    &\qquad 
    - 2c_\phi (-)^{p} 
    {*\big(\mu\wedge\tilde\mu
    - (-)^{q}\mu_\ell\wedge\tilde\mu_\psi \big)}
    {}^{(\mu} u^{\nu)}
    + {\cal T}^{\mu\nu}
    , \nn\\
    J
    &= u\wedge n - \tilde c_\phi\,{*\tilde\mu} + {\cal J}, \nn\\
    L
    &= u \wedge n_\ell + (-)^{p} \tilde c_\phi {*\tilde\mu_\psi}  
    + {\cal L},
    \label{eq:consti-ansatz-SF-Higgs}
\end{align}
These satisfy the adiabaticity equation with $\Delta=0$ when all the derivative corrections have been switched off. Note the additional terms compared to constitutive relations in the relaxed phase given in \cref{eq:consti-ansatz-SF}. These drop out if we take $\tilde\mu_\psi = m^2\tilde n_\psi$ to zero, i.e. turn off the mass $m$ of the pseudo-Goldstone fields.

Plugging the constitutive relations \eqref{eq:consti-ansatz-SF-Higgs} into the adiabaticity equation together with the Josephson equations \cref{eq:joseph-Higgs}, we can read of the constraint equation to be satisfied by the dissipative corrections 
\begin{align}
    &- \half {\cal T}^{\mu\nu} \delta_\scB g_{\mu\nu}
    - {\cal J}\cdot\delta_\scB A 
    - \ell {\cal L}\cdot\delta_\scB \Phi \nn\\
    &\qquad
    - {\cal\tilde J} \cdot \delta_\scB\tilde A 
    - {\cal\tilde J_\psi} \cdot \delta_\scB\tilde A_\psi
    + \nabla_\mu {\cal N^\mu}
    = \Delta  \geq 0,
    \label{eq:constraint-adiabaticity-sf-Higgs}
\end{align}
where we have defined 
\begin{align}
    \delta_\scB \tilde A 
    &= \df\frac{\tilde\mu}{T} 
    + \iota_\beta\tilde F - \ell\frac{\tilde\mu_\psi}{T}, \nn\\
    \delta_\scB \tilde A_\psi
    &= \df\frac{\tilde\mu_\psi}{T}
    + \iota_\beta\tilde F_\psi.
    \label{eq:modified-deltaA}
\end{align}
The Landau frame condition for various dissipative corrections is given as
\begin{equation}
    {\cal T}^{\mu\nu}u_\nu 
    = \iota_u {\cal J} = \iota_u {\cal L} 
    = \iota_u{\cal\tilde J}
    = \iota_u{\cal\tilde J_\psi}
    = 0.
\end{equation}

\subsubsection{Dual formulation}

We can also formulate U(1)$_q$ pseudo-superfluid in the pinned phase using the language of anomalous approximate $\rmU(1)_q\times\rmU(1)^\psi_{p+1}$ symmetry. The relevant free energy current can be defined similar to \cref{eq:freeE-dual-Coulomb} as
\begin{align}
    TN'^\mu
    &= TN^\mu   
    + (\tilde J \cdot \tilde\mu)^\mu
    + (\tilde J_\psi \cdot \tilde\mu_\psi )^\mu  \nn\\
    &= TS^\mu 
    + T^{\mu\nu} u_\nu 
    + (J\cdot\mu)^\mu 
    + (\tilde J \cdot \tilde\mu)^\mu \nn\\
    &\qquad 
    + (L\cdot\mu_\ell)^\mu
   + (\tilde J_\psi \cdot \tilde\mu_\psi )^\mu.
\end{align}
One can check that the respective adiabaticity equation takes a similar form to \cref{eq:adiabaticity} but generalised to two higher-form symmetries, i.e.
\begin{align}
    \nabla_\mu N'^\mu 
    &= \rmN_\rmH 
    + \half T^{\mu\nu} \delta_\scB g_{\mu\nu}
    + J\cdot\delta_\scB A 
    + \tilde J\cdot\delta_\scB \tilde A  \nn\\
    &\qquad 
    + \ell L\cdot\delta_\scB \Phi
    + \tilde J_\psi\cdot\delta_\scB \tilde A_\psi
    + \Delta.
    \label{eq:adiabaticity_sf-Higgs}
\end{align}
Here $\delta_\scB\tilde A$ and $\delta_\scB\tilde A_\psi$ are defined similarly to \cref{eq:deltaB-defs}.
We also get a new anomaly-induced ``Hall free energy'' contribution in the adiabaticity equation given by
\begin{align}
    \rmN_\rmH 
    &= - \frac{c_\phi}{T}
    \bigg( (-)^{pq+q}
    \mu\cdot{\star\tilde F}
    + (-)^q\tilde\mu\cdot{\star F} \nn\\
    &\qquad
    + (-)^{pq+p+q} \ell \mu_\ell \cdot {\star\tilde F_\psi}
    - \ell\mu_\psi\cdot{\star\Xi}
    \bigg).
\end{align}

Similarly, the canonical decomposition of the free energy current in \cref{eq:N-current-higgs} becomes
\begin{equation}
    N'^\mu 
    = P'\beta^\mu 
    - (-)^{p}\frac{c_\phi}{T}
    *\!\lb\mu\wedge\tilde\mu 
    - (-)^q \mu_\ell \wedge\tilde\mu_\psi
    \rb^\mu 
     + {\cal N}^\mu,
     \label{eq:Nmu-SF-Higgs}
\end{equation}
where the conjugate pressure is given as
\begin{align}
    P' 
    &= P + \tilde n \cdot \tilde\mu
    + \tilde n_\psi \cdot \tilde\mu_\psi.
\end{align}
The differential of the new pressure yields the thermodynamic relations
\begin{align}
    \delta P'
    &= s\delta T 
    + n\cdot\delta\mu
    + \tilde n\cdot\delta\tilde\mu
    + n_\ell\cdot\delta\mu_\ell 
    + \tilde n_\psi\cdot\delta\tilde\mu_\psi \nn\\
    &\qquad 
    - \half r'^{\mu\nu}\delta g_{\mu\nu}, \nn\\
    \epsilon
    &= -P' + Ts 
    + \mu\cdot n
    + \tilde\mu\cdot \tilde n \nn\\
    &\qquad 
    + \mu_\ell\cdot n_\ell
    + \tilde \mu_\psi\cdot \tilde n_\psi,
\end{align}
together with
\begin{align}
    r'^{\mu\nu}
    &= r^{\mu\nu}
    + P^{\mu\nu} \lb \tilde n \cdot \tilde\mu
    + \tilde n_\psi \cdot \tilde\mu_\psi \rb \nn\\
    &= \frac{1}{(q-1)!} n^\mu{}_{\rho_1\ldots}
    \mu^{\nu\rho_1\ldots} 
    + \frac{1}{(q-2)!} 
    (n_\ell)^\mu{}_{\rho_1\ldots}
    \mu_\ell^{\nu\rho_1\ldots} \nn\\
    &\quad 
    + \frac{1}{(p-1)!} \tilde n^\mu{}_{\rho_1\ldots}
    \tilde\mu^{\nu\rho_1\ldots} 
    + \frac{1}{p!} 
    (\tilde n_\psi)^\mu{}_{\rho_1\ldots}
    \tilde\mu_\psi^{\nu\rho_1\ldots}, \nn\\
    r'^{[\mu\nu]} &= 0.
\end{align}
Using these, together with the identifications in \cref{eq:current-identifications-higgs},  we can read off the constitutive relations in the dual formulation 
\begin{align}
    T^{\mu\nu}
     &= (\epsilon+P')u^{\mu}u^\nu + P'\, g^{\mu\nu}
     - r'^{\mu\nu} \nn\\
     &\qquad 
     - 2c_\phi (-)^{p} 
    {*\big(\mu\wedge\tilde\mu
    - (-)^{q}\mu_\ell\wedge\tilde\mu_\psi \big)}
    {}^{(\mu} u^{\nu)}
    + {\cal T}^{\mu\nu}, \nn\\
     J
     &= u \wedge n
     - \tilde c_\phi\,{*\tilde\mu}
     + {\cal J}, \nn\\
     \tilde J
     &= u \wedge \tilde n -
     c_\phi\,{*\mu}
     + {\cal\tilde J}, \nn\\
     L
     &= u \wedge n_\ell 
     + (-)^{p} \tilde c_\phi {*\tilde\mu_\psi}  
     + {\cal L}
     , \nn\\
     \tilde J_\psi
     &= u \wedge \tilde n_\psi
     - (-)^{q} c_\phi{*\mu_\ell}
     + \tilde{\cal J}_\psi.
     \label{eq:consti-Higgs-dual}
\end{align}
Note that the definitions of the dissipative corrections $\cal\tilde J$ and $\cal\tilde J_\psi$ coincide with \cref{eq:K-J-reln-Higgs}. Ignoring all derivative corrections, we can see that we have arrived at exactly the same form of the constitutive relations as derived using an equilibrium action in \cref{sec:appx-SF-Coulomb-eqb}.
The constraint equation for dissipative corrections in the higher-form formulation takes the same form given in \cref{eq:constraint-adiabaticity-sf-Higgs}.

\subsubsection{Dissipative corrections}

Similar to our discussion in \cref{sec:dissipation-Coulomb}, we will assume the constitutive relations to be parity and charge-conjugation invariant and ignore all anisotropic contributions. The constitutive relations for ${\cal T}^{\mu\nu}$ and $\cal J$ take the same form as in the temporally-spontaneously broken phase in \cref{eq:diss-consti-normal}. The constitutive relations for
$\cal\tilde J$ and $\cal L$ are the same in form as the relaxed phase in \cref{eq:tildeJ-L-diss}, but there are new $\tilde\mu_\psi$ contributions coming from the modified definition of $\delta_\scB\tilde A$ in \cref{eq:modified-deltaA}, i.e.
\begin{align}
    {\cal\tilde J}
    &= - T\tilde\sigma \,\rmP_u \delta_\scB \tilde A
    - (-)^{pq+p+q}
    T\ell\gamma_\times{*\delta_\scB\Phi} \nn\\
    &= -\tilde\sigma
    \lb T\,\rmP_u\df\frac{\tilde\mu}{T} 
    + \iota_u \tilde F - \ell\tilde\mu_\psi
    \rb \nn\\
    &\qquad 
    - (-)^{pq+p+q} \gamma_\times
    {*\!\lb 
    T\df\frac{\mu_\ell}{T} + \ell\lb \iota_u\Xi - \mu \rb 
    \rb } , \nn\\[0.5em]
    {\cal L}
    &= - T\ell\sigma_\ell\,\rmP_u\delta_\scB\Phi
    + (-)^p T\gamma'_\times
    {*\delta_\scB \tilde A} \nn\\
    &= 
    - \sigma_\ell
    \lb T\,\rmP_u\df\frac{\mu_\ell}{T} 
    + \ell\lb \iota_u\Xi - \mu \rb \rb \nn\\
    &\qquad 
    + (-)^p \gamma'_\times\, 
    {*\!\lb T\df\frac{\tilde\mu}{T} 
    + \iota_u \tilde F - \ell\tilde\mu_\psi \rb}.
\end{align}
Since $\tilde\mu_\psi = m^2\tilde n_\psi = m^2{*\psi}$ contains a term linear in $\phi$, the $\tilde\sigma$ contribution to $\cal\tilde J$ is now responsible is now responsible for relaxation of $\phi$ with relaxation rate $\tilde\Omega = \ell^2m^2\tilde\sigma$. This plays a role very similar to the vorticity-induced relaxation rate $\tilde\Gamma$ observed in the relaxed phase, but is qualitatively distinct. The $\tilde\mu_\psi$ contribution coupled to $\gamma'_\times$ in ${\cal L}$ renormalises the pinning term in \cref{eq:consti-Higgs-dual}. We need to supply new dissipative constitutive relations for $\cal\tilde J_\psi$, given by
\begin{align}
    {\cal\tilde J_\psi}
    &= -T\tilde\sigma_\psi\,\rmP_u\delta_\scB \tilde A_\psi \nn\\
    &= - \tilde\sigma_\psi
    \lb T\,\rmP_u\df\frac{\tilde\mu_\psi}{T} 
    + \iota_u \tilde F_\psi
    \rb.
\end{align}
Note that this object does not exist for $q=0$. In the context of 
$U(1)^\loc_{q-1}$ gauge theories, the coefficient $\tilde\sigma_\psi$ modifies the relation between the scalar gauge potential $\iota_\beta\phi$ and the electric chemical potential $\mu_\ell$, and will result in damping of the vector gauge potential $\rmP_u\phi$. 

One can check that that the dissipative corrections map to each other under the $q\leftrightarrow p+1$ self-duality transformation of the pinned phase, provided that we transform the dissipative transport coefficients as follows
\begin{gather}
    \sigma \leftrightarrow \tilde\sigma_\psi, \qquad 
    \sigma_\ell \leftrightarrow \tilde\sigma, \qquad
    \gamma_\times \leftrightarrow (-)^{pq+p}\gamma_\times.
\end{gather}
In particular, note that the conductivities $\sigma$ and $\tilde\sigma$ do not map to each other unlike the duality transformations in the relaxed phase.

\subsection{Linearised analysis and mode spectrum}

We will again focus on fluctuations around an equilibrium state with zero densities, so the energy-momentum fluctuations decouple and lead to the same longitudinal sound and shear diffusion modes in \cref{eq:hydro-sound}. The constitutive relations for the higher-form currents in the absence of background fields take the form
\begin{align}
    J
    &= u \wedge n - \tilde c_\phi {*\tilde\mu} 
    - \sigma\rmP_u\df\mu, \nn\\[0.5em]
    \tilde J
    &= u \wedge \tilde n
    - \lambda c_\phi {*\mu} 
    + \ell\tilde\sigma\tilde\mu_\psi
    - \tilde\sigma\rmP_u\df\tilde\mu \nn\\
    &\qquad 
    - (-)^{pq+p+q}\gamma_\times
    {*\df\mu_\ell}, \nn\\[0.5em]
    L
    &= u \wedge n_\ell 
    + (-)^p \lambda \tilde c_\phi {*\tilde\mu_\psi}
    + \ell\sigma_\ell \mu
    - \sigma_\ell \rmP_u\df\mu_\ell \nn\\
    &\qquad
    + (-)^p \gamma_\times {*\df\tilde\mu}, \nn\\[0.5em]
    \tilde J_\psi 
    &= u \wedge \tilde n_\psi 
    - (-)^{q} c_\phi{*\mu_\ell}
    - \tilde\sigma_\psi\rmP_u\df\tilde\mu_\psi.
\end{align}
The constitutive relations for $\tilde J$ and $\tilde J_\psi$ give rise to the Josephson equations 
\begin{align}
    \iota_u\xi 
    &= 
    \lambda c_\phi \mu
    - \ell^2 m^2\tilde\sigma\phi
    + (-)^{pq}\frac{\tilde\sigma}{\tilde\chi} {*\df{*\xi}} \nn\\
    &\qquad 
    + (-)^{pq+p+q}\gamma_\times {\df\mu_\ell}, \nn\\
    \ell\iota_u\phi 
    &=  - c_\phi\mu_\ell
    - (-)^{pq+p+q} \ell m^2\tilde\sigma_\psi {*\df{*\phi}}.
\end{align}
We have specialised to a gauge where $\phi_\ell = 0$, meaning that $\psi = \ell\phi$ in the absence of background fields. Compared to the relaxed phase expressions in \cref{eq:joseph-linear-coulomb}, $\iota_u\xi$ has a new pinning term proportional to $\phi$ controlled by $\ell^2m^2$. The second Josephson equation, which is specific to the pinned phase and is only non-trivial for $q>1$, provided the relation between the scalar potential $\iota_u\phi$ and the defect chemical potential $\mu_\ell$. 

Let us look at the linearised equations of motion. In the higher-form language, they are given as
\begin{align}
    (\dow_t + \Gamma) n
    &= (-)^{p}\frac{\lambda}{\tilde\chi} \tilde c_\phi {*\df\tilde n}
    - (-)^p \ell \lambda m^2 \tilde c_\phi {*\tilde n_\psi} \nn\\
    &\qquad 
    + D_n \dow^2 n
    - (-)^{pq+p+q} (D_\ell - D_n) {\df{*\df{*n}}}
    , \nn\\
    (\dow_t + \tilde\Omega) \tilde n_\psi
    &= -\frac{\lambda}{\chi_\ell} c_\phi {*\df n_\ell}
    + \ell \frac{\lambda}{\chi} c_\phi {* n} \nn\\
    &\qquad 
    + \tilde D_\psi \dow^2 \tilde n_\psi
    + (-)^{pq} (\tilde D_n - \tilde D_\psi) 
    {\df{*\df{*\tilde n_\psi}}}, \nn\\
    *\df{*n} 
    &= (-)^{pq+p+q+1} \ell n_\ell, \nn\\
    *\df{*\tilde n_\psi} 
    &= (-)^{pq} \ell\tilde n~,
\end{align}
where we have defined the diffusion and relaxation coefficients
\begin{gather}
    \tilde D_\psi = m^2\tilde\sigma_\psi, \qquad 
    \tilde\Omega = \ell^2 m^2 \tilde\sigma.
\end{gather}
In terms of the conventional formulation variables, we instead find 
\begin{align}
    (\dow_t + \Gamma) n
    &= (-)^{pq}\frac{\lambda}{\tilde\chi} c_\phi {*\df{*\xi}}
    + D_n \dow^2 n
    + \ell (D_\ell - D_n) {\df n_\ell} \nn\\
    &\qquad 
    - \ell^2 \lambda m^2 c_\phi {\phi}
    , \nn\\
    (\dow_t + \tilde\Omega) \phi
    &= -\frac{\lambda}{\ell\chi_\ell} c_\phi {\df n_\ell}
    + \tilde D_\psi \dow^2 \phi \nn\\
    &\qquad 
    + (-)^{pq} (\tilde D_n - \tilde D_\psi) {*\df{*\df\phi}} 
    + \frac{\lambda}{\chi} c_\phi n, \nn\\
    \rmP_u\df{*n} 
    &= (-)^{q+1} \ell{*n_\ell}, \nn\\
    \rmP_u\df\phi 
    &= \rmP_u\xi.
\end{align}
In particular, note that phase field $\phi$ is now directly relaxed with the characteristic rate $\tilde\Omega$, unlike the relaxed pseudo-superfluid phase in the presence of vortices, where the superfluid velocity is relaxed instead with rate $\tilde\Gamma$.

To obtain the respective mode spectrum, let us first look at the transverse U(1)$_q$ density sector, spanned by $n_\perp$ and $n^\|_\psi$, we find the dispersion relations
\begin{equation}
    \big(i\omega - D_n k^2 - \Gamma\big)
    \big(i\omega - \tilde D_n k^2 - \tilde\Omega\big)
    + \omega_0^2 + v_\perp^2k^2 = 0,
\end{equation}
where we have identified the pinning frequency
\begin{equation}
    \omega_0^2 
    = \frac{\ell^2m^2\lambda^2c_\phi^2}{\chi}.
\end{equation}
Taking $\ell\sim k$, this leads to a pinned and damped sound mode
\begin{align}
    \omega &= \pm \sqrt{\omega_0^2 + v_\perp^2k^2} \nn\\
    &\qquad
    - \frac{i}{2}
    \lb \Gamma + \tilde\Omega + (D_n+\tilde D_n) k^2 \rb
    + \ldots.
\end{align}
In the longitudinal U(1)$_q$ density channel, spanned by $n_\|$ and $\tilde n_\psi^\perp$, we find another set of dispersion relations
\begin{equation}
    \big(i\omega - D_\ell k^2 - \Gamma\big)
    \big(i\omega - \tilde D_\psi k^2 - \tilde\Omega\big)
    + \omega_0^2 + v_\|^2k^2 = 0,
\end{equation}
where we have defined 
\begin{equation}
    v_\|^2 = \frac{\lambda^2c_\phi^2m^2}{\chi_\ell}.
\end{equation}
This leads to another damped and pinned sound mode
\begin{align}
    \omega &= \pm \sqrt{\omega_0^2 + v_\|^2k^2} \nn\\
    &\qquad
    - \frac{i}{2}
    \lb \Gamma + \tilde\Omega + (D_\ell+\tilde D_\psi) k^2 \rb
    + \ldots.
\end{align}
The two sets of modes map to each other under $q\leftrightarrow p+1$ self-duality, which maps
\begin{gather}
    \chi \leftrightarrow \frac{1}{m^2}, \qquad 
    \chi_\ell \leftrightarrow \tilde\chi, \qquad 
    v_\perp \leftrightarrow v_\|, \qquad 
    k_0 \leftrightarrow k_0^\phi, \nn\\
    D_n \leftrightarrow D_\psi, \qquad
    D_\ell \leftrightarrow \tilde D_n, \qquad
    \Gamma \leftrightarrow \tilde\Omega.
\end{gather}
and leaves $\lambda$ and $\omega_0$ invariant.

If we increase the strength of explicit symmetry breaking by increasing the magnitude of $\ell$, both the relaxation rates $\Gamma$ and $\tilde\Omega$ will become finite, thereby gapping our both the sets of modes above and implementing the phase transition from the pinned superfluid phase to the neutral fluid phase. This is the transition from (e) to (h) in \cref{fig:chart}.

\section{Outlook}
\label{sec:outlook}
In this paper we have introduced a new framework to classify phases of matter according to their approximate higher-form symmetries. In particular, we focused on phases with a U(1)$_q$ generalised global symmetry and various patterns of symmetry breaking therein at zero and finite temperature. These patterns include temporal-(pseudo)-spontaneous symmetry breaking in higher-form fluids and complete-(pseudo)-spontaneous symmetry breaking in higher-form superfluids. We showed how these phases naturally incorporate dynamical $q$-form defects and $p$-form vortices, where \tightmath{$p=d-1-q$}. In all such phases of matter, we studied their out-of-equilibrium dynamics by formulating appropriate hydrodynamic theories that respect the symmetry breaking patterns. As we commented throughout the paper, this framework is applicable to many phases of matter including smectic crystals with topological defects, superfluids with vortices, polarised plasmas with free electromagnetic charges, magnetohydrodynamics with magnetic monopoles, spin ices, superconductors, among many others.

One of our main interests in this work was to understand potential phase transitions that can occur by the proliferation of $q$-form defects or $p$-form vortices. As we mentioned earlier, these transitions include the melting transition and the plasma phase transition. In order to address the physics of these transitions in a hydrodynamic framework, in sections \ref{sec:higher-form-hydro}-\ref{sec:higher-form-Higgs} we studied the linearised spectrum of various phases. We noted that in what we denoted by relaxed phases (or Coulomb phases) the higher-form charge is relaxed but the Goldstone is not, while in pinned phases (or Higgs phases) both the higher-form charge and the Golstone field is relaxed. By carefully studying the linearised spectrum of excitations in various regimes we could clearly identify features of phase transitions by proliferating defects or vortices. In particular, upon proliferation, certain modes become gapped and the spectrum of $q$-form (pseudo)-fluids or uncharged fluids is recovered. 

It should be noted, however, that our linearised mode analysis was not the most general one. We focused on isotropic phases by linearly perturbing around a state with vanishing chemical potential. While this is most relevant equilibrium state in certain contexts such as in 0-form superfluids, it is not the most physically relevant state in other contexts as in the case of smectic crystals or in magnetohydrodynamics with constant magnetic fields. The various correlation functions derived in appendix \ref{app:correlators} reflect the same choice of equilibrium states. As the main purpose of this work was to provide a new framework for describing phases of matter with higher-form symmetries, we have not performed an exhaustive analysis of the phenomenology of the mode spectra and correlators in the context of specific systems. In an upcoming series of papers, we will be using the formalism introduced here to study the dynamics of specific phases of matter.  

Our work has dealt with relativistic field theories, which is the natural language to describe electromagnetism and magnetohydrodynamics but also to describe the effects of elasticity in cosmological matter \cite{Zaanen:2021zqs}. However, many of the relevant applications of approximate higher-form symmetries are to non-relativistic systems, including spin ices and superconductors. Following the approach of \cite{Armas:2021vku, Armas:2022vpf} it is straightforward, though cumbersome, to couple these field theories to non-relativistic backgrounds. We plan on discussing this elsewhere. At the same time, having formulated this framework in the context of relativistic theories makes this work ideal to study phases of matter using holography. A few holographic studies of higher-form symmetries have been performed (see e.g. \cite{Hofman:2017vwr, Grozdanov:2017kyl, Grozdanov:2018ewh, Grozdanov:2018fic, Armas:2019sbe, Davison:2022vqh, Ahn:2022azl}) and it appears to be possible to extend them to approximate U(1)$_q$ symmetries more generally. 

Finally, we would like to note that we focused on continuous U(1) higher-form symmetries. Given the broad scope of applications of this single case we studied here, it would be extremely interesting to study other (discrete) higher-form symmetries and more exotic symmetries such as higher-group symmetries, subsystem symmetries and non-invertible symmetries in the context of hydrodynamics and phase transitions.

\begin{acknowledgments}
We would like to thank Diego Hofman, Felix Flicker and Jasper van Wezel for various helpful discussions. JA and AJ are partly supported by the Netherlands Organization for Scientific Research (NWO) and by the Dutch Institute for Emergent Phenomena (DIEP) cluster at the University of Amsterdam. AJ is funded by the European Union’s Horizon 2020 research and innovation programme under the Marie Sk{\l}odowska-Curie grant agreement NonEqbSK No. 101027527.
\end{acknowledgments}

\newpage

\appendix 

\section{Differential form conventions}
\label{app:form-conventions}
In this appendix we give details about our differential form conventions.
A $q$-rank differential form or $q$-form $\mu$ can be expressed in components as
\begin{equation}
    \mu
    = \frac{1}{q!} \mu_{\mu_1\ldots\mu_q} \df x^{\mu_1} \wedge\ldots\wedge \df x^{\mu_q}.
\end{equation}
The standard operations: Hodge dual, exterior derivative, interior product, and interior derivative
\begin{align}
    ({\star\mu})_{\mu_1\ldots\mu_{d+1-q}}
    &= \frac{1}{q!} \epsilon_{\nu_1\ldots\nu_q\mu_1\ldots\mu_{d+1-q}} \mu^{\mu_1\ldots\mu_q}, \nn\\
    (\df\mu)_{\mu_1\ldots\mu_{q+1}}
    &= (q+1) \dow_{[\mu_1}\mu_{\mu_2\ldots\mu_{q+1}]}, \nn\\
    (\iota_X\mu)_{\mu_1\ldots\mu_{q-1}}
    &= X^\mu\mu_{\mu\mu_1\ldots\mu_{q-1}}, \nn\\
    (\iota_\dow\mu)_{\mu_1\ldots\mu_{q-1}}
    &= \nabla^\mu\mu_{\mu\mu_1\ldots\mu_{q-1}}.
\end{align}
% \begin{align}
%     {\star\mu}_{\mu_1\ldots\mu_{d+1-q}}
%     &= \frac{1}{(d+1-q)!}
%     \lb \frac{1}{q!} \epsilon_{\nu_1\ldots\nu_q\mu_1\ldots\mu_{d+1-q}} \mu^{\mu_1\ldots\mu_q} \rb \nn\\
%     &\qquad\qquad\qquad
%     \df x^{\mu_1} \wedge\ldots\wedge \df x^{\mu_{d+1-q}}, \nn\\
%     %
%     \df\mu
%     &= \frac{1}{(q+1)!}
%     \lb (q+1) \dow_{[\mu_1}\mu_{\mu_2\ldots\mu_{q+1}]} \rb \nn\\
%     &\qquad\qquad\qquad
%     \df x^{\mu_1} \wedge\ldots\wedge \df x^{\mu_{q+1}}, \nn\\
%     \iota_X\mu
%     &= \frac{1}{(q-1)!} \lb 
%     X^\mu\mu_{\mu\mu_1\ldots\mu_{q-1}} \rb \nn\\
%     &\qquad\qquad\qquad 
%     \df x^{\mu_1} \wedge\ldots\wedge \df x^{\mu_{q-1}}, \nn\\
%     \iota_\dow\mu
%     &= \frac{1}{(q-1)!} \lb 
%     \nabla^\mu\mu_{\mu\mu_1\ldots\mu_{q-1}} \rb \nn\\
%     &\qquad\qquad\qquad 
%     \df x^{\mu_1} \wedge\ldots\wedge \df x^{\mu_{q-1}}
% \end{align}
Here $\epsilon_{\mu_1\ldots\mu_{d+1}}$ is the $(d+1)$-rank totally antisymmetric Levi-Civita tensor, with $\epsilon_{012\ldots} = \sqrt{\pm g}$. The upper sign corresponds to Euclidean spacetime while the lower one to Lorentzian spacetime. Also, $X^\mu$ is a vector field.
Given a $q$-form $\mu$ and another $r$-rank form $\nu$, we can define their exterior-product or wedge-product as
\begin{align}
    (\mu\wedge\nu)_{\mu_1\ldots\mu_{q+r}}
    &= \frac{(q+r)!}{q!r!} \mu_{[\mu_1\ldots\mu_{q}}\nu_{\mu_{q+1}\ldots\mu_{q+r}]}.
\end{align}
We also introduce another notation for contraction of differential forms. We define
\begin{alignat}{2}
    (\mu\cdot\nu)_{\mu_1\ldots\mu_{r-q}}
    &= \frac{1}{q!} \mu^{\nu_1\ldots\nu_q}
    \nu_{\nu_1\ldots\nu_q\mu_1\ldots\mu_{r-q}},
    &&\quad q\leq r, \nn\\
    (\mu\cdot\nu)_{\mu_1\ldots\mu_{q-r}}
    &= \frac{1}{r!} 
    \mu_{\mu_1\ldots\mu_{q-r}\nu_1\ldots\nu_r}
    \nu^{\nu_1\ldots\nu_r},
    &&\quad q\geq r.
\end{alignat}
Let us note a few useful identities
\begin{align}
    {\star{\star\mu}}
    &= \pm(-1)^{q(d+1-q)}\mu, \nn\\
    \iota_\dow{\star\mu}
    &= (-1)^q\, {\star \df\mu}, \nn\\
    \iota_\dow \df\mu
    &= \dow^2\mu - \df \iota_\dow\mu, \nn\\
    \df{\star\mu}
    &= (-1)^{q+1} {\star{\iota_\dow\mu}}, \nn\\
    \iota_X {\star\mu}
    &= (-1)^q\, {\star(X\wedge\mu)}, \nn\\
    {\star(\mu\wedge\nu)}
    &= \nu\cdot{\star\mu}, \nn\\
    {\star(\nu\wedge\star\mu)}
    &= \pm(-1)^{(q-r)(d+1-q)}\nu\cdot\mu.
\end{align}

A special differential form is the spacetime volume-form $\text{vol}_{d+1} = \star1$. It can be used to integrate a function $f(x)$ over spacetime 
\begin{align}
    \int {\star f}(x)
    = \int f(x) \text{vol}_{d+1} 
    = \int \df^{d+1}x\,\sqrt{\pm g}\,f(x).
\end{align}
Given a $(d-q)$-dimensional hypersurface $\Sigma_{d-q}$ in spacetime, with orthonormalised normalised normal vectors $n_1^\mu,\ldots,n^\mu_{q+1}$, we can define a volume-form on the hypersurface 
\begin{equation}
    \text{vol}_{d-q} = {\star(n_1\wedge\ldots\wedge n_{q+1})},
\end{equation}
which is understood to be zero everywhere in spacetime other than the hypersurface.
This can be used to integrate a $q$-form $J$ can be integrated as
\begin{align}
    \int_{\Sigma_{d-q}} \star J
    &= \int J\wedge\text{vol}_{d-q} \nn\\ 
    &= \int \df\Sigma_{\mu_1\ldots\mu_{q+1}} 
    J^{\mu_1\ldots\mu_{q+1}},
\end{align}
where 
$\df\Sigma_{\mu_1\ldots\mu_{q+1}} 
= \df^{q+1}x\sqrt{\pm g}\, n^1_{[\mu_1}\ldots n^{q+1}_{\mu_{q+1}]}$ is the integration measure on the hypersurface. A corollary of the setup above is the Green's theorem: given a hypersurface $\Sigma_{d-q}$ and a $(d-q-1)$-form $Y$, we have that
\begin{equation}
    \int_{\Sigma_{d-q}} \df Y
    = \int_{\dow\Sigma_{d-q}} Y,
\end{equation}
where $\dow\Sigma_{d-q}$ is the $(d-q-1)$-dimensional boundary of $\Sigma_{d-q}$. The volume-form on the boundary is defined with respect to the outward-pointing normal vector.

\section{Anomaly inflow for higher-form symmetries}
\label{app:anomaly-inflow}

In this appendix, we discuss the anomaly inflow mechanism for $\rmU(1)_q\times\rmU(1)_p$ and $\rmU(1)_q\times\rmU(1)_{p+1}^\psi$ mixed anomalies used in this work. We will set the anomaly coefficient $c_\phi = 1$ in this appendix for clarity. The anomaly coefficient can be restored by multiplying all the relevant expressions with $c_\phi$. 

Gauge anomalies in quantum field theories can be classified using the anomaly inflow mechanism. Herein, one posits that an anomalous \tightmath{$(d+1)$}-dimensional system, described by a symmetry-non-invariant effective action $S$, can be coupled to a \tightmath{$(d+2)$}-dimensional bulk with a symmetry-non-invariant action $S_{\text{bulk}}$, so that the full theory described by
\begin{equation}
    S_{\text{tot}} = S_{\text{bulk}} + S,
\end{equation}
is symmetry-invariant. The bulk action $S_{\text{bulk}}$ has to be such that the it is gauge-invariant up to a boundary term. It must also be a topological theory so that there are no propagating degrees of freedom residing in the bulk. In light of these requirements, anomalies can be characterised by an exact \tightmath{$(d+3)$}-form \emph{anomaly polynomial} made out of the wedge-product of gauge field strengths. The bulk Lagrangian is then given by a Chern-Simons form $I_{\text{CS}}$ related to the anomaly polynomial as ${\cal P} = \df I_{\text{CS}}$, so that
\begin{equation}
    S_{\text{bulk}} = \int_{\text{bulk}} I_{\text{CS}}.
\end{equation}

The explicit form of the anomaly polynomial depends on the dimensionality of the spacetime and the symmetry group under consideration. For a system with an anomalous $\rmU(1)_q\times\rmU(1)_p$ higher-form symmetry, where 
\tightmath{$p+q=d-1$}, the anomaly polynomial is given by
\begin{equation}
    {\cal P} = (-)^q F \wedge \tilde F~.
\end{equation}
The anomaly polynomial results in the Chern-Simons form for the bulk Lagrangian
\begin{equation}
    I_{\text{CS}}
    = F \wedge \tilde A.
    \label{eq:CS-form}
\end{equation}
We have chosen the bulk Lagrangian to respect the U(1)$_q$ part of the symmetry and violate U(1)$_p$, but we can also choose the conventions to be the other way round.

We can parametrise the variations of the boundary part of the action $S$ as
\begin{align}
    \delta S
    &= \int \delta A \wedge \star J_{\text{cons}}
    + \delta \tilde A \wedge \star \tilde J \nn\\
    &\qquad\qquad
    + \ell\delta \Phi \wedge \star L
    + \tilde\ell\delta \tilde\Phi \wedge \star \tilde L.
\end{align}
Note that we have used the conserved current $J_{\text{cons}}$, also often called the ``consistent current'' in the anomaly literature. Since we have chosen to manifest the U(1)$_q$ symmetry in \cref{eq:CS-form}, the associated current $J_{\text{cons}}$ obtained by varying the boundary part of the action $S$ will be exactly conserved, but will not be gauge-invariant; see \cref{eq:cons-currents}. Whereas, the U(1)$_p$ current $\tilde J$ will be gauge-invariant but anomalous; see \cref{eq:gauge-invariant-currents}.
We have allowed for symmetry violation terms in line with our discussion in the main text. Depending on the application in mind, they can be switched off by turning of $\ell$ or $\tilde\ell$ accordingly. We can vary the total action $S_{\text{tot}}$ with respect to the background fields to yield
\begin{align}
    \delta S_{\text{tot}}
    &= 
    \int_{\text{bulk}} (-)^q \delta A\wedge \tilde F
    + (-1)^{pq+q}\,\delta \tilde A \wedge F \nn\\
    &\quad 
    + \int
    \delta A \wedge \star J
    + \delta \tilde A \wedge \star \tilde J \nn\\
    &\qquad\qquad
    + \ell\delta \Phi \wedge \star L
    + \tilde\ell\delta \tilde\Phi \wedge \star \tilde L,
\end{align}
where the boundary variations of the total action define the ``covariant current'' $J$ as
\begin{equation}
    J = J_{\text{cons}} + (-)^{pq+p+q} {\star\tilde A}. 
\end{equation}
Since the full theory is meant to be invariant under both U(1)$_q$ and U(1)$_p$ symmetries, we are led to the conservation laws given in \cref{eq:conservation_explicitbroken}, \eqref{eq:conservation_defect}, or \eqref{eq:conservation-Coulomb-symmetric}, depending on the structure of explicit symmetry breaking.

Let us note the form of the bulk action in thermal equilibrium in the context of our discussion in \cref{sec:appx-SF-Coulomb-eqb}. Using the definitions of the higher-form chemical potentials form \eqref{eq:qform-mu}, we can see that 
\begin{align}
    S_{\text{bulk}}^{\text{eqb}}
    &= \int_{\Sigma_\beta^{\text{bulk}}} 
    \iota_u I_{\text{CS}} \nn\\
    &= \int_{\Sigma_\beta^{\text{bulk}}}  
    (-1)^q \mu \wedge \tilde F
    + (-1)^{pq+q} \tilde\mu \wedge F \nn\\
    &\qquad
    - \int_{\Sigma_\beta}
    T_0 F \wedge \tilde\varphi - \mu \wedge \tilde A.
    \label{eq:Sbulk-eqb}
\end{align}
The bulk part in this expression is completely gauge-invariant. Therefore, in equilibrium, the anomaly in the respective quantum field theory can be generated by the boundary part of the bulk action, which is precisely what we need to cancel the anomaly-induced terms in \cref{eq:lagrange-eqb-spatial}.

The construction for the $\rmU(1)_q\times\rmU(1)_{p+1}^\psi$ anomaly in the pinned phase of a U(1)$_q$ pseudo-superfluid follows similarly. The anomaly polynomial is given as
\begin{equation}
    {\cal P} = (-)^q F \wedge \tilde F
    - \ell \Xi \wedge \tilde F_\psi,
\end{equation}
where we have used the $\rmU(1)_{p+1}^\psi$-invariant definition of $\tilde F$ from \cref{eq:new-tildeF}. The relative coefficient between the two terms can be fixed by requiring that ${\cal P}$ is closed. The anomaly polynomial leads to the Chern-Simons form 
\begin{align}
    I_{\text{CS}}
    &= F \wedge \tilde A
    + (-)^q \ell \Xi \wedge \tilde A_\psi.
\end{align}
We have, again, chosen the bulk Lagrangian to be invariant under U(1)$_q$ part of the symmetry group. 

In this case, we can parametrise the variations of the boundary action $S$ in terms of the ``consistent currents'' according to
\begin{align}
    \delta S
    &= \int \delta A \wedge \star J_{\text{cons}}
    + \delta \tilde A \wedge \star \tilde J \nn\\
    &\qquad\qquad
    + \ell\delta \Phi \wedge \star L_{\text{cons}}
    + \tilde\ell\delta \tilde A_\psi \wedge \star \tilde J_\psi.
\end{align}
Combining this with the variations of the Chern-Simons form, we can read off the equivalent expression for the total Lagrangian $S_{\text{tot}}$ to give
\begin{align}
    \delta S_{\text{tot}}
    &= \int_{\text{bulk}}
    (-)^q \delta A \wedge \tilde F
    + (-)^{pq+q} \delta \tilde A \wedge F \nn\\
    &\qquad\qquad
    - \ell\delta\Phi \wedge \tilde F_\psi
    + (-)^{pq+p+q} \delta\tilde A_\psi \wedge \ell\Xi
    \nn\\
    &\qquad
    + \int 
    \delta A \wedge \star J
    + \delta \tilde A \wedge \star \tilde J \nn\\
    &\qquad\qquad
    + \ell\delta \Phi \wedge \star L
    + \tilde\ell\delta \tilde A_\psi \wedge \star \tilde J_\psi,
\end{align}
where we have defined the covariant currents
\begin{align}
    J 
    &= J_{\text{cons}} + (-)^{pq+p+q} {\star\tilde A}, \nn\\
    L
    &= L_{\text{cons}}
    - (-)^{pq+q} {\star\tilde A_\psi}.
\end{align}
This yields the respective set of conservation equations given in \cref{eq:cons-Higgs}.

Finally, we record the form of the bulk action in thermal equilibrium. In this instance, we find
\begin{align}
    S_{\text{bulk}}^{\text{eqb}}
    &= \int_{\Sigma_\beta^{\text{bulk}}} 
    \iota_u I_{\text{CS}} \nn\\
    &= \int_{\Sigma_\beta^{\text{bulk}}}  
    (-)^q \mu \wedge \tilde F
    + (-)^{pq+q} \tilde\mu \wedge F
    \nn\\
    &\qquad\qquad
    - \mu_\ell \wedge \tilde F_\psi
    + (-)^{pq+p+q} \tilde\mu_\psi \wedge \ell\Xi \nn\\
    &\qquad
    - \int_{\Sigma_\beta}
    T_0 F \wedge \tilde\varphi - \mu \wedge \tilde A \nn\\
    &\qquad\qquad
    + (-)^q  T_0 \ell\Xi \wedge\tilde\phi_\psi
    - \mu_\ell \wedge \tilde A_\psi.
    \label{eq:Sbulk-eqb-Higgs}
\end{align}
Again, the bulk part in this expression is completely gauge-invariant and the anomalous boundary part is precisely what is needed to cancel the anomaly-induced terms in \cref{eq:lagrange-eqb-spatial-Higgs}.

\section{Retarded correlation functions}
\label{app:correlators}

In this appendix, we compute the retarded correlation functions of various higher-form densities and fluxes using our hydrodynamic model.

\subsection{Approximate higher-form fluids}

Let us start with the U(1)$_q$ fluid with an approximate temporally-spontaneously broken U(1)$_q$ global symmetry. Split into longitudinal and transverse sectors, the higher-form conservation equations looks like
\begin{align}
    -i\omega {n_\perp}
    + ik {j_\|}
    &= - \ell {j_\ell^\perp}, \nn\\
    i\omega{n}_\| 
    &= \ell j_\ell^\|, \nn\\
    i k {n_\|}
    &= \ell {n^\perp_\ell}.
    \label{eq:hydro-eqs-perpparr}
\end{align}
Note that the longitudinal defect density $n^\|_\ell$ is identically zero.
The constitutive relations take the form
\begin{align}
    j_\|
    &= 
    - \sigma \lb ik{\mu_\perp} - ik A_t^\perp -i\omega A_\|\rb 
    , \nn\\
    j_\perp
    &= i\omega\sigma  A_\perp, \nn\\
    \ell j_\ell^\|
    &= \Gamma \big( n_\| - \chi A_t^\|
    + i\omega\chi\Phi_\| \big)
    - \ell\sigma_\ell
    \lb ik \mu_\ell
    - ik \ell\Phi_t^\perp \rb,
    \nn\\
    \ell j_\ell^\perp
    &= \Gamma\big( n_\perp - \chi A_t^\perp 
    + i\omega\chi\Phi_\perp
    \big).
\end{align}
We can use the constitutive relations together with the conservation equations to find the classical expectation values of the (approximately) conserved densities and fluxes in terms of the background fields.

This procedure can be used to compute the retarded correlation functions of hydrodynamic densities and fluxes. Denoting operators by ${\cal O}$ and respective sources by $J$, the retarded correlation functions are defined as
\begin{equation}
    G^R_{{\cal O}_1{\cal O}_2} 
    = - \frac{\delta\langle{\cal O}_1\rangle}{\delta J_2}.
\end{equation}
If the microscopic description underlying the effective theory is invariant under time-reversal symmetry, the correlation functions satisfy the so-called Onsager reciprocal relations
\begin{equation}
    G^R_{{\cal O}_1{\cal O}_2}(\omega,k)
    = \eta^T_1 \eta^T_2 G^R_{{\cal O}_2{\cal O}_1}(\omega,-k),
    \label{eq:onsager}
\end{equation}
where $\eta^T$ are the time-reversal eigenvalues of various operators given in \cref{tab:CPT}.

Let us start with the transverse density sector, spanned by the transverse density $n_\perp$, longitudinal flux $j_\|$, and transverse defect flux $j_\ell^\perp$. There are three independent retarded correlation functions in this sector, given as
\begin{align}
    G_{n_\perp n_\perp}^R 
    &= \chi \lb
    \frac{i\omega}{i\omega - D_n k^2 - \Gamma}
    - 1 \rb, \nn\\
    G^R_{n_\perp j_\|}
    &= \frac{\omega k \sigma}{i\omega - D_n k^2 - \Gamma}, \nn\\
    G^R_{j_\|j_\|}
    &= -i\omega\sigma \lb 
    \frac{D_n k^2}
    {i\omega - D_n k^2 - \Gamma} 
    + 1 \rb. 
    % \nn\\
    %
    % G_{n_\perp j^\ell_\perp}^R 
    % = - G_{j^\ell_\perp n_\perp}^R 
    % &= \chi \frac{-i\omega\Gamma}{i\omega - D_n k^2 - \Gamma}, \nn\\
    % %
    % G_{j_\| j^\ell_\perp}^R 
    % = - G_{j^\ell_\perp j_\|}^R 
    % &= 
    % \chi \frac{-\omega k D_n\Gamma}
    % {i\omega - D_n k^2 - \Gamma}, \nn\\
    % %
    % G^R_{j^\ell_\perp j^\ell_\perp}
    % &= \chi \frac{-i\omega (i\omega - D_n k^2)\Gamma}
    % {i\omega - D_n k^2 - \Gamma}.
    \label{eq:correlators-fluid-perp}
\end{align}
The correlation functions involving the defect flux $j_\ell^\|$ can be obtained using the Ward identities \eqref{eq:hydro-eqs-perpparr}.
In the longitudinal density sector, spanned by the longitudinal density $n_\|$, transverse flux $j_\perp$, defect density $n^\perp_\ell$, and the longitudinal defect flux $j_\ell^\|$. We find density and flux correlators
\begin{align}
    G_{n_\| n_\|}^R
    &=- \frac{i\chi\Gamma}{\omega + ik^2 D_\| + i\Gamma}, \nn\\
    G^R_{j_\perp j_\perp} 
    &= 
    -i\omega\sigma.
    \label{eq:correlators-fluid-parr}
\end{align}
Note that $G^R_{n_\|j_\perp}$ is trivially zero by isotropy. 
The defect density and flux correlators can be read off trivially from here using the Gauss constraints in \eqref{eq:hydro-eqs-perpparr}. Note that if we turn off explicit symmetry breaking, the longitudinal charge density $n_\|$ identically vanishes and so does its correlator. 

\subsection{Higher-form pseudo-superfluids: relaxed phase}

For a U(1)$_q$ superfluid, the discussion is slightly more involved. Since the U(1)$_q$ symmetry is now anomalous, the conservation equations \eqref{eq:hydro-eqs-perpparr} become
\begin{align}
    -i\omega {n_\perp}
    + ik {j_\|}
    &= - i (-)^{p}  \tilde c_\phi 
    *_k\!\lb k \tilde A^\perp_t
    +  \omega \tilde A_\| \rb
    - \ell {j^\ell_\perp}, \nn\\
    i\omega{n}_\| 
    &= - i\omega \tilde c_\phi {*_k\tilde A_\perp}
    + \ell j^\ell_\|, \nn\\
    i k {n_\|}
    &= -ik\tilde c_\phi {*_k\tilde A_\perp} 
    + \ell {n_\ell^\perp},
    \label{eq:hydro-eqs-perpparr-Coulomb}
\end{align}
where $*_k = *(k/|k|\wedge \circ) = \star(u\wedge k/|k|\wedge\circ)$ is the Hodge duality operation transverse to the wavevector and fluid velocity. 
The constitutive relations take the form
\begin{align}
    j_\|
    &= 
    - (-)^{p}\tilde\lambda \tilde c_\phi \,
    {*_k\tilde\mu_\perp}
    - \sigma \big(
    ik{\mu_\perp} - ik A_t^\perp -i\omega A_\|\big) \nn\\
    &\qquad 
    - (-)^{pq+q}\tilde\ell\tilde\gamma_\times
    *_k\!\big( \tilde A_t^\perp - i\omega \tilde\Phi_\perp \big)
    , \nn\\
    j_\perp
    &= -\tilde\lambda\tilde c_\phi
    {*_k \tilde\mu_\|}
    + i\omega\sigma  A_\perp \nn\\
    &\qquad 
    - (-)^{pq+p+q}\tilde\gamma_\times\!
    *_k\!
    \lb ik {\tilde\mu_\ell^\perp} 
    - i\omega \tilde\ell \tilde\Phi_\|
    - ik \tilde\ell \tilde\Phi_t^\perp
    + \tilde\ell {\tilde A^\|_t} \rb
    , \nn\\
    \ell j_\ell^\|
    &= \Gamma \big( n_\| - \chi A_t^\|
    + i\omega\chi\Phi_\| \big)
    - \ell\sigma_\ell
    \big(ik \mu_\ell^\perp
    - ik \ell\Phi_t^\perp \big) \nn\\
    &\qquad
    + \ell\gamma_\times\,
    i\omega {*_k\tilde A_\perp},
    \nn\\
    \ell j_\ell^\perp
    &= \Gamma\big( n_\perp - \chi A_t^\perp 
    + i\omega\chi\Phi_\perp
    \big) \nn\\
    &\quad 
    + (-)^p \ell\gamma_\times\! \lb
    ik{*_k\tilde\mu_\perp} - ik{*_k\tilde A^\perp_t}
    - i\omega {*_k \tilde A_\|} \rb.
\end{align}
The conservation equations and constitutive relations in the U(1)$_p$ sector follow similarly by performing a tilde-conjugation. 

While the Onsager relations \eqref{eq:onsager} are satisfied for most correlation functions, some mixed correlators between the U(1)$_q$ and U(1)$_p$ sectors now violate the Onsager relations due to the mixed anomaly. We find
\begin{align}
    G^R_{j_\|\tilde n_\perp}
    &= (-)^{pq} G^R_{\tilde n_\perp j_\|}
    + (-)^p \tilde c_\phi \frac{*k}{|k|}, \nn\\
    G^R_{j_\perp\tilde n_\|}
    &= (-)^{pq+p+q+1} G^R_{\tilde n_\| j_\perp}
    + \tilde c_\phi \frac{*k}{k},
\end{align}
and similarly for the tilde-conjugates.

\begin{widetext}

Let us now look at the explicit correlation functions. The transverse density sector correlation functions in 
\cref{eq:correlators-fluid-perp} modify for a superfluid to
\begin{align}
    G_{n_\perp n_\perp}^R 
    &= \chi \lb
    \frac{i\omega
    \big( i\omega - \tilde D_n k^2 - \tilde\Gamma \big)}{
    \big( i\omega - D_n k^2 - \Gamma \big)
    \big( i\omega - \tilde D_n k^2 - \tilde\Gamma \big)
    + v_\perp^2 k^2}
    - 1
    \rb, \nn\\
    G^R_{n_\perp j_\|}
    &= \omega k \frac{
    \sigma
    \big( i\omega - \tilde D_n k^2 - \tilde\Gamma \big)
    - (1+\tau)\chi v_\perp^2
    }
    {\big( i\omega - D_n k^2 - \Gamma \big)
    \big( i\omega - \tilde D_n k^2 - \tilde\Gamma \big)
    + v_\perp^2 k^2}, \nn\\
    G^R_{j_\|j_\|}
    &= -i\omega
    \frac{
    ( i\omega - \Gamma) \Big(\sigma
    \big( i\omega - \tilde D_n k^2 - \tilde\Gamma \big)
    - (1+\tau)^2\chi v_\perp^2 \Big)
    + \tau^2 \sigma v_\perp^2 k^2
    }
    {\big( i\omega - D_n k^2 - \Gamma \big)
    \big( i\omega - \tilde D_n k^2 - \tilde\Gamma \big)
    + v_\perp^2 k^2}.
    \label{eq:correlators-fluid-perp-Coulomb}
\end{align}
Here we have defined combinations of renormalisation coefficients for clarity
\begin{equation}
    \tau = \frac{\tilde\lambda}{\lambda_s} - 1
    = \frac{1-\lambda}{\lambda_s}, \qquad 
    \tilde\tau = \frac{\lambda}{\lambda_s} - 1
    = \frac{1-\tilde\lambda}{\lambda_s}.
\end{equation}
They are zero in the absence of the U(1)$_q$ defect coefficient $\gamma_\times$ and the U(1)$_p$ defect coefficient $\tilde\gamma_\times$ respectively. Note that $\tau+\tilde\tau = 1/\lambda_s - 1$.
The correlation functions for $\tilde n_\perp$, $\tilde\jmath_\|$, and $\tilde\jmath^\ell_\perp$ in the U(1)$_p$ sector can be obtained similarly by performing a tilde-conjugate of these expressions. The longitudinal density sector correlation functions in \cref{eq:correlators-fluid-parr} modify to
\begin{align}
    G_{n_\| n_\|}^R
    &= \frac{\chi\Gamma}{i\omega - k^2 D_\ell - \Gamma}, \nn\\
    G^R_{j_\perp j_\perp} 
    &= \frac{\lambda_s^2 c_\phi^2}{\tilde\chi}
    \frac{i\omega (1+\tau)^2
    + i\omega \tilde\tau^2 k^2/\tilde k_0^2
    - \tilde D_\ell k^2/\lambda_s^2
    }
    {i\omega -\tilde D_\ell k^2 - \tilde\Gamma} 
    -i\omega\sigma.
    \label{eq:correlators-fluid-parr-Coulomb}
\end{align}
Similar expressions hold in the U(1)$_p$ sector.

In addition to these, there are now also mixed correlation functions between the U(1)$_q$ and U(1)$_p$ sectors due to the mixed anomaly. In the transverse density sector, we find
\begin{align}
    \frac{(-)^p}{\lambda_s\tilde c_\phi} G_{n_\perp \tilde n_\perp}^R 
    &= 
    \frac{\omega k}{
    \big( i\omega - D_n k^2 - \Gamma \big)
    \big( i\omega - \tilde D_n k^2 - \tilde\Gamma \big)
    + v_\perp^2 k^2}
    \frac{*k}{|k|}
    , \nn\\
    \frac{(-)^p}{\lambda_s\tilde c_\phi}
    G^R_{n_\perp\tilde\jmath_\|}
    &= -
    \frac{i\omega\tilde D_n k^2
    + i\omega(1+\tilde\tau)
    \big( i\omega - \tilde D_n k^2 - \tilde\Gamma \big)
    }{\big( i\omega - D_n k^2 - \Gamma \big)
    \big( i\omega - \tilde D_n k^2 - \tilde\Gamma \big)
    + v_\perp^2 k^2} 
    \frac{*k}{|k|}, \nn\\
    \frac{(-)^p}{\lambda_s\tilde c_\phi}
    G^R_{j_\|\tilde\jmath_\|}
    &= \omega k
    \frac{(\tau\tilde\tau+1/\lambda_s) v_\perp^2
       - (1+\tau)\tilde D_n (i\omega - \Gamma)
     - D_n (1+\tilde\tau)(i\omega - \tilde\Gamma) 
    + D_n\tilde D_n k^2/\lambda_s
    }
    {\big( i\omega - D_n k^2 - \Gamma \big)
    \big( i\omega - \tilde D_n k^2 - \tilde\Gamma \big)
    + v_\perp^2 k^2}
    \frac{*k}{|k|}.
\end{align}
Finally, in the longitudinal density sector, we have
\begin{align}
    \frac{1}{\lambda_s\tilde c_\phi} G^R_{n_\|\tilde\jmath_\perp}
    &= 
    \frac{i\omega(1+\tilde\tau) - D_\ell k^2/\lambda_s}
    {i\omega - D_\ell k^2 - \Gamma} \frac{*k}{|k|}.
\end{align}
It is trivial to see that all the mixed correlators vanish if we turn off the anomaly coefficient $c_\phi$.
\end{widetext}

\subsection{Higher-form pseudo-superfluids: pinned phase}

We finally look at the correlation functions of a U(1)$_q$ pseudo-superfluid in the pinned phase. The conservation equations in the U(1)$_q$  sector modify from \cref{eq:hydro-eqs-perpparr-Coulomb} to include the new $\tilde A_\psi$ source terms
\begin{align}
    \omega {n_\perp}
    - k {j_\|}
    &= (-)^{p}  \tilde c_\phi 
    *_k\!\lb k {\tilde A^\perp_t}
    +  \omega {\tilde A_\|} 
    + i\ell {\tilde A^\|_{\psi,t}}
    \rb
    - i\ell {j_\ell^\perp}, \nn\\
    \omega{n}_\| 
    &= - \tilde c_\phi
    {*_k\!\lb\omega\tilde A_\perp
    + i\ell \tilde A^\perp_{\psi,t}
    \rb}
    - i\ell j_\ell^\|, \nn\\
    k {n_\|}
    &= - \tilde c_\phi {*_k \lb k\tilde A_\perp 
    - i\ell\tilde A_\psi^\|
    \rb} 
    - i\ell {n^\perp_\ell}, \nn\\
    n^\|_\ell
    &= (-)^{p}\tilde c_\phi {*_k \tilde A_\psi^\perp}.
\end{align}
The associated constitutive relations take the form
\begin{align}
    j_\|
    &= 
    (-)^{p+1}\tilde c_\phi \,
    {*_k\tilde\mu_\perp}
    - \sigma \big(
    ik{\mu_\perp} - ik A_t^\perp -i\omega A_\|\big), \nn\\
    j_\perp
    &= - \tilde c_\phi
    {*_k \tilde\mu_\|}
    + i\omega\sigma  A_\perp, \nn\\
    \ell j^\ell_\|
    &= 
    - \ell \lambda \tilde c_\phi {*_k\tilde\mu^\perp_\psi}
    + \Gamma \big( n_\| - \chi A_t^\|
    + i\omega\chi\Phi_\| \big) \nn\\
    &\qquad
    - \ell\sigma_\ell
    \big(ik \mu_\ell
    - ik \ell\Phi_t^\perp \big) \nn\\
    &\qquad
    + \ell\gamma_\times\!
    *_k\!\lb i\omega{\tilde A_\perp }
    - \ell {\tilde A_{\psi,t}^\perp}
    \rb,
    \nn\\
    \ell j^\ell_\perp
    &= 
    \ell (-)^p \lambda \tilde c_\phi {*_k\tilde\mu^\|_\psi}
    + \Gamma\big( n_\perp - \chi A_t^\perp 
    + i\omega\chi\Phi_\perp
    \big) \nn\\
    &\quad 
    + (-)^p \ell\gamma_\times\! *_k\!\lb
    ik{\tilde\mu_\perp} - ik{\tilde A^\perp_t}
    - i\omega {\tilde A_\|}
    + \ell {\tilde A_{\psi,t}^\|}
    \rb.
\end{align}
The conservation equations and constitutive relations in the U(1)$_p$ sector can be obtained by performing the $q\leftrightarrow p+1$ self-duality transformation of the pinned phase.

\begin{widetext}

The correlators in the transverse U(1)$_q$ density sector in the pinned phase modify from their relaxed phase expressions in \cref{eq:correlators-fluid-perp-Coulomb} to 
\begin{align}
    G^R_{n_\perp n_\perp}
    &= \chi
    \lb \frac{
    i\omega
    \big(i\omega - \tilde D_n k^2 - \tilde\Omega\big) 
    }
    {\big(i\omega - D_n k^2 - \Gamma\big)
    \big(i\omega - \tilde D_n k^2 - \tilde\Omega\big)
    + \omega_0^2 + v_\perp^2k^2} 
    - 1
    \rb, \nn\\
    G^R_{n_\perp j_\|}
    &= \omega k 
    \frac{
    \sigma (i\omega - \tilde D_n k^2 - \tilde\Omega)
    - (1+\tau)\chi v_\perp^2
    }
    {{\big(i\omega - D_n k^2 - \Gamma\big)
    \big(i\omega - \tilde D_n k^2 - \tilde\Omega\big)
    + \omega_0^2 + v_\perp^2k^2}}, \nn\\
    G^R_{j_\|j_\|}
    &=
    - i\omega \frac{
    (i\omega - \Gamma)
    \lb \sigma
    (i\omega - \tilde D_n k^2 -\tilde\Omega)
    - (1+\tau)^2\chi v_\perp^2
    \rb
    + \tau^2\sigma v_\perp^2 k^2
    }
    {\big(i\omega - D_n k^2 - \Gamma\big)
    \big(i\omega - \tilde D_n k^2 - \tilde\Omega\big)
    + \omega_0^2
    + v_\perp^2k^2} \nn\\
    &\qquad 
    - \chi\omega_0^2
    \frac{ i\omega D_n
    + (1+\tau)^2 \big(i\omega - D_n k^2 - \Gamma\big) \tilde D_n
    - (1+\tau)^2 v_\perp^2
    }
    {\big(i\omega - D_n k^2 - \Gamma\big)
    \big(i\omega - \tilde D_n k^2 - \tilde\Omega\big)
    + \omega_0^2
    + v_\perp^2k^2}~.
    \label{eq:correlators-fluid-perp-Higgs}
\end{align}
We see that these expressions are almost identical to the relaxed phase correlators in the presence of vortices (magnetic monopoles) in \cref{eq:correlators-fluid-perp-Coulomb}, except that the vortex-induced relaxation $\tilde\Gamma$ gets exchanged by pinning-induced relaxation $\tilde\Omega$ and the pole has an additional $\omega_0^2$ pinning term. The flux correlator also gets an entirely new term proportional to $\omega_0^2$. Since the pinned phase is not self-dual in $q\leftrightarrow p$, the correlators in the transverse U(1)$_p$ density sector take a qualitatively different form
\begin{align}
    G^R_{\tilde n_\perp \tilde n_\perp}
    &= \tilde\chi
    \lb \frac{
    \big(i\omega - \tilde\Omega \big)
    \big(i\omega - D_n k^2 - \Gamma\big)
    + \omega_0^2}
    {\big(i\omega - D_n k^2 - \Gamma\big)
    \big(i\omega - \tilde D_n k^2 - \tilde\Omega\big)
    + \omega_0^2 + v_\perp^2k^2}
    - 1
    \rb, \nn\\
    G^R_{\tilde n_\perp \tilde\jmath_\|}
    &= \omega k\frac{
    \tilde\sigma (i\omega - D_n k^2 - \Gamma)
    - \tilde\chi v_\perp^2}
    {\big(i\omega - D_n k^2 - \Gamma\big)
    \big(i\omega - \tilde D_n k^2 - \tilde\Omega\big)
    + \omega_0^2 + v_\perp^2k^2}, \nn\\
    G^R_{\tilde\jmath_\| \tilde\jmath_\|}
    &= -i\omega\frac{
    i\omega \lb
    \tilde\sigma (i\omega - D_n k^2 - \Gamma)
    - \tilde\chi v_\perp^2 \rb}
    {\big(i\omega - D_n k^2 - \Gamma\big)
    \big(i\omega - \tilde D_n k^2 - \tilde\Omega\big)
    + \omega_0^2 + v_\perp^2k^2}.
    \label{eq:correlators-fluid-perp-Higgs-tilde}
\end{align}
Physically, the qualitative differences between \cref{eq:correlators-fluid-perp-Higgs,eq:correlators-fluid-perp-Higgs-tilde} can be ascribed to the fact that U(1)$_p$ charges are conserved in the pinned phase while the U(1)$_q$ charges are not.

Next, in the longitudinal U(1)$_q$ density sector, we find the correlation functions
\begin{align}
    G^R_{n_\|n_\|}
    &= \frac{
    \chi\Gamma (i\omega - \tilde D_\psi k^2 - \tilde\Omega)
    - \chi\omega_0^2}
    {\big(i\omega - D_\ell k^2 - \Gamma\big)
    \big(i\omega - \tilde D_\psi k^2 - \tilde\Omega \big)
    + \omega_0^2 + v_\|^2k^2}, \nn\\
    G^R_{j_\perp j_\perp}
    &= \frac{c_\phi^2}{\tilde\chi} - i\omega\sigma.
\end{align}
These should be compared to \cref{eq:correlators-fluid-parr-Coulomb} from the relaxed phase. On the other hand, in the longitudinal U(1)$_p$ density sector we find
\begin{align}
    G^R_{\tilde n_\|\tilde n_\|} 
    &= 0, \nn\\
    G^R_{\tilde\jmath_\perp\tilde\jmath_\perp}
    &= \frac{\lambda^2c_\phi^2}{\chi}
    \frac{
    (i\omega - \tilde D_\psi k^2) i\omega(1+\tau^2k^2/k_0^2)
    - (1+\tau)^2(i\omega - \tilde D_\psi k^2 - \tilde\Omega)D_\ell k^2
    + (1+\tau)^2v_\|^2 k^2}
    {\big(i\omega - D_\ell k^2 - \Gamma\big)
    \big(i\omega - \tilde D_\psi k^2 - \tilde\Omega \big)
    + \omega_0^2 + v_\|^2k^2} \nn\\
    &\qquad 
    - i\omega \tilde\sigma  \frac{
    (i\omega - \tilde D_\psi k^2) 
    (i\omega - D_\ell k^2 - \Gamma) 
    + (1+\tau)^2v_\|^2 k^2
    }
    {\big(i\omega - D_\ell k^2 - \Gamma\big)
    \big(i\omega - \tilde D_\psi k^2 - \tilde\Omega \big)
    + \omega_0^2 + v_\|^2k^2}~.
\end{align}
Note that U(1)$_p$ charges are exactly conserved (up to anomaly), so $\tilde n_\|$ is not dynamical. 

Finally we have the mixed correlators; in the transverse density sector
\begin{align}
    \frac{(-)^p}{\lambda \tilde c_\phi}
    G^R_{n_\perp \tilde n_\perp}
    &= \frac{\omega  k}
    {\big(i\omega - D_n k^2 - \Gamma\big)
    \big(i\omega - \tilde D_n k^2 - \tilde\Omega\big)
    + \omega_0^2 + v_\perp^2k^2}
    \frac{*k}{|k|}, \nn\\
    \frac{(-)^p}{\lambda \tilde c_\phi}
    G^R_{n_\perp \tilde\jmath_\|}
    &= \frac{\omega^2}
    {\big(i\omega - D_n k^2 - \Gamma\big)
    \big(i\omega - \tilde D_n k^2 - \tilde\Omega\big)
    + \omega_0^2 + v_\perp^2k^2}
    \frac{*k}{|k|}, \nn\\
    \frac{(-)^p}{\lambda \tilde c_\phi}
    G^R_{j_\| \tilde n_\perp}
    &= - k^2 \frac{
    i\omega D_n
    +
    (i\omega - D_n k^2 - \Gamma) \tilde D_n/\lambda 
    - v_\perp^2/\lambda}
    {\big(i\omega - D_n k^2 - \Gamma\big)
    \big(i\omega - \tilde D_n k^2 - \tilde\Omega\big)
    + \omega_0^2 + v_\perp^2k^2}
    \frac{*k}{|k|}, \nn\\
    \frac{(-)^p}{\lambda\tilde c_\phi}G^R_{j_\|\tilde\jmath_\|}
    &=
    - \omega k \frac{
    i\omega D_n
    +
    (i\omega - D_n k^2 - \Gamma) \tilde D_n/\lambda 
    - v_\perp^2/\lambda}
    {\big(i\omega - D_n k^2 - \Gamma\big)
    \big(i\omega - \tilde D_n k^2 - \tilde\Omega\big)
    + \omega_0^2 + v_\perp^2k^2}
    \frac{*k}{|k|}, 
\end{align}
and in the longitudinal density sector
\begin{align}
    \frac{1}{\lambda\tilde c_\phi} G^R_{n_\| \tilde\jmath_\perp}
    &= \frac{i\omega
    (i\omega - \tilde D_\psi k^2)
    - (i\omega - \tilde D_\psi k^2 - \tilde\Omega)
    D_\ell k^2/\lambda
    + v_\|^2 k^2/\lambda}
    {\big(i\omega - D_\ell k^2 - \Gamma\big)
    \big(i\omega - \tilde D_\psi k^2 - \tilde\Omega \big)
    + \omega_0^2 + v_\|^2k^2}
    \frac{*k}{|k|}, \nn\\
    \frac{1}{\tilde c_\phi}G^R_{j_\perp \tilde n_\|}
    &= 0.
\end{align}
All other correlators not reported here can be obtained by $q\leftrightarrow p+1$ self-duality of the pinned phase.
    
\end{widetext}

\bibliography{mySpiresCollaboration_JAAJ}

\end{document}